\documentclass[
amsmath,
amssymb,
aps, 
pra,
twocolumn, 
groupedaddress,
nobalancelastpage,
floatfix]
{revtex4-2}
\usepackage{mathtools}
\usepackage[colorlinks]{hyperref}
\usepackage{lipsum}
\usepackage{bbold}
\usepackage{bm}
\usepackage{dsfont}
\usepackage{enumerate}
\usepackage{graphicx}
\usepackage{float}
\usepackage {xcolor}
\usepackage{cleveref}
\usepackage{amsmath}
\newcommand{\orcid}[1]{%
  ~\href{https://orcid.org/#1}{\includegraphics[width=8pt]{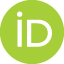}}%
  }

\DeclareMathOperator{\e}{e}					% Exponential e
\newcommand{\nullarg}{{}\cdot{}}

\renewcommand{\vec}[1]{\bm{#1}}	% Bold and upright vectors without arrow

\DeclarePairedDelimiterX{\mean}[1]{\langle}{\rangle}{
	\ifblank{#1}{\nullarg}{#1}
}
\DeclarePairedDelimiterX{\abs}[1]{\lvert}{\rvert}{
	\ifblank{#1}{\nullarg}{#1}
}
\DeclarePairedDelimiterX{\norm}[1]{\lVert}{\rVert}{
	\ifblank{#1}{\nullarg}{#1}
}
\DeclarePairedDelimiterX{\bra}[1]{\langle}{\rvert}{#1}

\DeclarePairedDelimiterX{\ket}[1]{\lvert}{\rangle}{#1}
\DeclarePairedDelimiterX{\braket}[2]{\langle}{\rangle}{
	\ifblank{#1}{\nullarg}{#1} \delimsize\vert \ifblank{#2}{\nullarg}{#2}
}
\DeclarePairedDelimiterX{\sandwich}[3]{\langle}{\rangle}{
	\ifblank{#1}{\nullarg}{#1} \delimsize\vert \ifblank{#2}{\nullarg}{#2} \delimsize\vert \ifblank{#3}{\nullarg}{#3}
}
\DeclarePairedDelimiterX{\inner}[2]{\langle}{\rangle}{
	\ifblank{#1}{\nullarg}{#1} , \ifblank{#2}{\nullarg}{#2}
}

\begin{document}
	
	\title{Electron-correlation induced nonclassicallity of light from high-harmonic generation}

	\author{Christian Saugbjerg Lange\orcid{0009-0007-8698-8919}}
	\author{Thomas Hansen\orcid{0000-0003-0990-2732}}
	\author{Lars Bojer Madsen\orcid{0000-0001-7403-2070}}
	\affiliation{Department of Physics and Astronomy, Aarhus University, Ny Munkegade 120, DK-8000 Aarhus C, Denmark}
	\date{\today}
	\begin{abstract}
	We study the effect of electron-electron correlations on the quantum state of the light emitted from high-harmonic generation (HHG). The quantum state of the emitted light is obtained by using a fully quantum mechanical description of both the optical modes as well as the electronic system. This is different from the usual semiclassical description of HHG, which only treats the electronic target system quantum mechanically. Using the generic Fermi-Hubbard model, the strength of the electron-electron correlation can be treated as a parameter enabling us to investigate the two limiting cases of a completely uncorrelated phase and a correlated Mott-insulating phase. In the completely uncorrelated phase, the model reduces to a single-band tight-binding model in which only intraband currents contribute to the spectrum. In this limit, we analytically find that the emitted light is in a classical coherent state. In the Mott-insulating phase, a consideration of the photon statistics and squeezing of the emitted photonic state shows that the inter-Hubbard-subband current generates nonclassical light. In this sense, we show that electron-electron correlation can induce the generation of nonclassical states of light.
	\end{abstract}
	\maketitle
	%%% Introduction %%%
	\section{Introduction} \label{sec:introduction}
	High-harmonic generation (HHG) is a well-studied process in which a system (atom, molecule, solid) is driven by an intense laser field resulting in the emission of higher harmonics at integer values times the frequency of the laser.
	% \cite{Eden2004}. 
	HHG has been observed in numerous experiments and has led to the field of attosecond physics \cite{Krausz2009} that was awarded the Nobel Prize in physics in 2023 \cite{LHuillier1988, Agostini2001, Krausz2001}. The experimental progress has opened the door to study electron dynamics at their natural times scales 
	%\cite{Li2008, Lein2003, Schubert2014} 
	as well as using the HHG spectrum to study the generating process within the HHG sample \cite{Krausz2009}. HHG in atomic and molecular gases is typically rationalized in terms of the three-step model \cite{Schafer1993, Corkum1993, Lewenstein1994}, with ionization, propagation and recombination and associated emission of light as essential elements. HHG in solids \cite{Ghimire2011, Ghimire2019, Goulielmakis2022} can often be understood as originating from two different but coupled kinds of currents, namely the inter- and intraband currents. Similarly to HHG in atoms, the interband current can be explained semiclassically by the three-step-model for HHG in bandgap materials \cite{Golde2008, Vampa2014, Vampa2015, Vampa2017}.
	%[Golde, D., Meier, T. and Koch, S. W. High harmonics generated in semiconductor nanostructures by the coupled dynamics of optical inter- and intraband excitations. Phys. Rev. B 77, 075330 (2008), OriginalVampaPRL2014,Vampa2015, VampaandBrabecJPBreview]
	%\cite{CorkumLewensteinHHG1994, Krause1992, Vampa2015}.  
	In this model, (i) an electron is promoted to the conduction band due to the interaction with the driving field leaving a hole in the valance band. (ii) Both the electron and the hole are accelerated in their respective bands by the strong laser field. (iii) The electron and hole can recombine resulting in the emission of light. This process typically occurs in semiconductors with a valence band and at least one conduction band which is why such a process cannot be included in one-band models. Intraband currents, on the other hand, can exist also in one-band models and emerge from the acceleration of an electron in the non-parabolic band of the solid. The two mechanisms are intrinsically coupled as the intraband mechanism is the second step in the three-step model for interband current generation.
	
	Though the above mentioned physical pictures have allowed for accurate descriptions of the emitted photonic spectrum and have found a wide range of applications, they are not entirely complete as they cannot account for a possible quantum mechanical nature of the emitted light. In recent years, much attention has been given to the inclusion of a quantized electromagnetic field in the context of HHG both theoretically \cite{Bogastkaya2017, Gombkoto2021, KaminerNatCom2020, KaminerManyBody2023, StammerLewenstein_tutorial_Quantum_state_engineering,Riveradean2023entanglementHHG_SolidState, GonoskovGrafe2022nonclassical, Tzur2023_photon_statistics_force, GolarchKaminerHHGWithQuantumLight2022, Stammer2023entanglement, Yangaliev2020_OlegToltikhin, Stammer2022_mmt_paper, Tzur2023generation_of_squeezed_HHG, LewensteinCatState2021} and experimentally \cite{Tsatrafyllis2017, Tsatrafyllis2019, LewensteinCatState2021, RiveraDean2022}, merging the fields of strong-field physics and quantum optics. In this fully quantum mechanical setting it was found that nonclassical states of light can be generated when including transitions between the initial and different final electronic states in atomic gasses, e.g., in a gas of helium atoms \cite{KaminerNatCom2020}. It was also found that a nonclassical cat-like state, a coherent state superposition (CSS), can be generated if one performs a conditional measurement subsequent to the HHG process as shown for both atomic gasses \cite{LewensteinCatState2021,RiveraDean2022, StammerLewenstein_tutorial_Quantum_state_engineering, Stammer2022_mmt_paper} and solids \cite{Riveradean2023entanglementHHG_SolidState, GonoskovGrafe2022nonclassical}. Furthermore, there has been investigations in the direction of a nonclassical driving field \cite{Tzur2023_photon_statistics_force, GolarchKaminerHHGWithQuantumLight2022, Tzur2023generation_of_squeezed_HHG} which is also not possible within a conventional semiclassical framework. Investigating the quantum features of the emitted HHG light is not only of interest to fundamental research and understanding of the HHG process itself but is also important with regards to quantum information science, as a fully quantized theory could enable HHG to be a feasible way to reliably create nonclassical states of light (e.g., squeezed states, Fock states, CSS \cite{Ourjoumtsev2006, Ourjoumtsev2007, Zavatta2004}). These could serve as a great resource in quantum technology \cite{Acin2018_QTRoadmap, DeutschSecondQuantumRevolution} and quantum metrology \cite{Polino2020, Barbieri2022} linking the fields of attosecond science and quantum information \cite{StammerLewenstein_tutorial_Quantum_state_engineering}.
	
	In parallel with these developments, there has recently been an increasing interest in HHG in correlated materials, that is, materials with a beyond mean-field electron-electron repulsion. These materials, including cuprates and high-temparature superconductors, are of great interest and they can lead to a signal enhancement for certain harmonics \cite{Silva2018, Murakami2018, Murakami2018_2,Hansen22,Hansen22_2,  Murakami2021}. HHG is hence relevant as a spectroscopic tool to resolve the dynamics of electrons in correlated materials. Mott-insulators in particular have recently been investigated both theoretically \cite{Silva2018, Murakami2018, Murakami2018_2,Murakami2021, Lysne2020, Tancogne-Dejean2018, Imai2020, Chinzei2020, Orthodoxou2021, Shao2022, Hansen22, Hansen22_2, Masur2022, Udono22, Murakami2022} and experimentally \cite{Uchida2022, Granas2022, Bionta2021}, although, not with quantum optical considerations. Generally, a correlated quantum system exhibits nonclassical behavior motivating the use of correlations to generate nonclassical light in HHG. This is also indicated by recent theoretical studies where it has been shown that by preparing a gas of atoms in a highly correlated (superradiant) state, nonclassical light was emitted from HHG \cite{KaminerManyBody2023}. It was also shown that including correlations between atoms can generate entangled and squeezed light \cite{Stammer2023entanglement} or entangled photon pairs \cite{Sloan2023entangling}. In this paper, we focus on correlated materials and ask the question of how electron-electron correlations affect the generated photonic state. Specifically, we ask how the HHG spectrum in a fully quantum mechanical calculation differs from semiclassical calculations when including electron correlations and how the presence of these correlations affect the photon statistics and squeezing of the emitted light; the latter two of which would not be possible to address in a usual semiclassical HHG setting.
	
	To capture generic effects of electron correlation we use the prototypical Fermi-Hubbard model \cite{Hubbard_Essler}. This model has been shown to capture important physical properties of real materials \cite{Lee06, Imada98}. In this model the on-site electron-electron is included via the so called Hubbard $U$-term and the correlation strength can thus be treated as a changeable parameter allowing us to study how the quantum mechanical nature of the generated light changes when increasing the correlation strength, $U$. This allows us to study both the uncorrelated phase and the highly correlated Mott-insulating phase where the electrons are highly real-space localized with only one electron per site as doubly occupied sites are energetically unfavorable.
	
	The paper is organized as follows. In Sec. \ref{sec:theory} the general fully quantum mechanical theory and the Fermi-Hubbard model are presented as well as the measures of interest. Then in Sec. \ref{sec:reuslts} we present the results for the two limiting cases of an uncorrelated phase and a Mott-insulating phase followed by a discussion in Sec. \ref{sec:discussion} before summarizing and concluding in Sec. \ref{sec:conclusion}.  In the appendices further details on the derivation and application of the formalism are given. Throughout this paper atomic units ($\hbar = m_e = 4 \pi \epsilon_0 = -e = 1 $) are used unless explicitly stated otherwise.

	%%% Theory %%%
	
	\section{Theory} \label{sec:theory}
	
	\subsection{Quantum optical description of HHG}
	In this section, we describe a fully quantum mechanical framework that includes the quantum optical state of the light emitted from an electronic system driven by an intense time-dependent electromagnetic field. A similar approach is found in Refs. \cite{KaminerNatCom2020, StammerLewenstein_tutorial_Quantum_state_engineering}. These theories go beyond usual semiclassical HHG theory by considering the quantum state of both the emitted light and the coherent state of the driving laser. This quantum optical calculation consists of three parts. In the first step, the state of the laser is transformed into vacuum \cite{Mollow1975, Cohen_Tannoudji1998_atomphoton}. This in consequence separates the vector potential into a classical and a quantum part and  is accompanied by a transformation of the  time-dependent Schrödinger equation (TDSE). In the second step, the TDSE is solved for electrons driven only by the classical field. This is only possible due to the transformation performed in the first step and enables one to use conventional TDSE solvers as only the classical field is considered. Here we specifically calculate all transition currents and not only the time dependent expectation value of the current. The third and final step is to integrate an equation of motion for the photonic state, which is coupled to the current generated by the response of the electrons to the classical part of the vector potential. From the photonic degrees of freedom one can calculate all expectation values of interest. This protocol will now be explained in greater detail.
	
	We investigate a system driven by a multimode coherent state laser, $\ket{\psi_{laser}(t)} = \otimes_{\vec{k}, \sigma} \ket{\alpha_{\vec{k} \sigma}  \e^{-i \omega_k t} }$ where $\omega_k = c \lvert \vec{k} \rvert$ is the angular frequency with the wave number $\vec{k}$, $c$ being the speed of light in vacuum, and $\sigma$ the polarization. The distribution of the complex coherent state parameters $\{ \alpha_{\vec{k} \sigma} \}$ determines the properties of the field. We note that the coherent state amplitude vanishes for modes far away from the laser mode, i.e., $\alpha_{\vec{k} \sigma} = 0$ for $\vec{k}, \sigma \gg \vec{k}_L, \sigma_L$ where $\vec{k}_L, \sigma_L$ is the wave number and polarization of the laser mode, respectively.
	
	We consider a case, where, under field-free conditions, $N$ electrons move subject to an electrostatic potential $\hat{U}$ generated by other electrons and static nuclei. The state of the combined electronic and photonic system, denoted by $\ket{\Psi(t)}$, satisfies the TDSE
	\begin{equation}
		i \dfrac{\partial}{\partial t} \ket{\Psi(t)} = \hat{H} \ket{\Psi(t)}, \label{eq:TDSE_full_system}
	\end{equation}
	where the full Hamiltonian is given by $\hat{H} = \dfrac{1}{2} \sum_{j=1}^N (\hat{\vec{p}}_j+ \vec{A})^2 + \hat{U} + \hat{H}_F$, where $\hat{\vec{p}}_j $ is momentum operator for electron $j$, $\hat{U}$ is the potential accounting for the Coulomb interaction between the particles, 
	\begin{equation}
		\hat{\vec{A}} =  \sum_{\vec{k}, \sigma}  \dfrac{g_0}{\sqrt{\omega_k}} \big( \hat{\vec{e}}_\sigma \hat{a}_{\vec{k}, \sigma} \e^{i \vec{k} \cdot \vec{r}} +  \hat{\vec{e}}_\sigma^* \hat{a}^\dagger_{\vec{k}, \sigma} \e^{-i \vec{k} \cdot \vec{r}} \big) \label{eq:A_quantized}
	\end{equation}
	is the quantized vector potential, and 
	\begin{equation}
		\hat{H}_F = \sum_{\vec{k}, \sigma}  \omega_k \hat{a}_{\vec{k}, \sigma}^\dagger \hat{a}_{\vec{k}, \sigma} \label{eq:H_free_field}
	\end{equation}
	is the Hamiltonian for the free photonic field.	In Eqs. (\ref{eq:A_quantized}) and (\ref{eq:H_free_field}) the effective coupling strength is denoted $g_0 = \sqrt{2 \pi/V}$ with quantization volume $V$, $\vec{r}$ is the position, $\Sigma_{\vec{k}, \sigma}$ is the sum over all photon momenta and polarizations with photon unit vector $\hat{\vec{e}}_\sigma$, and $\hat{a}_{\vec{k}, \sigma}$ and $ \hat{a}^\dagger_{\vec{k}, \sigma}$ are the photonic annihilation and creations operators, respectively. 
	In order to solve Eq. (\ref{eq:TDSE_full_system}) we follow the same procedure as in Refs. \cite{KaminerNatCom2020, StammerLewenstein_tutorial_Quantum_state_engineering} for a quantized electromagnetic field. 
	The first step follows a transformation originally introduced by Ref. \cite{Mollow1975} and later used in the context of HHG \cite{KaminerNatCom2020, LewensteinCatState2021, Riveradean2023entanglementHHG_SolidState, StammerLewenstein_tutorial_Quantum_state_engineering, RiveraDean2022} which is to transform away the laser field by considering the time-dependent unitary displacement operator
%	\begin{equation}
%		\hat{D}(t) = \otimes_{\vec{k}, \sigma} \exp( \alpha_{\vec{k} \sigma} \e^{-i\omega_k t} \hat{a}_{\vec{k}, \sigma}^\dagger - \alpha_{\vec{k} \sigma}^* \e^{i \omega_k t} \hat{a}_{\vec{k}, \sigma})
%	\end{equation}
	\begin{equation}
		\hat{D}(t) = \otimes_{\vec{k}, \sigma} \hat{\mathcal{D}}(\alpha_{\vec{k} \sigma}(t)) \label{eq:displacement_unitary}
	\end{equation}
	with
	\begin{equation}
		\hat{\mathcal{D}}(\alpha_{\vec{k}, \sigma}(t)) =  \exp( \alpha_{\vec{k} \sigma}(t) \hat{a}_{\vec{k}, \sigma}^\dagger - \alpha^*_{\vec{k} \sigma}(t) \hat{a}_{\vec{k}, \sigma}) \label{eq:displacement_single_k},
	\end{equation}
	with $\alpha_{\vec{k}, \sigma}(t) = \alpha_{\vec{k}, \sigma} \e^{-i \omega_k t}$ such that $\hat{D}^\dagger(t) \ket{\psi_{laser}(t)} = \ket{0}$. Due to this transformation the vector potential $\hat{\vec{A}}$ now splits up into a classical part $\vec{A}_{cl}(t) = \bra{\psi_{laser}(t)} \hat{\vec{A}} \ket{\psi_{laser}(t)}$ as well as a quantum part
	\begin{equation}
		\hat{\vec{A}}_Q = \Sigma_{\vec{k}, \sigma}  \dfrac{g_0}{\sqrt{\omega_k}} \big( \hat{\vec{e}}_\sigma \hat{a}_{\vec{k}, \sigma} \e^{i \vec{k} \cdot \vec{r}} +  \hat{\vec{e}}_\sigma^* \hat{a}^\dagger_{\vec{k}, \sigma} \e^{-i \vec{k} \cdot \vec{r}} \big).
	\end{equation}

	Consequently, the Hamiltonian is now separated into three parts, 
	\begin{equation}
		\hat{\tilde{H}}(t) = \hat{H}_{TDSE}(t) + \hat{V}(t) + \hat{H}_F, \label{eq:Hamiltonian_separated}
	\end{equation}
	where $\hat{H}_{TDSE}(t) = \sum_{j=1}^N \dfrac{1}{2} (\hat{\vec{p}}_j + \vec{A}_{cl}(t))^2 + \hat{U}$ is the Hamiltonian governing the electronic system subject to a classical driving, $\hat{V}(t) = \sum_{j=1}^N \hat{\vec{A}}_Q \cdot (\hat{\vec{p}}_j + \vec{A}_{cl}(t))$ is the electronic interaction with the quantum field, and $\hat{H}_F$ is the free field Hamiltonian given in Eq. (\ref{eq:H_free_field}). In Eq. (\ref{eq:Hamiltonian_separated}) a tilde has been used on the left hand site to denote the fact that the Hamiltonian is transformed due to the application of the displacement operator in Eq. (\ref{eq:displacement_unitary}). A full derivation of the result in Eq. (\ref{eq:Hamiltonian_separated}) can be found in App. \ref{App:Interaction_with_a_qauntized_field}. 
	
	The full Hamiltonian in Eq. (\ref{eq:Hamiltonian_separated}) is now further transformed by going to a rotating frame with respect to $\hat{H}_{TDSE}(t)$ and $\hat{H}_F$, i.e.,  by applying the unitary time evolution operator $\hat{\mathcal{U}}^\dagger_0(t,t_0) = \hat{\mathcal{U}}^\dagger_{TDSE}(t, t_0) \cdot \hat{\mathcal{U}}^\dagger_F(t, t_0)$ with $t_0$ being the initial time such that
	\begin{equation}
		\ket{\tilde{\Psi}(t)}_I = \hat{\mathcal{U}}_0^\dagger(t, t_0) \ket{\tilde{\Psi}(t)}_S, \label{eq:Psi_I_transformation}
	\end{equation}
	where the subscripts refer to the frame of reference with $S$ denoting the Schrödinger picture and $I$ the interaction picture.
	The transformation in Eq. (\ref{eq:Psi_I_transformation}) separates the dynamics of the classical and quantum parts of the electromagnetic potential and transforms the TDSE into
	\begin{equation}
		i \frac{\partial}{\partial t} \ket{\tilde{\Psi}(t)}_I = \hat{V}_I(t)  \ket{\tilde{\Psi}(t)}_I , \label{eq:TDSE_transformed}
	\end{equation}
	with
	\begin{align}
		\hat{V}_I(t) =~& \hat{\mathcal{U}}^\dagger_F(t, t_0) ~ \hat{\vec{A}}_Q ~ \hat{\mathcal{U}}_F(t, t_0) \nonumber \\
		&   \cdot  \hat{\mathcal{U}}^\dagger_{TDSE}(t,t_0) ~ \vec{\hat{j}}(t) ~\hat{\mathcal{U}}_{TDSE}(t, t_0) \label{eq:V_I_first}
	\end{align}
	where we have defined the current operator $\hat{\vec{j}}(t) = \Sigma_{j=1}^N  (\hat{\vec{p}}_j + \vec{A}_{cl}(t))$ and where the tilde denotes that the state is displaced with respect to the laser modes. The quantized vector potential in the rotating frame within the dipole approximation is given by
	\begin{align}
		\hat{\vec{A}}_{Q,I}(t)& =\hat{\mathcal{U}}^\dagger_0(t, t_0) ~ \hat{\vec{A}}_Q ~ \hat{\mathcal{U}}_0(t, t_0) =  \hat{\mathcal{U}}^\dagger_F(t, t_0) ~ \hat{\vec{A}}_Q ~ \hat{\mathcal{U}}_F(t, t_0)  \nonumber \\
		&= \sum_{\vec{k}, \sigma} \dfrac{g_0}{\sqrt{\omega_k}} [\hat{\vec{e}}_\sigma \hat{a}_{\vec{k},\sigma} \e^{-i \omega_k (t-t_0)} + \hat{\vec{e}}_\sigma^* \hat{a}_{\vec{k},\sigma}^\dagger \e^{i \omega_k (t-t_0)} ]. \label{eq:A_q_interaction_picture}
	\end{align} 
	We rewrite the right hand side of Eq. (\ref{eq:V_I_first}) by utilizing that an electronic state $\ket{\phi_m}$ is time evolved via
	\begin{equation}
		\ket{\phi_m(t)} = \hat{\mathcal{U}}_{TDSE}(t, t_0) \ket{\phi_m (t_0)}. \label{eq:phi_m_time_evolved_definition}
	\end{equation}
	By inserting identity operators $\mathbb{1}= \Sigma_m \ket{\phi_m}\bra{\phi_m}$ into Eq. (\ref{eq:V_I_first}) and use the definition in Eq. (\ref{eq:phi_m_time_evolved_definition}) we then express the interaction as
	
	\begin{equation}
	 	\hat{V}_I(t) = \sum_{m,n} \hat{\vec{A}}_{Q,I}(t) \cdot \vec{j}_{m,n} (t) \ket{\phi_m}\bra{\phi_n}, \label{eq:V_I}
	\end{equation}
	where we have defined the matrix element
	\begin{align}
		\vec{j}_{m,n}(t) = \bra{\phi_m(t)} \hat{\vec{j}}(t) \ket{\phi_n(t)}, \label{eq:j_mn_def}
	\end{align}
	which is referred to as the transition current between two electronic states. Thus, the second step in the quantum optical protocol is to obtain a set of electronic states $\{\ket{\phi_m(t)}\}$ for all times using standard TDSE solving techniques to solve 
	\begin{equation}
		i \dfrac{\partial}{\partial t} \ket{\phi_m(t)} = \hat{H}_{TDSE}(t) \ket{\phi_m(t)}, \label{eq:TDSE_electronic_state}
	\end{equation}
	where $\ket{\phi_m(t)}$ is a wavepacket starting out in the $m$th state for the electronic part of the problem. The choice of basis, $\{\ket{\phi_m(t)}\}$, is in principle completely arbitrary. We have chosen the basis to be the eigenstates of the field-free Hamiltonian of the electronic system, as it enables the use of conventional numerical tools to solve the TDSE.
		
	In the present work, we investigate how correlations in a generic correlated material described by the Fermi-Hubbard model affect the emitted photonic state produced by HHG. To that end, we now specify the electronic Hamiltonian to be the field-driven Fermi-Hubbard Hamiltonian, i.e., $\hat{H}_{TDSE} (t)\rightarrow \hat{H}_{FH}(t)$. Specifically, we consider the one-dimensional, one-band Fermi-Hubbard model at half filling with periodic boundary conditions and with an equal number of spin-up and spin-down electrons to keep the whole system spin neutral. This model is chosen as it allows us to treat the correlation strength $U$ as a parameter to study the effects of correlations. The system is driven by a classical laser pulse such that the system within the dipole approximation is described by the time-dependent Hamiltonian \cite{Hubbard_Essler}
	\begin{align}
		\hat{H}_{FH}(t) &= \hat{H}_{hop}(t) + \hat{H}_U, \label{eq:Hamiltonian_FH}
	\end{align}
	with
	\begin{align}
		\hat{H}_{hop}(t) =& -t_0 \sum_{j, \mu} (\e^{i a A_{cl}(t)} \hat{c}_{j, \mu}^\dagger \hat{c}_{j+1, \mu} + \text{H. c.}), \label{eq:H_hop}\\
		\hat{H}_U =&~ U \sum_j (\hat{c}^\dagger_{j, \uparrow}\hat{c}_{j, \uparrow}) (\hat{c}_{j, \downarrow}^\dagger \hat{c}_{j, \downarrow}), \label{eq:H_U}
	\end{align}
	where $t_0$ is the hopping matrix element for an electron to move to the nearest neighboring sites, i.e., from site $j$ to site $j \pm 1$, $A_{cl}(t)$ is the classical vector potential of the driving field along the lattice dimension, $a$ is the lattice spacing, and $U$ is the onsite electron-electron repulsion. The fermionic creation (annihilation) operator for an electron on site $i$ with spin $\mu \in \{\uparrow, \downarrow \}$ is denoted $\hat{c}_{j,\mu}^\dagger$ ($\hat{c}_{j, \mu}$). Note that each term in the sum in Eq. (\ref{eq:H_U}) counts the number of electrons on site $j$.	
	%For convenience, we take the set of electronic states $\{\ket{\phi_m(t)}\}$ to be the propagated field free eigenstates, i.e. the set of states that satisfy $\frac{\partial}{\partial t} \ket{\phi_m(0)} = \hat{H}_{FH}(0)\ket{\phi_m(0)}$. 
	In the Fermi-Hubbard model, the current operator is explicitly given as
	\begin{equation}
		\hat{\vec{j}}(t) = - i a t_0  \sum_{j, \mu} \big(\e^{iaA_{cl}(t)} \hat{c}_{j, \mu}^\dagger \hat{c}_{j+1, \mu} - \text{H.c.} \big)  \hat{\vec{x}}, \label{eq:current_opeator}
	\end{equation}
	which is in the direction of the Fermi-Hubbard chain taken to be the $x$-direction; see App. \ref{App:derivation_of_current_operator}. 

	% i.e. $\vec{A}_{q,I}(\vec{r},t) \simeq \vec{A}_{q,I}(t)$. 
	
	We are now ready to solve Eq. (\ref{eq:TDSE_transformed}). To that end, it is convenient to expand the full state of the electronic and photonic degrees of freedom in terms of field-free electronic eigenstates
	\begin{equation}
		\ket{\tilde{\Psi}(t)}_I = \sum_m \ket{\tilde{\chi}^{(m)}(t)} \ket{\phi_m}, \label{eq:psi_tilde_expansion}
	\end{equation}
	where $m$ is the index corresponding to the $m$'th electronic state and $\ket{\tilde{\chi}^{(m)}(t)}$ is the photonic state associated with the electronic state $\ket{\phi_m}$. Note that normalization of the state in Eq. (\ref{eq:psi_tilde_expansion}) requires $\Sigma_m \langle \tilde{\chi}^{(m)}(t) \ket{\tilde{\chi}^{(m)}(t)} = 1$. The choice of basis in Eq. (\ref{eq:psi_tilde_expansion}) can be completely arbitrary but it is convenient to expand in the same basis as the one used in Eq. (\ref{eq:V_I}) as we will exploit the orthogonality between basis states in the following. 
	We now insert Eq. (\ref{eq:psi_tilde_expansion}) into the transformed TDSE in Eq. (\ref{eq:TDSE_transformed}), project onto $\bra{\phi_m}$, write out $ \hat{\vec{A}}_{Q,I}(t)$ explicitly, and obtain the following equation of motion for the emitted photonic field
	\begin{align}
		i  \frac{\partial}{\partial t} \ket{\tilde{\chi}^{(m)}(t)} =&  \sum_{\vec{k}, \sigma} \dfrac{g_0}{\sqrt{\omega_k}} [\hat{\vec{e}}_\sigma \hat{a}_{\vec{k},\sigma} \e^{-i \omega_k t} +\hat{\vec{e}}_\sigma^* \hat{a}_{\vec{k},\sigma}^\dagger \e^{i \omega_k t} ] \nonumber \\
		& \cdot \sum_n \vec{j}_{m,n}(t) \ket{\tilde{\chi}^{(n)}(t)}, \label{eq:photonic_EOM_general}
	\end{align}
	where we have taken $t_0=0$ for convenience. 
	Equation \eqref{eq:photonic_EOM_general} has the same structure as the equations derived for the  atomic case in  Refs. \cite{KaminerNatCom2020, Riveradean2023entanglementHHG_SolidState, Stammer2023entanglement, StammerLewenstein_tutorial_Quantum_state_engineering, RiveraDean2022} and it is in general difficult to solve as it couples all frequency modes and all electronic states. To proceed we now assume that each frequency component can be treated independently (that is, assuming no coupling between the different harmonics \cite{KaminerNatCom2020}) yielding the equation
	\begin{align}
		i  \frac{\partial}{\partial t} \ket{\tilde{\chi}^{(m)}_{\vec{k}\sigma}(t)} = &  \dfrac{g_0}{\sqrt{\omega_k}} [\hat{\vec{e}}_\sigma \hat{a}_{\vec{k},\sigma} \e^{-i \omega_k t} +\hat{\vec{e}}_\sigma^* \hat{a}_{\vec{k},\sigma}^\dagger \e^{i \omega_k t} ] \nonumber \\
		& \cdot \sum_n \vec{j}_{m,n}(t) \ket{\tilde{\chi}^{(n)}_{\vec{k}\sigma}(t)},  \label{eq:photonic_EOM_independent_harmonics}
	\end{align}
	where $\ket{\tilde{\chi}^{(m)}_{\vec{k}\sigma}(t)}$ is the frequency component with wave number $\vec{k}$ and polarization $\sigma$ of the photonic state associated to the field-free electronic eigenstate $\ket{\phi_m}$. The integration of Eqs. (\ref{eq:photonic_EOM_general}) or (\ref{eq:photonic_EOM_independent_harmonics}) is the third step in the quantum optical description of HHG, and by using the full state of the system in Eq. (\ref{eq:psi_tilde_expansion}) expectation values of interest can be calculated. Details on the calculation of expectation values are given in App. \ref{App:Explicit_expectation_values}.
	
\begin{figure}
		\centering
		\includegraphics[width=1.0\linewidth]{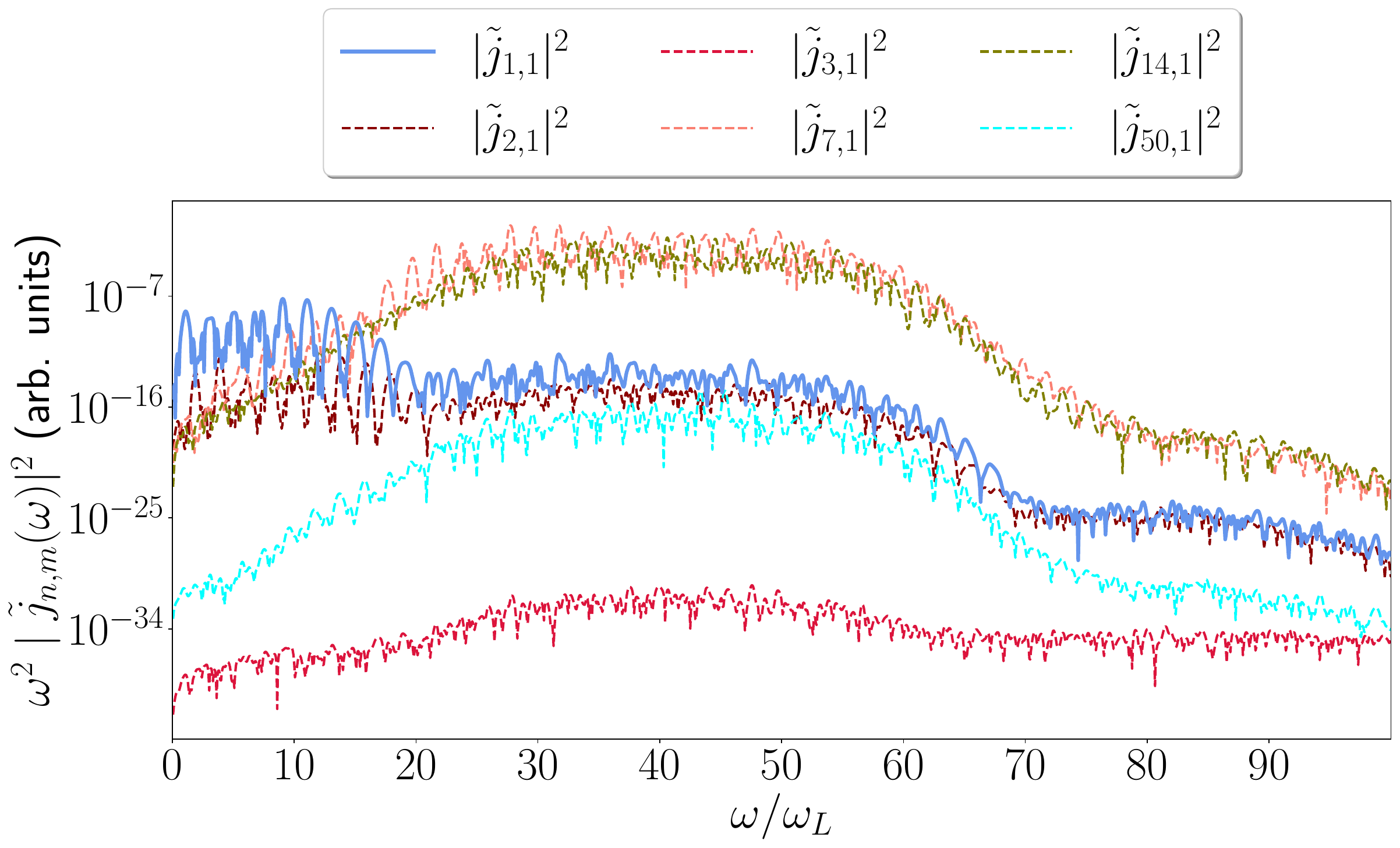}
		\caption{The norm squared of the time derivative of the Fourier transformed current, $\tilde{j}_{n,m}(\omega)$, for various transition currents. Note that some elements are comparable or even larger than the classical current element, $\tilde{j}_{1,1}(\omega)$ (solid blue line). The results shown are for a Fermi-Hubbard model with $U = 10 t_0$ and the parameters specified in the main text.}
		\label{fig:classical_currents_L8_U10}
	\end{figure}

	We emphasize that Eq. (\ref{eq:photonic_EOM_general}) (or equivalently Eq. (\ref{eq:photonic_EOM_independent_harmonics})) are the central equations in this fully quantum mechanical framework highlighting that it is not only the classical current, $\vec{j}_{i,i}(t)$ with $i$ being the initial state [see Eq. (\ref{eq:j_mn_def})], that contributes to the photonic quantum state. 
	In fact, for the system considered in this paper, we find that other elements of this transition-current, $\vec{j}_{n,m}(t)$, are significantly greater than the classical contribution as seen in Fig. \ref{fig:classical_currents_L8_U10} where Fourier-transformed transition currents, $\tilde{\vec{j}}_{n,m}(\omega)$, are shown. At first glance, this may indicate that many transition currents contribute more than $\langle \vec{j}(t) \rangle = \vec{j}_{1,1}(t)$. However, as seen from Eqs. (\ref{eq:photonic_EOM_general}) and (\ref{eq:photonic_EOM_independent_harmonics}) the transition currents (together with $g_0$) determine how much population is transferred to a given state. As we take the initial condition to be the vacuum state associated with the electronic ground state, i.e., $\ket{\tilde{\chi}_{\vec{k}\sigma}^{(m)}(t_0)} = \delta_{m,i} \ket{0}$, and the coupling is in general weak (determined by $g_0$) most population remains in the photonic state associated with the initial electronic state.  Note also, that this formalism can be used to calculate the quantum features of any system once $\vec{j}_{n,m}(t)$ is known for all $n$ and $m$.
	
	In other works further approximations have been made on the transition currents. For instance, the strong-field approximation (SFA) has been made to neglect transitions between excited states (or continuum-continuum transitions for atoms) \cite{StammerLewenstein_tutorial_Quantum_state_engineering, RiveraDean2022} or by performing a Markov-type approximation on the photonic state \cite{Stammer2023entanglement} allowing for an analytical solution of Eq. (\ref{eq:photonic_EOM_general}). However, we do not pursue this further here.
	
%	For clarity, we now summarize the protocol of this fully quantum mechanical theory. First, a unitary displacement transforms the photonic state such that the vacuum state is now "on top" of the laser, which separates all the dynamics of the emitted HHG from the driving laser. This also separates the photonic field into a classical and a quantum part. The set of electronic eigenstates, $\{ \ket{\phi_m(t)}\}$, can be obtained by solving the TDSE for the electronic system subject to a classical driving field in Eq. (\ref{eq:TDSE_electronic_state}).
%	The emitted photonic state can then be obtained by solving Eq. (\ref{eq:photonic_EOM_general}) (or Eq. (\ref{eq:photonic_EOM_independent_harmonics})), allowing for expectation values to be calculated using the state in Eq. (\ref{eq:psi_tilde_expansion}).

	In the results presented in this work, we use a one-dimensional lattice of $L=8$ sites with periodic boundary conditions, a lattice spacing of $a = 7.5589$ a.u., and $t_0 = 0.0191$ a.u picked to fit those of Sr$_2$CuO$_3$ \cite{Tomita2001} as done previously in Refs. \cite{Silva2018, Hansen22, Hansen22_2}. We use a linearly polarized pulse with polarization along the lattice dimension with $N_c=10$ cycles and a $\sin^2$ envelope function in the dipole approximation
	
	\begin{equation}
		A_{cl}(t) = A_0 \sin(\omega_L t + N_c \pi) \sin^2 \big(\dfrac{\omega_L t }{2 N_c} \big).
	\end{equation}
	Here, the potential of vector potential is $A_0 = F_0/\omega_L = 0.194$ a.u. with the angular frequency $\omega_L = 0.005 \text{ a.u. }=33$ THz. With these choices the field strength, $F_0$, corresponds to a peak intensity of $3.3 \times 10^{10}$ W/cm$^2$. To study the effects of correlations, we treat the strength of the on-site electron-electron correlation, $U$, as a parameter. We solve the TDSE in Eq. (\ref{eq:TDSE_electronic_state}) for all states $\{\ket{\phi_m(t)}\}$ using the Arnoldi-Lancoz algorithm \cite{Park1986, Smyth1998, Guan2007, Frapiccini2014} with a Krylov subspace dimension of $4$. As this fully quantum mechanical theory requires all time evolved states of the electronic system,$\{ \ket{\phi_m(t)}\}$, it is computationally more demanding than a usual semiclassical calculation, limiting us to consider at most $L=8$ electrons in the model.	We found consistent results for smaller system with $L=4,6$ electrons. Furthermore, to limit the dimensionality of the system, we utilize that the Hamiltonian in Eq. (\ref{eq:Hamiltonian_FH}) possesses spin-flip symmetry and is invariant under translations of the entire system corresponding to conservation of the total crystal momentum \cite{Hubbard_Essler}. The initial state is the spin-symmetric ground state with vanishing total crystal momentum and due to the symmetries of the Hamiltonian, only couplings to states within that subspace are needed.
	
	\subsection{Measures}
	Using a fully quantum mechanical approach to calculate the quantum state of the emitted HHG allows one to calculate other measures than just the spectrum which is the primary observable in usual HHG approaches. Particularly, since the photon distribution in each mode is obtained within this theory, the photon statistics of the emitted state can be calculated enabling the study of nonclassicallity in HHG. In this paper, we follow Ref. \cite{KaminerNatCom2020} calculate the photonic spectrum, the Mandel $Q$-parameter, and the squeezing of the photonic state which we will explain in the following.
	
	The spectrum can be calculated in the following way. The total energy of the emitted field after the HHG process is given by $\epsilon = \Sigma_{\vec{k}, \sigma} \omega_k  \langle \hat{n}_{\vec{k}, \sigma} \rangle$, where $ \hat{n}_{\vec{k}, \sigma} = \hat{a}_{\vec{k}, \sigma}^\dagger \hat{a}_{\vec{k}, \sigma}$ is the photonic counting operator, and where we for simplicity have denoted $\langle \hat{n}_{\vec{k}, \sigma} \rangle = \,_I\bra{\tilde{\Psi}(\infty)}\hat{\mathcal{U}}_0^\dagger ~ \hat{n}_{\vec{k}, \sigma} ~ \hat{\mathcal{U}}_0 \ket{\tilde{\Psi}(\infty)}_I$ and include $\hat{\mathcal{U}}_0$ as the state is in a rotating frame as seen from Eq. (\ref{eq:Psi_I_transformation}).
	By taking the continuum limit, $\sum_{\vec{k}} \rightarrow V/(8 c^3 \pi^3) \int d\Omega d\omega \omega^2$, where $V$ is the quantization volume, one can calculate the energy per frequency per solid angle
	\begin{equation}
		S(\omega)  \equiv \dfrac{d \epsilon}{d \omega d \Omega} = \dfrac{\omega^3}{g_0^2 (2\pi)^2 c^3} \langle \hat{n}_{\vec{k}, \sigma} \rangle, \label{eq:spectrum_general}
	\end{equation}
	which we will refer to as the spectrum.	To illustrate the effect of the quantum description of the photonic degrees of freedom, we also consider the conventional semiclassical prediction of the spectrum \cite{Baggesen2011},
	\begin{equation}
		S_{cl}(\omega) = \omega^2 \lvert \tilde{\vec{j}}_{i,i}(\omega) \rvert^2 \label{eq:spectrum_classical}
	\end{equation}
	where $\tilde{\vec{j}}_{i,i}(\omega)$ is the Fourier transform of the classical current which in the present case is obtained from the time-evolved ground state $\ket{\phi_1(t)}$ used in Eq. (\ref{eq:j_mn_def}).
	
	We note that the spectrum in Eq. (\ref{eq:spectrum_general}) only contains the first moment of the photon counting operator which does not reveal all the statistical properties of the underlying distribution from which it was calculated. Information about the photon statistics of the generated light is hence not found in the spectrum.
	The statistical properties of the generated light can instead be quantified via the so-called Mandel $Q$ parameter. For a single mode it is given by \cite{Gerry_knight_2004}
	\begin{equation}
		Q_{\vec{k}, \sigma} =   \dfrac{\langle \hat{n}_{\vec{k}, \sigma}^2 \rangle - \langle \hat{n}_{\vec{k}, \sigma} \rangle^2}{\langle \hat{n}_{\vec{k}, \sigma} \rangle}  - 1. \label{eq:Q_param}
	\end{equation}
	If $Q_{\vec{k}, \sigma} > 0$ the photon statistics is called super-Poissonian while for $Q_{\vec{k}, \sigma} < 0$ it is called sub-Poissonian referring to a broader and narrower distribution than a Poissonian distribution, respectively. Note that a classical coherent state will have $Q_{\vec{k}, \sigma} = 0$ as it has Poissonian statistics. While only a nonclassical state can produce sub-Poissonian statistics, both classical mixtures of coherent states, thermal states as well as nonclassical states can have super-Poissonian photon statistics. In other words, a super-Poissonian statistics is not necessarily a quantum feature \cite{Gerry_knight_2004}.
	The last measure of interest in this paper is the degree of squeezing which is a clear nonclassical feature of light. The degree of squeezing in the unit of dB is given as \cite{Scully_zubairy_1997, Braunstein2005}
	\begin{equation}
		\eta_{\vec{k}, \sigma} = -10 \log (4\underset{\theta \in [0, \pi)}{\min}(\Delta X_{\vec{k}, \sigma}(\theta))^2 ), \label{eq:squeezing_func}
	\end{equation}
	where the minimum is found over angles $0 \leq \theta < \pi$ that minimizes the variance in the generalized quadrature operator $\hat{X}_{\vec{k}, \sigma}(\theta) = (\hat{a}_{\vec{k}, \sigma} \e^{-i \theta} +\hat{a}^\dagger_{\vec{k}, \sigma} \e^{i \theta})/2$. We note that classical coherent light is not squeezed, i.e., $\eta_{\vec{k}, \sigma} = 0$ for all modes and polarizations for coherent light.
	
	Experimentally, one can measure the degree of squeezing with a homodyne detection scheme \cite{Gerry_knight_2004, Slusher1985_squeezed_light_homodyne, Breitenbach1997_squeezed_light_homodyne} and the Mandel Q-parameter can be obtained by photon counting \cite{Short1983_photon_counting_experiment, Bartley2013_sub_binom_light}.
	
	%%% Results %%%

	\section{Results} \label{sec:reuslts}
	We now use this fully quantum mechanical framework to study the harmonics generated by the field-driven Fermi-Hubbard model in two limiting cases, namely the completely uncorrelated phase ($U=0$) and in a Mott-insulating phase ($U=10t_0$). These cases are chosen as the dynamics is governed by two different mechanisms and the limiting cases allow us to present qualitatively physical pictures to explain the results. Additionally, the Mott-insulating phase will highlight the role of electron-electron correlations in the generation of nonclassical light when compared to the uncorrelated phase.
	
	\subsection{Uncorrelated phase}
	
	\subsubsection{Qualitative physical picture}
	As a limiting case we consider a system without correlations, i.e., $U=0$, as it allows for an exact analytic solution. In this case, the Fermi-Hubbard Hamiltonian reduces to a simple one-band tight-binding model described by the hopping Hamiltonian, $\hat{H}_{hop}(t)$ in Eq. (\ref{eq:H_hop}). 
	Equation (\ref{eq:H_hop}) can be diagonalized by transforming the creation operators as follows 
	\begin{align} \label{eq:transformation_wannier_bloch}
		\hat{c}_{j, \mu}^\dagger &= \dfrac{1}{\sqrt{L}} \sum_{q} \e^{-i q R_j} \hat{c}_{q, \mu}^\dagger,
	\end{align}
	with a similar expression for the annihilation operators. Here $R_j$ denotes the $j$th lattice position and $q$ is the crystal momentum of the particle. The transformation reflects a shift from real-space to momentum space corresponding to a shift from an underlying localized Wannier basis to a spatially delocalized Bloch basis. 
	The result for $\hat{H}_{hop}(t)$ in terms of the new operators reads
	\begin{equation}
		\hat{H}_{hop}(t) = \sum_{q, \mu} \mathcal{E}(q + A_{cl}(t)) \hat{c}_{q, \mu}^\dagger \hat{c}_{q, \mu},  \label{eq:H_hop_k_space}
	\end{equation}
	with the dispersion relation describing the Bloch band given as
	\begin{equation}
		\mathcal{E}(q) = -2t_0 \cos(a q). \label{eq:dispersion_tight_binding}
	\end{equation}
	That is, the individual crystal momenta of the electrons are conserved and have a time-dependent dispersion determined by $A_{cl}(t)$. Note that the energy bandwidth is 
	\begin{equation}
		\Delta_{band} = 4 t_0 \label{eq:delta_band}
	\end{equation}
	as seen in Eq. (\ref{eq:dispersion_tight_binding}).
	\subsubsection{Results}
		\begin{figure}
		\centering
		\includegraphics[width=1.0\linewidth]{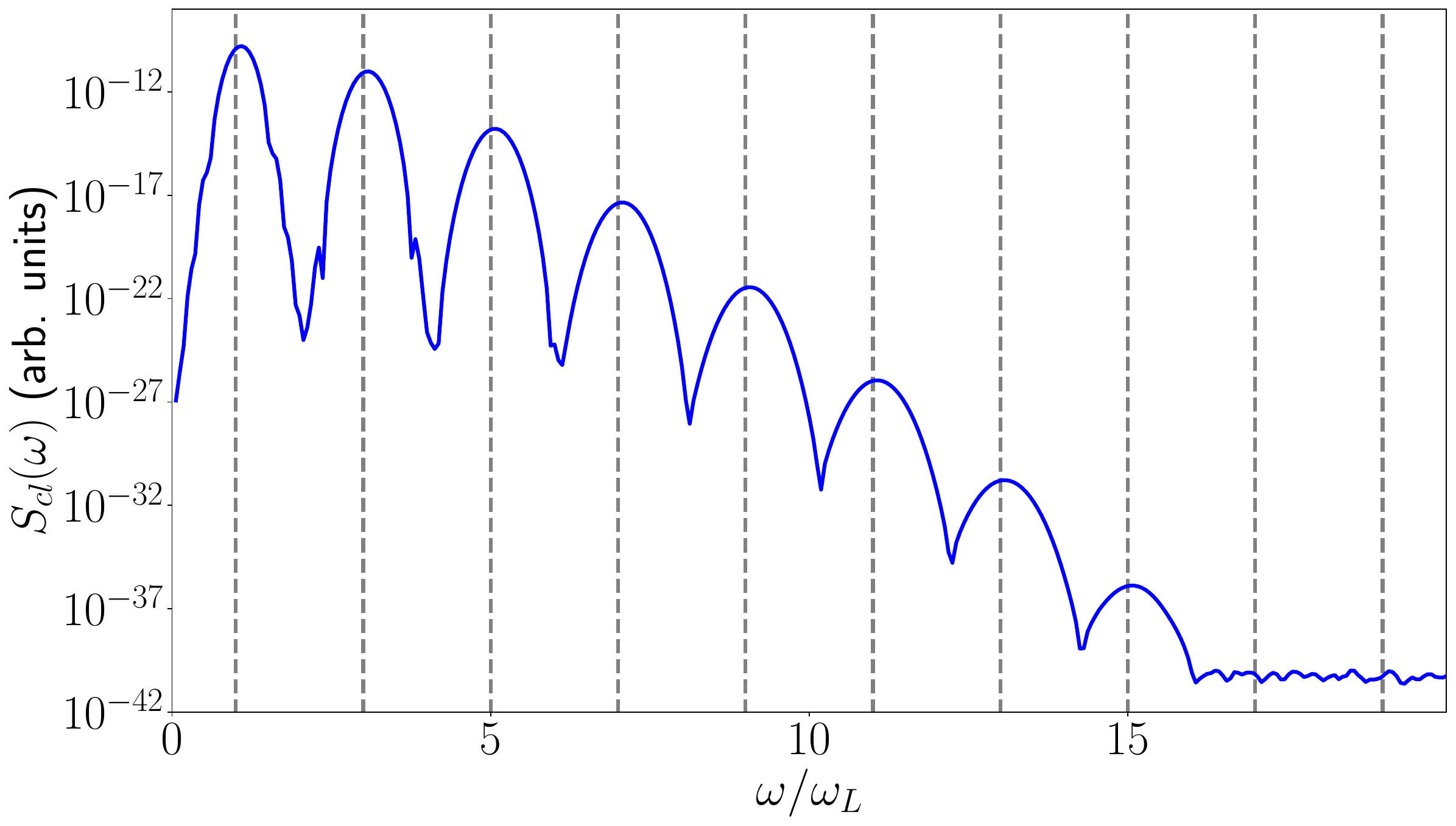}
		\caption{HHG spectrum for the uncorrelated $U=0$ case calculated via Eq. (\ref{eq:spectrum_U0}). Parameters for the system is given in the main text. The dashed lines are placed at odd harmonics to guide the eye. See text in Sec. II.A for laser and system parameters.}
		\label{fig:classicalcurrentsju0}
	\end{figure}
	One can show that $[\hat{H}_{hop}(t), \hat{\vec{j}}(t)] = 0$ which allows for a shared set of eigenstates. As the Hamiltonian also commutes with itself at different times, the time evolution operator is simply given as $\hat{\mathcal{U}}_{hop}(t, t_0) = \exp(- i \int_{t_0}^t \hat{H}_{hop}(t') dt')$ and it then follows that $[\hat{\mathcal{U}}_{hop}(t, t_0), \hat{\vec{j}}(t)] = 0$. Because of these relations, all off-diagonal elements of the current vanish, i.e.,
	\begin{align}
		\vec{j}_{m,n}(t) &= \bra{\phi_m(t_0)} \hat{\mathcal{U}}_{hop}^\dagger(t, t_0) \hat{\vec{j}}(t) \hat{\mathcal{U}}_{hop}(t, t_0) \ket{\phi_n(t_0)} \nonumber \\
		&= \vec{j}_{n,n}(t) \delta_{m,n}, \label{eq:current_diagonal}
	\end{align}
	where it was used that $\ket{\phi_n(t)}$ is an eigenstate for the current operator. Eq. (\ref{eq:current_diagonal}) consequently decouples all the electronic states in Eq. (\ref{eq:photonic_EOM_general}). As the initial state prior to interaction with the laser is the field-free ground state, $\ket{\phi_i(t_0)}$, and there is no coupling to states $\ket{\phi_m(t)}$ with $m\neq i$, only $\ket{\phi_i(t)}$ is populated throughout the dynamics. That is, the equation of motion for the photonic state is now simply given as
	\begin{align}
		i  \frac{\partial}{\partial t} \ket{\tilde{\chi}^{(i)}(t)} =& \sum_{\vec{k}, \sigma} \dfrac{g_0}{\sqrt{\omega_k}} [ \hat{\vec{e}}_\sigma  \hat{a}_{\vec{k},\sigma} \e^{-i \omega_k t} +  \hat{\vec{e}}_\sigma^* \hat{a}_{\vec{k},\sigma}^\dagger \e^{i \omega_k t} ]  \nonumber \\
		&\qquad  \cdot \vec{j}_{i,i}(t) \ket{\tilde{\chi}^{(i)}(t)}. \label{eq:photonic_EOM_diagonal}
	\end{align}
	As Eq. (\ref{eq:photonic_EOM_diagonal}) is linear in the photonic creation and annihilation operators, it can be solved analytically  \cite{Scully_zubairy_1997}
	\begin{equation}
		\ket{\tilde{\chi}^{(i)}(t)} = \otimes_{\vec{k}, \sigma} \hat{\mathcal{D}}(\beta_{\vec{k}, \sigma}^{(i)}(t)) \ket{0},
	\end{equation}
	where $\hat{\mathcal{D}}$ is the unitary displacement operator for the photonic field given in Eq. (\ref{eq:displacement_single_k}) and $\beta_{\vec{k}, \sigma}^{(i)}(t)$ is the time-dependent coherent state amplitude given by
	\begin{equation}
		\beta_{\vec{k}, \sigma}^{(i)}(t) = - i \frac{g_0}{\sqrt{\omega_k}} \int_{t_0}^t \e^{i \omega_k t'} \vec{j}_{i,i}(t') \cdot \hat{\vec{e}}_\sigma^* dt'. \label{eq:beta_analytical_expression}
	\end{equation}
	This result is to be expected  as the emitted light is generated by a classical current, $\langle \hat{\vec{j}} \rangle = \vec{j}_{i,i}$, and consequently the emitted HHG is a multimode coherent state. The mean photon number for a coherent state is readily evaluated $\langle \hat{n}_{\vec{k}, \sigma} \rangle = \lvert\beta_{\vec{k}, \sigma}^{(i)}(t) \rvert^2$ and using Eq. (\ref{eq:spectrum_general}), it is found that the HHG spectrum for 	$U=0$ is given by (taking $t \rightarrow \infty$) 
	\begin{equation}
		S(\omega) = \dfrac{\omega^2}{(2 \pi)^2 c^3} \lvert\tilde{j}_{i,i}(\omega) \rvert^2, \label{eq:spectrum_U0}
	\end{equation}
	where $\tilde{j}_{i,i}(\omega)$ is the Fourier-transformed current. This result is identical to the classical spectrum in Eq. \eqref{eq:spectrum_classical}, proving that no quantum optical considerations are necessary when studying the uncorrelated phase. The spectrum can be seen in Fig. \ref{fig:classicalcurrentsju0} where peaks are found at only odd harmonics as expected. We note that since the emitted HHG is in a coherent state it shows no nonclassical nature, and hence $Q_{\vec{k}, \sigma} = \eta_{\vec{k}, \sigma} = 0$ for all wave numbers and polarizations. We can thus conclude that in the uncorrelated phase, no quantum optical considerations are needed as usual semiclassical calculations are exact.

	A few additional notes for the uncorrelated phase are worth making. First, we find that a perturbative calculation to first order in $\hat{\vec{A}}_Q$, limiting the total state to contain at most one photon as done in Ref. \cite{KaminerNatCom2020}, interestingly yields the exact analytical result for the spectrum though the photon distribution is a superposition of vacuum and $1$-photon modes and is hence not Poissonian. See App. \ref{App:perturbative_calculation} for details. 
	Moreover, we find that even though the spectrum is peaked only at odd harmonics, we observe that during the interaction with the laser $\langle \hat{n}_{\vec{k}, \sigma} \rangle (t) \neq 0$ also for even harmonics. Of course, due to the symmetry of the electronic system and laser, no peaks at even harmonics are observed in the final spectrum. See App. \ref{App:non_zero_signal_even_harmonics} for details. 
	As a final note, one can perform a subsequent condition measurement of the emitted HHG signal. Due to the quantum back-action of this measurement, the state is no longer a single multimode coherent state but is a superposition of coherent states (similar to a cat-state) as found in Refs. \cite{StammerLewenstein_tutorial_Quantum_state_engineering, RiveraDean2022, LewensteinCatState2021, GonoskovGrafe2022nonclassical}. We will not discuss conditional measurement schemes  further in the present work.

	\subsection{Mott-insulating phase}
	
	\subsubsection{Qualitative physical picture}
\begin{figure}
			\centering
			\includegraphics[width=1\linewidth]{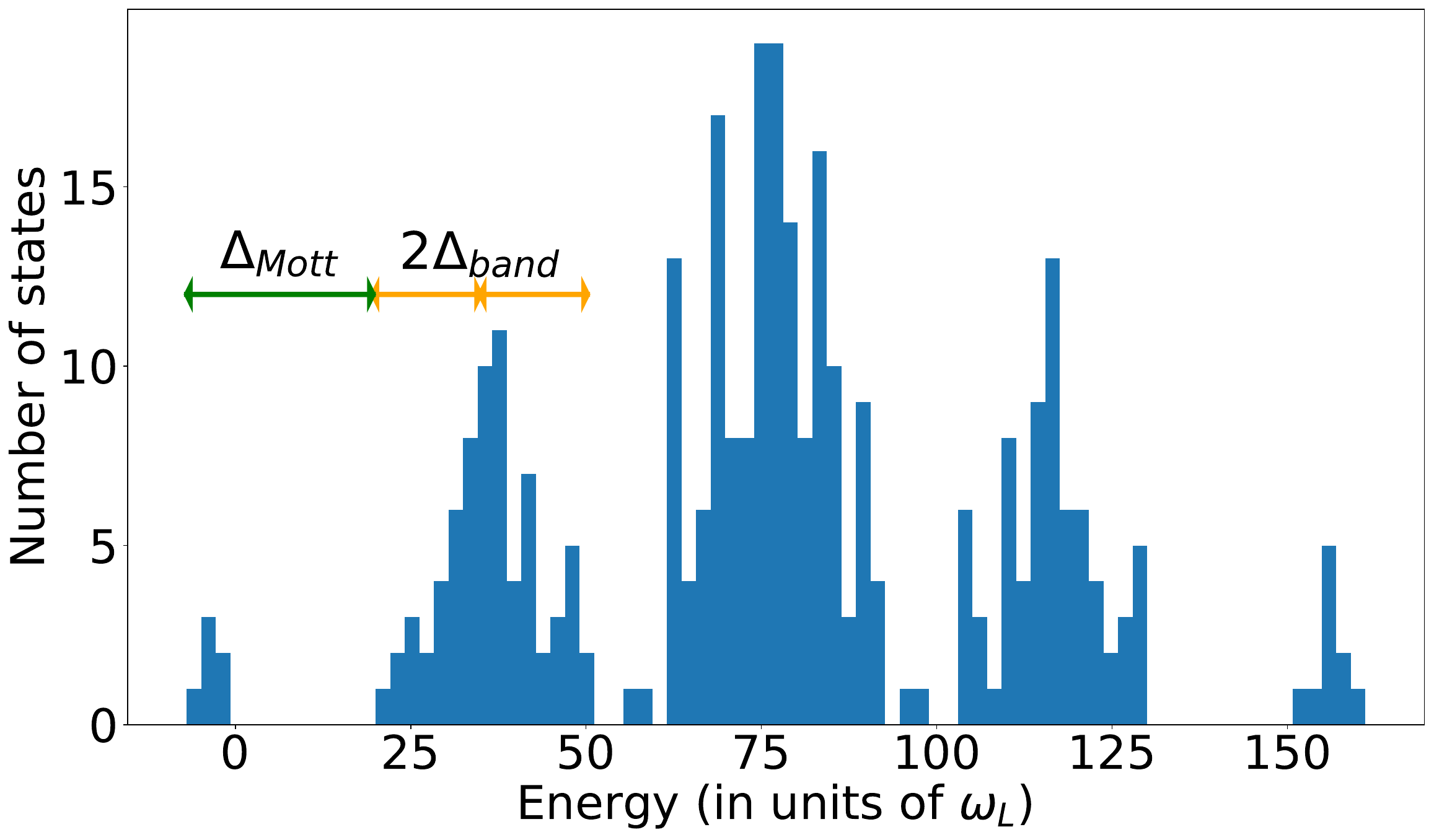}
			\caption{The spectrum of the field-free Fermi-Hubbard model in the Mott-insulating phase with $U=10 t_0$ and parameters in the main text. We note how the spectrum is separated into so-called Hubbard subbands. The characteristic energies are given by the Mott gap $\Delta_{Mott} = 26.7 \omega_L$ [Eq. (\ref{eq:Mott_gap})] (green arrow) and the band width $\Delta_{band} = 4t_0 =  15.3 \omega_L$ [Eq. (\ref{eq:delta_band})] (orange arrows).}
			\label{fig:number_of_states}
		\end{figure}

	We now investigate the so-called Mott-insulating phase where $U \gg t_0$ and where we use $U=10 t_0$ in the calculations. In Fig. (\ref{fig:number_of_states}) all the eigenenergies for the field-free Hamiltonian with $U=10 t_0$ are shown. Here we see how the spectrum is grouped into the so-called Hubbard subbands \cite{Hubbard_Essler}. The energy difference between the ground state and the second subband is given by the Mott-gap (green arrow) \cite{Hubbard_Essler, MMIT97, TMO04}
	\begin{equation} \label{eq:Mott_gap}
		\Delta_{Mott} = E_{GS}^{L+1} + E_{GS}^{L-1} - 2E_{GS}^L
	\end{equation}
	Here $E_{GS}^n$ denotes the ground state energy containing $n$ electrons on a lattice with $L$ sites.  Further discussion can be found in Ref. \cite{Hansen22_2}. The ground state of the system is dominated largely by configurations with only single site occupations as it requires a relatively large energy to have a doubly occupied site \cite{Murakami2018, Murakami2018_2, Murakami2021, Hansen22_2}. The Mott-insulating phase is therefore best analyzed in a quasiparticle picture with doublons (doubly occupied sites) and holons (empty sites). 
	The states in the lowest-lying subband (most left band in Fig. (\ref{fig:number_of_states})) contain virtually no doublon-holon pairs while the states in the first subband are dominated by configurations that contain a single doublon-holon pair. Hence, the Mott gap in Eq. (\ref{eq:Mott_gap}) approximates how much energy is needed to create a doublon-holon pair from the ground state. Furthermore, we note that the width of the first subband is approximately $2 \Delta_{band}$ (orange arrows in Fig. (\ref{fig:number_of_states})) which is due to the fact, that both the doublon and the holon can propagate within the band, each with an identical bandwidth similar to $\Delta_{band}$. Details on the spectrum of the Mott-insulating phase is discussed further in Refs. \cite{Oka2012, Murakami2021, Hubbard_Essler}.
	
	Neglecting scattering between doublons and holons, we can qualitatively describe the HHG process in the Mott-insulating phase in a similar three-step model as the one for multiband models presented in Sec. \ref{sec:introduction}, where doublons now are analogous to electrons and holons are analogous to electron holes \cite{Murakami2021}. This three-step model for Mott-insulators is described by (i) the creation of a doublon-holon pair as the external laser now transfers population from the ground state in the first subband to a state in the second subband. The two bands are energetically separated by $\Delta_{Mott}$ as seen in Fig. (\ref{fig:number_of_states}). The doublon-holon pair then (ii) propagates within the second subband by coupling to different states in that subband whose eigenenergies are closely spaced when compared to $\Delta_{Mott}$ as seen in Fig. (\ref{fig:number_of_states}). Finally, the doublon-holon pair (iii) recombines (annihilates) emitting a high-harmonic photon populating one of the states in the first subband.

	%As in the general case for multiple band models, HHG can be created in both step (ii) which generates intraband current from both doublons and holons and in step (iii) which generates doublon-holon interband current. 
	The intraband current in step (ii) originates from both the doublon and holon propagating in their respective bands with a width similar to $\Delta_{band}$ and hence intraband harmonics can at most have an energy of $\Delta_{band}$. The width of $2\Delta_{band}$ in the second subband in Fig. (\ref{fig:number_of_states}) comes from the fact, that it contains both the holon- and doublon band. Transitions from states in the top to the bottom of the second Hubbard subband in Fig. (\ref{fig:number_of_states}) are multi-electron transitions which are not the dominant contributions to the intraband current. Interband harmonics originating from step (iii) are restricted to have an energy between $\Delta_{Mott}$ and $\Delta_{Mott} +  2\Delta_{band}$ as this is the smallest and largest possible energy when coupling from a state within the second subband to the ground state, respectively.

	\begin{figure*}[t]
		\begin{center}
			\includegraphics[width = 18cm]{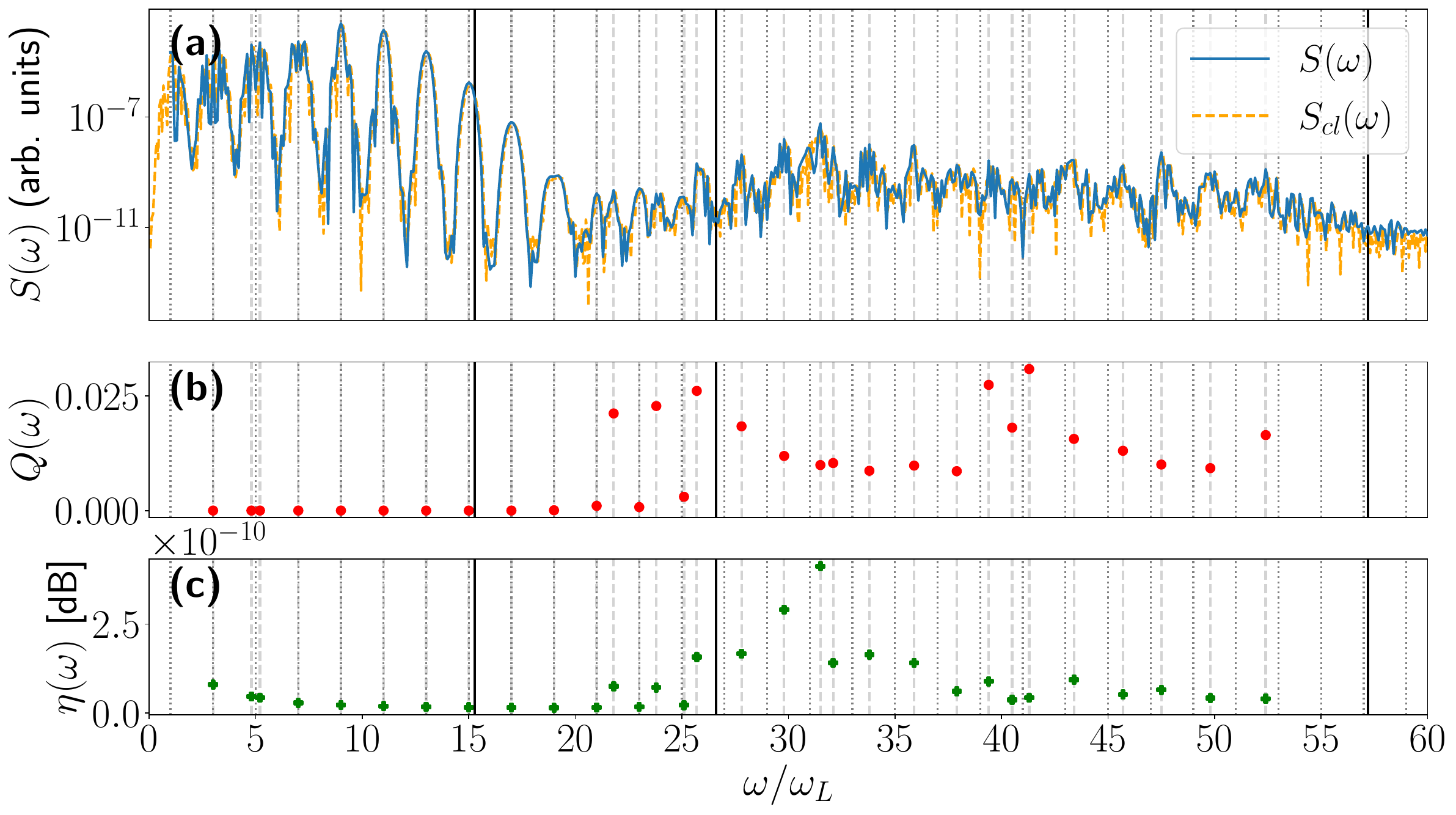}
		\end{center}
		\caption{Measures obtained for a system of $L=8$ sites (periodic boundary conditions) with a correlation strength of $U=10~t_0$. The vertical lines indicate odd harmonics while the dashed lines show some selected frequencies of interest. The vertical black lines are placed at $\Delta_{band}, \Delta_{Mott}$, and $\Delta_{Mott} + 2 \Delta_{band}$, respectively (see text). (a) The HHG spectrum calculated using the fully quantum mechanical theory [Eq. (\ref{eq:spectrum_general})] (blue). We see atransition between intraband harmonics and interband harmonics at $\omega/\omega_L \sim 21$. For comparison the spectrum based on a semiclassical calculation [Eq. \eqref{eq:spectrum_classical}] is shown (orange) which does not differ notably from the quantum optical spectrum. The classical spectrum is normalized to lie on top of the spectrum obtain by the fully quantum mechanical theory. (b)The Mandel $Q$ parameter [Eq. (\ref{eq:Q_param})]. We see a clear increase in $Q$ for interband harmonics showing clear non-Poissonian photon statistics for the emitted Harmonics. (c) The squeezing parameter [Eq. (\ref{eq:squeezing_func})]. Note that the scaling is in units of $10^{-10}$ dB. We note an increased squeezing for higher harmonics showing that the light is slightly squeezed, i.e., nonclassical.} 
		\label{fig:fulldatal8u10}
	\end{figure*}
	
	\subsubsection{Results}
	
	With this physical picture in mind, we are now ready to look at the spectrum, the Mandel $Q$ parameter and the squeezing parameter for the Mott-insulating phase. We obtain the following results by integrating Eq. (\ref{eq:photonic_EOM_independent_harmonics}) for each harmonic mode with a frequency spacing of $\omega/\omega_L=0.1$ with a fourth-order Runge-Kutta routine. In Fig. \ref{fig:fulldatal8u10}(a) the spectrum [Eq. (\ref{eq:spectrum_general})] is shown (blue line). The vertical dashed lines indicate selected frequencies of interest and the vertical dotted lines are placed at odd harmonic orders. The vertical solid black lines are placed at $\Delta_{band}$, $\Delta_{Mott}$ and $\Delta_{Mott} + 2 \Delta_{band}$, respectively.

	In Fig. \ref{fig:fulldatal8u10}(a), we find well-defined peaks at odd harmonics up until $\omega/\omega_L \sim 21$. These harmonics originate from  the doublon-holon intraband current. For harmonics above this value we note an increasing signal which we attribute mainly to the interband current. We also note that the higher-lying peaks are not as clear as the lower ones and this is  due to scattering events between both doublons and holons. These scatterings and the less clear peaks are not a consequence of the quantum optical considerations as it has previously been seen in usual semiclassical HHG calculations \cite{Murakami2021, Hansen22_2, hansen2023effects_of_lattice_imperfections}, and are indeed also present in the semiclassical result [Eq. (\ref{eq:spectrum_classical})] shown by the dashed curve in Fig. \ref{fig:fulldatal8u10}(a). These peaks at non-integer harmonics are due to scattering events between the quasi-particles which allows for exchange of energy that are not integer multiples of the laser frequency though the total momentum is conserved; a comprehensive analysis of these processes will be given elsewhere. The good agreement between the spectrum obtained by including quantum optical theory and the semiclassical spectrum shows that the quantum nature of the generated light can not be seen clearly from HHG spectra.
	
	%We note that no window function has been used in obtaining the spectrum, which is why the spectrum is not as well-resolved as it would be in a usual HHG spectra. For comparison such a classical spectrum (again without window function) is also shown in Fig. (\ref{fig:fulldatal8u10}) (orange dashed line) and it shows no significant difference to the SFQED-spectrum. We thus conclude that the spectrum from a SFQED-calculation does not differ from usual HHG-spectra.
	
	Fortunately, using a quantum optical description allows one to investigate nonclassical properties through the calculation of,  e.g., the Mandel $Q$ parameter and squeezing defined in Eqs. (\ref{eq:Q_param}) and (\ref{eq:squeezing_func}), respectively. In Figs. \ref{fig:fulldatal8u10} (b) and (c) the results for $Q(\omega)$ and $\eta(\omega)$, respectively, are shown.  These have been calculated at frequencies of interest, $\omega'/\omega_L$, and are averaged over a small region of $\omega'/\omega_L \pm 0.2 \omega/\omega_L$, where $\omega'$ is the central frequency. We see in both Figs. \ref{fig:fulldatal8u10} (b) and (c) that the values drastically increase around and above the Mott-gap (middle vertical solid black line). That is, the nonvanishing values of $Q$ and $\eta$ show that the harmonics created by the Hubbard-interband current have non-Poissonian photon statistics as well as non-zero squeezing. In this sense, the results show that correlation between electrons affect the HHG generation process and the photon statistics of the emitted light.
	This is a key result of the present work and it is, to the best of our knowledge, the first time nonclassical light from HHG in a solid has been predicted (without any condition measurement). In this sense, the present findings open the door to the study of correlation induced nonclassicallity of light in the context of strong-field processes.
	
	%%% Discusison and outlook %%%
	
	\section{Discussion} \label{sec:discussion}
	In the results presented we use a coupling strength of $g_0 = 4 \times 10^{-8}$ a.u. which corresponds to a quantization volume with a side length of few wavelengths and is similar to the value of $g_0= 10^{-8}$ a.u. used in Ref. \cite{LewensteinCatState2021} (in length gauge) for a single electron. The value of $g_0$ does not affect spectra in the $U=0$ case. This is because the $g_0$ dependency in Eq. (\ref{eq:beta_analytical_expression}) cancels the $g_0$ dependency in Eq. (\ref{eq:spectrum_general}) leading to the expression in Eq. (\ref{eq:spectrum_U0}).	However, the value $g_0$ quantitatively affects the $U\neq0$ results shown in Fig. \ref{fig:fulldatal8u10}. Specifically, we find much smaller values of both $Q$ and $\eta$ than the ones presented for similar work in atoms in Ref. \cite{KaminerNatCom2020}. However, we emphasize that the present results are based on a single Fermi-Hubbard chain with $L=8$ electrons. In Ref. \cite{LewensteinCatState2021} the number of atoms participating in the HHG process is estimated to be on the order of $N_p \sim 10^{13}$, while in Ref. \cite{KaminerNatCom2020} results are presented with up to $N_p = 5 \times 10^4$ phase matched atoms which in both cases drastically increases the effective coupling. In our work we find that both $Q$ and $\eta$ increases significantly when increasing $g_0$. 
	However, we do not pursue phase matching of multiple Fermi-Hubbard models since a clear assessment would require a consideration of propagation effects of the light, which is beyond the scope of this work. Still, we can conclude that the emitted light from HHG in a correlated material is nonclassical at certain wavelengths and only show weak squeezing for a small system size. 
	%We found qualitatively the same behavior for smaller system size with $L= 4, 6$ electrons.

	We also note that in Ref. \cite{KaminerNatCom2020} sub-Poissonian statistics was found at lower harmonics and particular at transition resonances within a single atom. However, for many atoms only super-Poissonian statistics was found. In the present work only super-Poissonian statistics was found across all frequencies.
	
	Another point worth discussing is how well Eq. (\ref{eq:photonic_EOM_independent_harmonics}) approximates Eq. (\ref{eq:photonic_EOM_general}), i.e., how good of an approximation it is to decouple the different harmonics. This is at present unclear. It is in general not feasible to solve Eq. (\ref{eq:photonic_EOM_general}) and hence approximations are needed. Instead of neglecting couplings between different harmonics as done here and in Ref. \cite{KaminerNatCom2020}, one could also consider making approximations on $\vec{j}_{n,m}(t)$. In Ref. \cite{Stammer2023entanglement}, e.g.,  this approach is followed by neglecting dipole transitions from continuum states to continuum states. In the Fermi-Hubbard model this would amount to only include contributions from $\vec{j}_{m, i}(t) \neq 0$ with $i$ referring to the ground state. This approximation can be justified on grounds of small matrix elements or, as in our case, that population transfer from $\vec{j}_{n,m}(t)$ with both $n, m \neq i$ is second order in $g_0$ and hence negligible. Following this approach, Ref. \cite{Stammer2023entanglement} finds that in the case of atoms, all harmonics are indeed squeezed by including the dipole-dipole coupling. However, in doing so a Markov-type approximation has to be made on the photonic quantum state, the implications of which also calls for further investigation.
	
	%%% Summary and conclusion %%%
	\section{Summary and conclusion} \label{sec:conclusion}
	In this work, we studied how electron-electron correlations affect the quantum state of the emitted HHG light using the prototypical Fermi-Hubbard model. We simulated this within a fully quantum mechanical setting where both the driving and emitted electromagnetic fields are quantized, different to usual semiclassical HHG calculations where only the electronic system is described quantum mechanically. We studied the two limiting cases of an uncorrelated phase ($U=0$) and a Mott-insulating phase ($U=10t_0$), the latter of which may be rationalized in a quasiparticle picture of doublons and holons.
	
	We set out to investigate how the HHG spectrum differs when using a quantized field description, how correlations affect the photons statistics of the emitted light  as well as its squeezing, investigating if the emitted light is nonclassical. With respect to the spectrum, we find that in general it does not differ notably when performing a fully quantum mechanical calculation compared to a conventional semiclassical approach. This clearly shows that the spectrum is largely dominated by the classical current, setup by the oscillating electrons starting out in the field-free ground state. Furthermore, we find that in the uncorrelated phase, the analytical formula for the spectrum matches that of a semiclassical calculation proving that no quantum optical considerations are needed in this case, and that the generated light is coherent, i.e., no quantum features were found in the uncorrelated phase. 
	In contrast to this situation, we  find, using the fully quantum mechanical approach,  that in the Mott-insulating phase nonclassical states of light are generated. Specifically, we find that the Hubbard interband current yields nonclassical light while the doublon-holon intraband current does not show clear nonclassical features; highlighting the importance of accounting for electron-electron interactions for predicting the quantum properties of HHG radiation. 
	
	To the best of our knowledge, our work reports for the first time the generation of nonclassical light in a generic condensed matter model without any subsequent condition measurements, that is, that a correlated material is shown to generate nonclassical light. This opens the door to study other types of interactions in other kinds of systems as a source for obtaining different nonclassical states of light. To this end one can benefit from the generality of this quantum optical approach.
	It requires one to calculate the time-dependent transition currents (or dipoles) between all quantum states of the system. In order to obtain the transition currents, one can benefit from conventional TDSE solvers.  Once the transition currents are obtained, one can directly study the quantum nature of the emitted light. The formalism also allows for the possibility of a nonclassical driving field and it would be worth investigating the interplay between nonclassical driving and correlations with regard to generation nonclassical light with this approach. Such studies would further bridge the gap between strong-field physics and quantum information science by considering strong-field processes as a reliable source for nonclassical light which is central in quantum information and quantum technology \cite{Braunstein2005, Acin2018_QTRoadmap, Walmsley2015}.

	\begin{acknowledgments}
		This work was supported by the Independent Research Fund Denmark (Grant No. 1026-00040B) and by the Danish National Research Foundation through the Center of Excellence for Complex
Quantum Systems (Grant Agreement No. DNRF156).
	\end{acknowledgments}
	\appendix
	%%% Appendix A %%%
	\section{Interaction with a quantized field} \label{App:Interaction_with_a_qauntized_field}
	Here we derive the Hamiltonian [Eq. (\ref{eq:Hamiltonian_separated})] of the main text for the full system of electronic and photonic degrees of freedom when considering a fully quantized field. The presentation builds on the supplementary material in Ref. \cite{KaminerNatCom2020}, and is included here for completeness and reference.
	In vacuum the free electromagnetic field is given as
	\begin{align}
		\hat{\vec{A}} &=  \sum_{\vec{k}, \sigma} \dfrac{g_0}{\sqrt{\omega_k}}   \big( \hat{\vec{e}}_\sigma \hat{a}_{\vec{k}, \sigma} \e^{i \vec{k} \cdot \vec{r}} +  \hat{\vec{e}}_\sigma^* \hat{a}^\dagger_{\vec{k}, \sigma} \e^{-i \vec{k} \cdot \vec{r}} \big),
	\end{align}
	where $\vec{k}$ is the wavevector with related frequency $\omega_k$, $\sigma$ denotes the  polarization, $\hat{\vec{e}}_\sigma$ is a unit vector, $\hat{a}_{\vec{k}, \sigma}$ ($\hat{a}_{\vec{k}, \sigma}^\dagger$) is the annihilation (creation) operator, and  $g_0 = \sqrt{2\pi/V}$ is the coupling constant with quantization volume $V$. In this paper we use $g_0 = 4 \times10^{-8}$ a.u. which corresponds to a quantization volume with a side length of a few wavelengths, and which is similar to the value used in Ref. \cite{LewensteinCatState2021}.
	
	A general many-body Hamiltonian of $N$ identical electrons for the system of interest reads
	\begin{equation}
		\hat{H} = \dfrac{1}{2} \sum_{j=1}^N (\hat{\vec{p}}_j +  \hat{\vec{A}})^2 + \hat{U} + \hat{H}_F,
	\end{equation}
	where $\hat{\vec{p}}_j$ is the momentum operator for particle $j$, $\hat{U}$ is a interaction (Coulomb) within the system, and $\hat{H}_F$ is the free-field Hamiltonian.
	
	Before any interaction between system and laser, the state of the combined system is simply
	\begin{equation}
		\ket{\Psi_i(t)} = \ket{\phi_i(t)} \ket{\psi_{laser}(t)}, \label{eq:psi_phi_psi_laser}
	\end{equation}
	where $\ket{\phi_i(t)}= e^{-i E_i t} \ket{\phi_i} $ is the initial electronic state with trivial time evolution and
	\begin{equation}
		\ket{\psi_{laser}(t)} = \otimes_{k, \sigma} \ket{\alpha_{\vec{k}, \sigma} \e^{-i \omega_k t}},
	\end{equation}
	is the state of the laser which has a frequency broadening around the laser frequency, $\omega_L$. Note that $\alpha_{\vec{k}, \sigma}$ determines the coherent-state amplitude for the given frequency component. For frequencies far away from $\omega_L$ there is no amplitude, i.e. $\alpha_{\vec{k}, \sigma} = 0$ for $\omega_k \gg \omega_L$.
	
	We now define the displacement operator %corresponding to the modes populated in Eq. \eqref{eq:psi_phi_psi_laser} 
%	\begin{equation}
%		\hat{D}(t) = \otimes_{\vec{k}, \sigma} \exp( \alpha_{\vec{k} \sigma} \e^{-i\omega_k t} \hat{a}_{\vec{k}, \sigma}^\dagger - \alpha_{\vec{k} \sigma}^* \e^{i \omega_k t} \hat{a}_{\vec{k}, \sigma}).
%	\end{equation}
	\begin{equation}
	\hat{D}(t) = \otimes_{\vec{k}, \sigma} \hat{\mathcal{D}}(\alpha_{\vec{k} \sigma}(t))
	\end{equation}
	with
	\begin{equation}
		\hat{\mathcal{D}}(\alpha_{\vec{k}, \sigma}(t)) =  \exp( \alpha_{\vec{k} \sigma}(t) \hat{a}_{\vec{k}, \sigma}^\dagger - \alpha^*_{\vec{k} \sigma}(t) \hat{a}_{\vec{k}, \sigma}),
	\end{equation}
	where we take the coherent state parameters to be those of the laser, i.e., $\alpha_{\vec{k}, \sigma}(t) = \alpha_{\vec{k}, \sigma} \e^{-i\omega_k t}$. The displacement operator satisfies the following relations
	\begin{subequations}  \label{eq:displacement_operator_properties}
		\begin{align}
			\hat{D}(t)  \ket{0} &=   \otimes_{k, \sigma}  \ket{\alpha_{\vec{k}, \sigma} \e^{-i \omega_k t}}  \\
			\hat{D}(t) \hat{D}^\dagger(t) &= \mathbb{1}  \\
			\hat{D}^\dagger(t) \hat{a}_{\vec{k}, \sigma} \hat{D}(t) &= \hat{a}_{\vec{k},\sigma} + \alpha_{\vec{k}, \sigma} \e^{-i \omega_k t}  \\
			\hat{D}(t) \hat{a}_{\vec{k}, \sigma} \hat{D}^\dagger(t) &= \hat{a}_{\vec{k}, \sigma} - \alpha_{\vec{k}, \sigma} \e^{-i \omega_k t}, 
		\end{align} 
	\end{subequations}
	where it has been used that $[\hat{a}_{\vec{k}, \sigma}, \hat{a}^\dagger_{\vec{k}', \sigma'}] = \delta_{\vec{k}, \vec{k}'} \delta_{\sigma, \sigma'}$ and $[\hat{a}_{\vec{k}, \sigma}, \hat{a}_{\vec{k}', \sigma'}] = [\hat{a}^\dagger_{\vec{k}, \sigma}, \hat{a}^\dagger_{\vec{k}', \sigma'}] = 0$.
	
	We now displace the laser state such that it is transformed into vacuum, i.e.,
	\begin{equation}
		\hat{D}^\dagger (t)  \ket{\psi_{laser}} = \ket{0}. \label{eq:laser_displaced}
	\end{equation}

	\noindent Using the displacement operator on the state in Eq. (\ref{eq:psi_phi_psi_laser}) we obtain
	\begin{equation}
		\ket{\tilde{\Psi}_i(t)} = \hat{D}^\dagger (t) \ket{\Psi_i(t)} = \ket{\phi_i(t)} \ket{0} ,
	\end{equation}
	where $\ket{\Psi_i(t)}$ satisfies the time-dependent Schrödinger equation, $i  \frac{\partial}{\partial t} \ket{\Psi_i(t)} = \hat{H}(t) \ket{\Psi_i(t)}$. It then follows that the displaced state, $\ket{\tilde{\Psi}_i(t)\textbf{}}$, satisfies
	\begin{equation}
		i  \dfrac{\partial \ket{\tilde{\Psi}_i(t)}}{\partial t} = \hat{\tilde H}(t) \ket{\tilde{\Psi_i}(t)}
		\label{eq:Psi_tilde_TDSE}
	\end{equation}
	with 
	\begin{equation}
		\hat{\tilde H}(t) =  \hat{D}^\dagger(t) \hat{H}(t) \hat{D}(t) - i  \hat{D}^\dagger(t)\dfrac{\partial \hat{D}(t) }{\partial t}
		\label{eq:H_tilde_TDSE}
	\end{equation}
	%where we have suppressed the explicit time-dependence and will continue to do so in the following. 
	Using the properties of $\hat{D}(t)$ from Eq. (\ref{eq:displacement_operator_properties}) one can find that
	
	\begin{widetext}
		\begin{align}
			i  \hat{D}^\dagger  ~ \dfrac{\partial \hat{D}(t)}{\partial t} &= \sum_{\vec{k}, \sigma} \omega_k (\alpha_{\vec{k}, \sigma} \e^{-i \omega_k t} \hat{a}_{\vec{k}, \sigma}^\dagger + \alpha^*_{\vec{k}, \sigma} \e^{i \omega_k t} \hat{a}_{\vec{k}, \sigma} + \lvert \alpha_{\vec{k}, \sigma} \rvert^2 ) \\
			\hat{D}^\dagger(t)  \hat{H}  \hat{D}(t) - i   \hat{D}^\dagger(t)  ~ \dfrac{\partial \hat{D}(t)}{\partial t} &=  \sum_{j = 1}^N \dfrac{ \hat{\vec{p}}_j^2}{2}  + \hat{U} + \sum_{j=1}^{N} \dfrac{1}{2} (\hat{D}(t)^\dagger \hat{\vec{A}}\hat{D}(t))^2 + \dfrac{1}{2}\sum_{j=1}^{N} \big( \hat{\vec{p}}_j \cdot \hat{D}^\dagger (t)\hat{\vec{A}} \hat{D}(t) + \hat{D}^\dagger(t) \hat{\vec{A}} \hat{D}(t) \cdot \hat{\vec{p}}_j \big) \nonumber \\
			& \quad+ \big(\hat{D}^\dagger(t) \hat{H}_F \hat{D}(t) -	i  \hat{D}^\dagger(t)  ~ \dfrac{\partial \hat{D}(t)}{\partial t} \big )  \\
			\hat{D}^\dagger(t) \hat{H}_F \hat{D}(t) -	i  \hat{D}^\dagger(t) ~ \dfrac{\partial \hat{D}(t)}{\partial t} &= \hat{H}_F = \sum_{\vec{k}, \sigma} \omega_k \hat{a}_{\vec{k}, \sigma}^\dagger \hat{a}_{\vec{k}, \sigma}.
		\end{align}
	\end{widetext}
	
	The action of the displacement operator on the quantized electromagnetic field is the following
	\begin{equation}
		\hat{D}^\dagger(t) \hat{\vec{A}} \hat{D}(t)= \vec{A}_{cl}(t) + \hat{\vec{A}}_Q,
	\end{equation}
	where 
	\begin{align}
		\vec{A}_{cl}(t)& = \bra{\psi_{laser}(t)} \hat{\vec{A}} \ket{\psi_{laser}(t)} \nonumber \\
		& = \sum_{\vec{k}, \sigma} \dfrac{g_0}{\sqrt{\omega_k}} \big( \hat{\vec{e}}_\sigma \alpha_{\vec{k}, \sigma} \e^{i \vec{k} \cdot \vec{r} - i \omega_k t} + \text{H. c.} \big) \nonumber  \\
		\hat{\vec{A}}_Q& = \sum_{\vec{k}, \sigma}  \dfrac{g_0}{\sqrt{\omega_k}} \big( \hat{\vec{e}}_\sigma \hat{a}_{\vec{k}, \sigma} \e^{i \vec{k} \cdot \vec{r}} +  \text{H. c.}\big),
	\end{align}	
	is a classical and a quantized field, respectively. Note that the interval of $\vec{k}$ and $\sigma$ in the first equation is only over the populated coherent states in the laser.
	It then follows that the Hamiltonian of Eq. \eqref{eq:H_tilde_TDSE} transforms as 
	\begin{align}
		\hat{\tilde H}(t) &=  \sum_{j=1}^{N} \dfrac{1}{2} (\hat{\vec{p}}_j +  \vec{A}_{cl}(t))^2 + \hat{U} \nonumber \\
		&+ \sum_{i=j}^N \vec{\hat{A}}_Q \cdot (\hat{\vec{p}}_j + \vec{A}_{cl}(t)) + \hat{H}_F
	\end{align}
	where we work in the Coulomb gauge and also discarded the term $\hat{\vec{A}}_Q^2/2$ as it assumed to be of much smaller magnitude. 
	%We have also assumed that the quantized radiation field commutes with the momentum operator, since the quantized field is divergence less in free space.
	
	%In total, we can write the transformation as
	%\begin{equation}
	%	i  \dfrac{\partial \ket{\tilde{\Psi}}(t)}{\partial t} = \hat{\tilde H}(t) \ket{\tilde{\Psi}(t)}, \label{eq:TDSE_displaced}
	%\end{equation}
	We can order the terms in the Hamiltonian as 
	\begin{equation}
		\hat{\tilde H}(t) = \hat{H}_{TDSE}(t) + \hat{H}_F + \hat{V}(t), \label{eq:H_tilde_app}
	\end{equation}
	with
	\begin{align}
		\hat{H}_{TDSE}(t)&=   \sum_{j=1}^{N} \dfrac{1}{2} (\hat{\vec{p}}_j +  \vec{A}_{cl}(t))^2 + \hat{U}, \label{eq:H_TDSE_general}  \\
		\hat{V}(t) &= \sum_{j=1}^N \vec{\hat{A}}_Q \cdot (\hat{\vec{p}}_j +  \vec{A}_{cl}(t)).  \label{eq:app_interaction_term_P}
	\end{align}
	The net result of this transformation is that the laser driving on the electronic system only needs to be accounted for on a classical level as seen in $\hat{H}_{TDSE}$. However, the electronic system then interacts with a quantized field via $\hat{V}(t)$ which then affects the photon emission. Equation (\ref{eq:H_tilde_app}) corresponds to Eq. (\ref{eq:Hamiltonian_separated}) of the main text.

	%%% Appendix B %%%
	
\section{Adaptation to the Fermi-Hubbard model} \label{App:derivation_of_current_operator}
In this appendix we show how we go from the general TDSE Hamiltonian in Eq. (\ref{eq:H_TDSE_general}) to the Hamiltonian of the Fermi-Hubbard model. 
For notational convenience, we define the operator $\hat{\vec{P}}(t)$
\begin{equation}
	\hat{\vec{P}}(t) = \sum_{l=1}^N (\hat{\vec{p}}_l + \vec{A}_{cl}(t)),
\end{equation}
which is the quantity that interacts with the quantum field  in Eq. (\ref{eq:app_interaction_term_P}).

We calculate the commutator $[\hat{H}_{TDSE}(t), \sum_{n = 1}^N \vec{r}_n] $ and obtain
%As the system is a $1$D-system the commutator between $\hat{\vec{p}}$ and $\vec{r}$ is just $[\hat{\vec{p}}, \vec{r}] = [\hat{p} \hat{\vec{x}}, r  \hat{\vec{x}}] = -i$, as we take the direction of the chain to be in the $x$-direction (polariztion direction) and hence fix the coordinate system by doing so. Calculating this commutator yields

\begin{align}
	[\hat{H}_{TDSE}(t), \sum_{n = 1}^N \vec{r}_n] =  -i \hat{\vec{P}}(t). \label{eq:current_appendix_commutator}
\end{align}
We now specify that $\hat{H}_{TDSE}(t) \rightarrow \hat{H}_{FH}(t)$ with the $H_{FH}(t)$ given in Eq.(\ref{eq:Hamiltonian_FH}). Likewise, the dipole-operator will be adapted to the discrete Fermi-Hubbard model
\begin{equation}
	\sum_{n=1}^N \vec{r}_n  \rightarrow \hat{\vec{R}} = \sum_{l =1}^L \vec{R}_l (\hat{c}_{l, \uparrow}^\dagger \hat{c}_{l, \uparrow} + \hat{c}_{l, \downarrow}^\dagger \hat{c}_{l, \downarrow}) \label{eq:dipole_quantized_app} 
\end{equation}
where $N$ is the number of electrons, $L$ is the number of sites in the system.
Inserting $\hat{H}_{FH}$ and Eq. (\ref{eq:dipole_quantized_app}) into Eq. (\ref{eq:current_appendix_commutator}) gives
\begin{equation}
	\hat{\vec{P}}(t) = i [\hat{H}_{FH}, \hat{\vec{R}}].
	\label{P}
\end{equation}
Since the right hand side of Eq. \eqref{P} is the current operator \cite{Mahan2000}
\begin{align}
	i [\hat{H}_{FH}, \hat{\vec{R}}] = \frac{\partial}{\partial t} \hat{\vec{R}} \equiv \hat{\vec{j}}(t), \label{eq:commutator_to_current}
\end{align}
we obtain
\begin{equation}
	\hat{\vec{P}}(t) = \hat{\vec{j}}(t).
\end{equation}
The explicit expression of the current operator, $\hat{\vec{j}}(t)$, is found by calculating the commutator on the left hand side of Eq. (\ref{eq:commutator_to_current}) and we obtain
\begin{equation}
	\hat{\vec{j}} (t)= - i a t_0 \sum_{j, \mu}\big( \e^{i a A_{cl}(t)}\hat{c}_{j, \mu}^\dagger \hat{c}_{j+1, \mu} - \text{H.c.}\big) \hat{\vec{x}},
\end{equation}
which is the same as Eq. (\ref{eq:current_opeator}) of the main text. Here we have used that the $j$th position in the Fermi-Hubbard chain is $R_j = j \cdot a$ and taken the direction of the chain to be along the $x$-axis (polarization direction) without loss of generality.

	%%% Appendix C %%%
	
	\section{Explicit calculation of expectation values} \label{App:Explicit_expectation_values}
	
	This appendix is dedicated to the calculation of expectation values from the state in Eq. (\ref{eq:psi_tilde_expansion}) which we restate here
	\begin{equation}
		\ket{\tilde{\Psi}(t)}_I = \sum_{m=1}^M  \ket{\tilde{\chi}^{(m)}(t)} \ket{\phi_m}. \label{eq:app_interaction_picture_state_expansion}
	\end{equation}
	By assuming that all modes can be treated independently, i.e., neglecting correlations between different modes (going from Eq. (\ref{eq:photonic_EOM_general}) to Eq. (\ref{eq:photonic_EOM_independent_harmonics})) we can expand the photonic state as a product state where each product state is further expanded in terms of Fock-states
	\begin{align}\label{eq:chi_expansion}
		\ket{\tilde{\chi}^{(m)}} = \otimes_{\vec{k}, \sigma} \ket{\tilde{\chi}^{(m)}_{\vec{k}, \sigma}} \quad \text{with}\quad \ket{\tilde{\chi}^{(m)}_{\vec{k}, \sigma}} = \sum_{n_{\vec{k},\sigma}} c_{n_{\vec{k},\sigma}}^{(m)} \ket{n_{\vec{k}, \sigma}}. 
	\end{align}
	In our calculations, the Hilbert space was truncated to contain at most $100$ photons, though we did not find any population for Fock-states with more than $n_{\vec{k}, \sigma} \leq 15$. This cutoff depends on the choice of $g_0$ as a larger value would give a larger population when applying the photon operators in Eq. (\ref{eq:photonic_EOM_general}) meaning that higher photon numbers can be reached.
	
	We are particularly interested in calculating moments of the number operator $\langle (\hat{n}_{\vec{k}, \sigma})^l \rangle$ used for photon statistics and will use this as an example for a general expectation value. As the full state in Eq. (\ref{eq:app_interaction_picture_state_expansion}) is in the interaction picture we have to transform accordingly. To this end it is useful to use the following relations (with $\hat{\mathcal{U}}_0(t) = \hat{\mathcal{U}}_{TDSE}(t) \cdot \hat{\mathcal{U}}_F(t)$)
	
	\begin{align}
		\hat{\mathcal{U}}_0^\dagger(t) \hat{a}_{\vec{k}, \sigma} \hat{\mathcal{U}}_0(t) =& \hat{a}_{\vec{k}, \sigma} \e^{-i \omega_k t}, \nonumber \\
		\quad 	\hat{\mathcal{U}}_0^\dagger(t) \hat{a}^\dagger_{\vec{k}, \sigma} \hat{\mathcal{U}}_0(t) =&  \hat{a}^\dagger_{\vec{k}, \sigma} \e^{i \omega_k t}.
	\end{align}
	
	We now calculate the expectation value of the number operator to the power of $l$. We find that
	
	\begin{align}
		\langle (\hat{n}_{\vec{k}', \sigma'})^{l} \rangle &=  \,_I\bra{\tilde{\Psi}(t)}  \hat{\mathcal{U}}^\dagger_0(t) ~(\hat{n}_{\vec{k}', \sigma'})^{l} \hat{\mathcal{U}}_0(t) \ket{\tilde{\Psi}(t)}_I \nonumber \\
		& = \sum_{m,m'} \bra{\phi_{m'}} \phi_m \rangle  \bra{\tilde{\chi}^{(m')}} (\hat{n}_{\vec{k}', \sigma}')^{l} \ket{\tilde{\chi}^{(m)}} \nonumber \\
		&= \sum_m  \bra{\tilde{\chi}^{(m)}} (\hat{n}_{\vec{k}', \sigma'})^{l} \ket{\tilde{\chi}^{(m)}},
	\end{align}
	where the orthogonality of the electronic states has been exploited. Now expanding into product states via Eq. (\ref{eq:chi_expansion}) we find
	\begin{align}
		\langle (\hat{n}_{\vec{k}', \sigma'})^{l} \rangle &= \sum_m   \bra{\chi^{(m)}_{\vec{k'}, \sigma'}} (\hat{n}_{\vec{k}', \sigma'})^l \ket{\chi^{(m)}_{\vec{k'}, \sigma'}}                      \prod_{\substack{\vec{k}, \sigma\\ \neq \vec{k}', \sigma'}} \bra{\tilde{\chi}^{(m)}_{\vec{k}, \sigma}} \tilde{\chi}^{(m)}_{\vec{k}, \sigma} \rangle.
	\end{align}
	Expanding further in terms of Fock states via Eq. (\ref{eq:chi_expansion}), the two factors can readily be calculated
	\begin{align}
		&\bra{\chi^{(m)}_{\vec{k'}, \sigma'}} (\hat{n}_{\vec{k}', \sigma'})^l \ket{\chi^{(m)}_{\vec{k'}, \sigma'}} \nonumber \\
		&= \sum_{n'_{\vec{k}', \sigma'}} \sum_{n''_{\vec{k}', \sigma'}}  \big( c_{n'_{\vec{k}', \sigma'}}^{(m)} \big)^*c_{n'_{\vec{k}', \sigma'}}^{(m)} \bra{n'_{\vec{k}', \sigma'}} (\hat{n}_{\vec{k}', \sigma'})^l \ket{\hat{n}{''}_{\vec{k}', \sigma'}} \nonumber \\
		&= \sum_{n_{\vec{k}', \sigma'}} \lvert c_{n_{\vec{k}', \sigma'}}^{(m)} \rvert^2 (n_{\vec{k}', \sigma'})^l,
	\end{align}
	and
	\begin{equation}
		\bra{\tilde{\chi}_{\vec{k}, \sigma}^{(m)}} \tilde{\chi}^{(m)}_{\vec{k}, \sigma} \rangle = \sum_{n_{\vec{k}, \sigma}} \lvert c_{n_{\vec{k}, \sigma}}^{(m)} \rvert^2,
	\end{equation}
	and hence the expectation value is given by
	\begin{align}
		\langle (\hat{n}_{\vec{k}', \sigma'})^{l} \rangle =& \sum_m \bigg[ \big(  \sum_{n_{\vec{k}', \sigma'}} \lvert c_{\vec{k}', \sigma'}^{(m)} \rvert^2 (n_{\vec{k}', \sigma'})^l \big) \nonumber \\
		&\times \prod_{\substack{\vec{k}, \sigma\\ \neq \vec{k}', \sigma'}} \big(  \sum_{n_{\vec{k}, \sigma}} \lvert c_{n_{\vec{k}, \sigma}}^{(m)} \rvert^2 \big) \bigg],
	\end{align}
	which allows all moments of the number operator to be calculated. A similar calculation is done for expectation values of other operators, e.g., $\hat{a}_{\vec{k}, \sigma}$ and $\hat{a}_{\vec{k}, \sigma}^\dagger$.

	%%% Appendix D %%%
	
	\section{Perturbative calculation} \label{App:perturbative_calculation}
	
	\begin{figure}[t]
		\centering
		\includegraphics[width=1\linewidth]{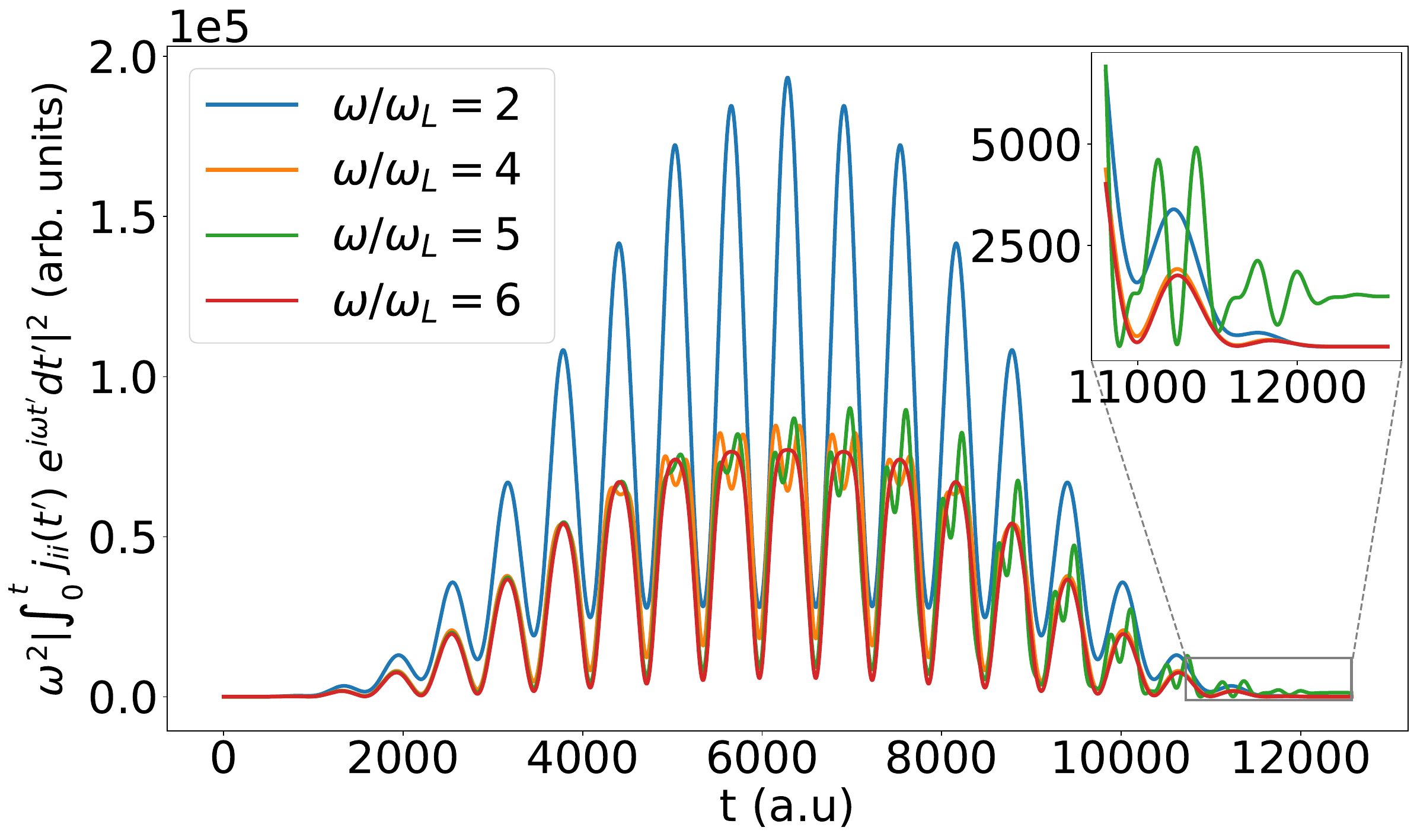}
		\caption{The occupation of the lowest even harmonics during the dynamics of the pulse. We see a clear non-zero occupation at most times except at the end where all even harmonics have a vanishing occupation and hence a vanishing contribution to the signal. For comparison the fifth harmonic (green line) does not vanish after the interaction with the pulse as seen in the inset.}
		\label{fig:meanphotonnumberevenl8u0}
	\end{figure}
	
	Here we derive the state of the system in the perturbative limit of small coupling to the quantum field and show that this yields the exact spectrum in the uncorrelated ($U=0$) phase despite the fact that the perturbative calculation gives a wrong description of the generated light statistics (see below). A perturbative calculation is also found in Supplementary Note $2$ in Ref. \cite{KaminerNatCom2020}.
	
	Starting from Eq. (\ref{eq:TDSE_transformed}) in the main text, we have
	\begin{equation}
		i \frac{\partial}{\partial t} \ket{\tilde{\Psi}(t)}_I = \hat{V}_I(t)  \ket{\tilde{\Psi}(t)}_I, \label{eq:Psi_tilde_EOM_appendix}
	\end{equation}
	with 
	\begin{align}
		\hat{V}_I(t) =  \hat{\vec{A}}_{Q,I}(t) \cdot \hat{\mathcal{U}}^\dagger_{FH}(t, t_0) \hat{\vec{j}}(t) \hat{\mathcal{U}}_{FH}(t, t_0),
	\end{align}
	with $\hat{\vec{A}}_{Q,I}(t)$ given in Eq. (\ref{eq:A_q_interaction_picture}) of the main text.
	The general solution to Eq. (\ref{eq:Psi_tilde_EOM_appendix}) is given by
	\begin{equation}
		\ket{\tilde{\Psi}(t)}_I = \hat{T} \exp \big[ -i \int_{t_0}^t  \hat{V}_I(t') ~ dt' \big] 	\ket{\tilde{\Psi}(t)}_I, \label{eq:Psi_T_ordered}
	\end{equation}
	where $\hat{T}$ is the time ordering operator and $t_0$ is the initial time prior to any interaction between the laser and the electronic system. We now investigate the case of a weak interaction with the quantized field and expand Eq. (\ref{eq:Psi_T_ordered}) to first order in $\hat{V}_I(t)$
	
	\begin{equation}
		\ket{\tilde{\Psi}(t)}_I = \left[1  - i \int_{t_0}^t   \hat{V}_I(t')~ dt' \right] \ket{\tilde{\Psi}(t_0)}_I.
	\end{equation}
	We now take the electronic state to initially be its ground state prior to interaction with the laser. Due to the coherent displacement done in Eq. (\ref{eq:laser_displaced}), the photonic state is initially in the vacuum state for all modes. That is, we take $\ket{\tilde{\Psi}(t_0)}_I = \ket{0}\ket{\phi_i}$. Now writing out $\hat{V}_I(t)$ and letting the photonic operators act on $\ket{\tilde{\Psi}(t_0)}_I$ yields
	\begin{widetext}
		\begin{align}
			\ket{\tilde{\Psi}(t)}_I =& \ket{0}\ket{\phi_i}  -i \sum_{\vec{k}, \sigma} \dfrac{g_0}{\sqrt{\omega_k}} \left[ \int_{t_0}^t  \e^{i \omega_k t'}   \hat{\mathcal{U}}_{FH}^\dagger (t') \hat{\vec{j}}(t') \cdot \vec{e}_\sigma^*    \hat{\mathcal{U}}_{FH}(t') dt' \right] \ket{\vec{k}, \sigma} \ket{\phi_i} \nonumber \\
			=& \ket{0}\ket{\phi_i}  -i \sum_{m} \ket{\phi_m} \sum_{\vec{k}, \sigma} \dfrac{g_0}{\sqrt{\omega_k}} \left[ \int_{t_0}^t \e^{i \omega_k t'}  \vec{j}_{m,i}(t') \cdot \vec{e}_\sigma^* dt \right] \ket{\vec{k}, \sigma}, \label{eq:Psi_perturbative_general}
		\end{align}
	\end{widetext}
	where we in the second line have inserted $\mathbb{1} = \Sigma_m \ket{\phi_m}\bra{\phi_m}$ and defined $\vec{j}_{m,i}(t) = \bra{\phi_m(t_0)} \hat{\mathcal{U}}_{FH}^\dagger (t,t_0) \hat{\vec{j}}(t) \hat{\mathcal{U}}_{FH}(t,t_0) \ket{\phi_i(t_0)}$. In the weak-coupling limit Eq. (\ref{eq:Psi_perturbative_general}) is valid for all choices of $U$ within the electronic system. 
	
	We now investigate the uncorrelated phase of $U=0$. In this case the transition current is diagonal, i.e. $\vec{j}_{m,i}(t) = \vec{j}_{i,i}(t) \delta_{m,i}$. Inserting this into Eq. (\ref{eq:Psi_perturbative_general}) we obtain
	\begin{equation}
		\ket{\tilde{\Psi}(t)}_I = \left[\ket{0}  + \sum_{\vec{k}, \sigma} \beta_{\vec{k},\sigma}^{(i)}(t) \ket{\vec{k}, \sigma} \right] \ket{\phi_i}, \label{eq:Psi_tilde_perturnative_beta}
	\end{equation}
	where we have used the definition of $\beta_{\vec{k}, \sigma}^{(i)}$ in Eq. (\ref{eq:beta_analytical_expression}) in the main text. Calculating the expectation value of the number operator yields $\langle \hat{n}_{\vec{k}, \sigma} \rangle = \lvert \beta_{\vec{k}, \sigma} \rvert^2$, which yields the exact spectrum in Eq. (\ref{eq:spectrum_U0}) when inserted into Eq. (\ref{eq:spectrum_general}) for $t \rightarrow \infty$. However, higher moments such as $\langle (\hat{n}_{\vec{k}, \sigma})^2 \rangle$ are not exact in this perturbative approach as the state in Eq. (\ref{eq:Psi_tilde_perturnative_beta}) is truncated to contain at most one photon and is hence not a coherent state. We have found that this perturbative approach in general does not match the non-perturbative results for a finite value of $U \neq 0$ which is why the perturbative approach is not pursued further.

	%%% Appendix E %%%
	\section{Non-zero signal for even harmonics} \label{App:non_zero_signal_even_harmonics}
		
	Here we briefly show that even harmonics have a non-zero signal during the dynamics though they vanish in the spectrum after the end of the pulse. We will use the case of $U=0$ as example, as we have an exact analytical solution namely $\langle \hat{n}_{\vec{k}, \sigma} \rangle = \lvert \beta_{\vec{k}, \sigma}^{(i)}(t) \rvert^2$ with the coherent state amplitude given in Eq. (\ref{eq:beta_analytical_expression}). In Fig. (\ref{fig:meanphotonnumberevenl8u0}) we show the quantity
	\begin{equation}
		\omega^2 ~\big\lvert \int_0^t  j_{ii}(t) \e^{i \omega t} \big\rvert^2 = \dfrac{\omega^3}{g_0^2}  \lvert \beta_{\vec{k}, \sigma}^{(i)}(t) \rvert^2,
	\end{equation}
	which is  directly related to the classical spectrum as seen from Eqs. (\ref{eq:spectrum_classical}) and (\ref{eq:spectrum_U0}). We see from Fig. (\ref{fig:meanphotonnumberevenl8u0}) that there is a non-zero occupation for all harmonics during most of the interaction with the laser pulse but only the odd harmonics (here the fifth harmonic as an example) are visible in the final spectrum as seen in the inset. If one could very rapidly stop the dynamics during the pulse, even harmonics would be observed since such a rapid change of the system  breaks the inversion symmetry that usually permits only odd harmonics.

	\bibliography{sources_self3}

\begin{thebibliography}{81}%
\makeatletter
\providecommand \@ifxundefined [1]{%
 \@ifx{#1\undefined}
}%
\providecommand \@ifnum [1]{%
 \ifnum #1\expandafter \@firstoftwo
 \else \expandafter \@secondoftwo
 \fi
}%
\providecommand \@ifx [1]{%
 \ifx #1\expandafter \@firstoftwo
 \else \expandafter \@secondoftwo
 \fi
}%
\providecommand \natexlab [1]{#1}%
\providecommand \enquote  [1]{``#1''}%
\providecommand \bibnamefont  [1]{#1}%
\providecommand \bibfnamefont [1]{#1}%
\providecommand \citenamefont [1]{#1}%
\providecommand \href@noop [0]{\@secondoftwo}%
\providecommand \href [0]{\begingroup \@sanitize@url \@href}%
\providecommand \@href[1]{\@@startlink{#1}\@@href}%
\providecommand \@@href[1]{\endgroup#1\@@endlink}%
\providecommand \@sanitize@url [0]{\catcode `\\12\catcode `\$12\catcode
  `\&12\catcode `\#12\catcode `\^12\catcode `\_12\catcode `\%12\relax}%
\providecommand \@@startlink[1]{}%
\providecommand \@@endlink[0]{}%
\providecommand \url  [0]{\begingroup\@sanitize@url \@url }%
\providecommand \@url [1]{\endgroup\@href {#1}{\urlprefix }}%
\providecommand \urlprefix  [0]{URL }%
\providecommand \Eprint [0]{\href }%
\providecommand \doibase [0]{https://doi.org/}%
\providecommand \selectlanguage [0]{\@gobble}%
\providecommand \bibinfo  [0]{\@secondoftwo}%
\providecommand \bibfield  [0]{\@secondoftwo}%
\providecommand \translation [1]{[#1]}%
\providecommand \BibitemOpen [0]{}%
\providecommand \bibitemStop [0]{}%
\providecommand \bibitemNoStop [0]{.\EOS\space}%
\providecommand \EOS [0]{\spacefactor3000\relax}%
\providecommand \BibitemShut  [1]{\csname bibitem#1\endcsname}%
\let\auto@bib@innerbib\@empty
%</preamble>
\bibitem [{\citenamefont {Krausz}\ and\ \citenamefont
  {Ivanov}(2009)}]{Krausz2009}%
  \BibitemOpen
  \bibfield  {author} {\bibinfo {author} {\bibfnamefont {F.}~\bibnamefont
  {Krausz}}\ and\ \bibinfo {author} {\bibfnamefont {M.}~\bibnamefont
  {Ivanov}},\ }\bibfield  {title} {\bibinfo {title} {Attosecond physics},\
  }\href {https://doi.org/10.1103/RevModPhys.81.163} {\bibfield  {journal}
  {\bibinfo  {journal} {Rev. Mod. Phys.}\ }\textbf {\bibinfo {volume} {81}},\
  \bibinfo {pages} {163} (\bibinfo {year} {2009})}\BibitemShut {NoStop}%
\bibitem [{\citenamefont {Ferray}\ \emph {et~al.}(1988)\citenamefont {Ferray},
  \citenamefont {L'Huillier}, \citenamefont {Li}, \citenamefont {Lompre},
  \citenamefont {Mainfray},\ and\ \citenamefont {Manus}}]{LHuillier1988}%
  \BibitemOpen
  \bibfield  {author} {\bibinfo {author} {\bibfnamefont {M.}~\bibnamefont
  {Ferray}}, \bibinfo {author} {\bibfnamefont {A.}~\bibnamefont {L'Huillier}},
  \bibinfo {author} {\bibfnamefont {X.~F.}\ \bibnamefont {Li}}, \bibinfo
  {author} {\bibfnamefont {L.~A.}\ \bibnamefont {Lompre}}, \bibinfo {author}
  {\bibfnamefont {G.}~\bibnamefont {Mainfray}},\ and\ \bibinfo {author}
  {\bibfnamefont {C.}~\bibnamefont {Manus}},\ }\bibfield  {title} {\bibinfo
  {title} {Multiple-harmonic conversion of 1064 nm radiation in rare gases},\
  }\href {https://doi.org/10.1088/0953-4075/21/3/001} {\bibfield  {journal}
  {\bibinfo  {journal} {Journal of Physics B: Atomic, Molecular and Optical
  Physics}\ }\textbf {\bibinfo {volume} {21}},\ \bibinfo {pages} {L31}
  (\bibinfo {year} {1988})}\BibitemShut {NoStop}%
\bibitem [{\citenamefont {Paul}\ \emph {et~al.}(2001)\citenamefont {Paul},
  \citenamefont {Toma}, \citenamefont {Breger}, \citenamefont {Mullot},
  \citenamefont {Augé}, \citenamefont {Balcou}, \citenamefont {Muller},\ and\
  \citenamefont {Agostini}}]{Agostini2001}%
  \BibitemOpen
  \bibfield  {author} {\bibinfo {author} {\bibfnamefont {P.~M.}\ \bibnamefont
  {Paul}}, \bibinfo {author} {\bibfnamefont {E.~S.}\ \bibnamefont {Toma}},
  \bibinfo {author} {\bibfnamefont {P.}~\bibnamefont {Breger}}, \bibinfo
  {author} {\bibfnamefont {G.}~\bibnamefont {Mullot}}, \bibinfo {author}
  {\bibfnamefont {F.}~\bibnamefont {Augé}}, \bibinfo {author} {\bibfnamefont
  {P.}~\bibnamefont {Balcou}}, \bibinfo {author} {\bibfnamefont {H.~G.}\
  \bibnamefont {Muller}},\ and\ \bibinfo {author} {\bibfnamefont
  {P.}~\bibnamefont {Agostini}},\ }\bibfield  {title} {\bibinfo {title}
  {Observation of a train of attosecond pulses from high harmonic generation},\
  }\href {https://doi.org/10.1126/science.1059413} {\bibfield  {journal}
  {\bibinfo  {journal} {Science}\ }\textbf {\bibinfo {volume} {292}},\ \bibinfo
  {pages} {1689} (\bibinfo {year} {2001})},\ \Eprint
  {https://arxiv.org/abs/https://www.science.org/doi/pdf/10.1126/science.1059413}
  {https://www.science.org/doi/pdf/10.1126/science.1059413} \BibitemShut
  {NoStop}%
\bibitem [{\citenamefont {Hentschel}\ \emph {et~al.}(2001)\citenamefont
  {Hentschel}, \citenamefont {Kienberger}, \citenamefont {Spielmann},
  \citenamefont {Reider}, \citenamefont {Milosevic}, \citenamefont {Brabec},
  \citenamefont {Corkum}, \citenamefont {Heinzmann}, \citenamefont {Drescher},\
  and\ \citenamefont {Krausz}}]{Krausz2001}%
  \BibitemOpen
  \bibfield  {author} {\bibinfo {author} {\bibfnamefont {M.}~\bibnamefont
  {Hentschel}}, \bibinfo {author} {\bibfnamefont {R.}~\bibnamefont
  {Kienberger}}, \bibinfo {author} {\bibfnamefont {C.}~\bibnamefont
  {Spielmann}}, \bibinfo {author} {\bibfnamefont {G.~A.}\ \bibnamefont
  {Reider}}, \bibinfo {author} {\bibfnamefont {N.}~\bibnamefont {Milosevic}},
  \bibinfo {author} {\bibfnamefont {T.}~\bibnamefont {Brabec}}, \bibinfo
  {author} {\bibfnamefont {P.}~\bibnamefont {Corkum}}, \bibinfo {author}
  {\bibfnamefont {U.}~\bibnamefont {Heinzmann}}, \bibinfo {author}
  {\bibfnamefont {M.}~\bibnamefont {Drescher}},\ and\ \bibinfo {author}
  {\bibfnamefont {F.}~\bibnamefont {Krausz}},\ }\bibfield  {title} {\bibinfo
  {title} {Attosecond metrology},\ }\href {https://doi.org/10.1038/35107000}
  {\bibfield  {journal} {\bibinfo  {journal} {Nature}\ }\textbf {\bibinfo
  {volume} {414}},\ \bibinfo {pages} {509} (\bibinfo {year}
  {2001})}\BibitemShut {NoStop}%
\bibitem [{\citenamefont {Schafer}\ \emph {et~al.}(1993)\citenamefont
  {Schafer}, \citenamefont {Yang}, \citenamefont {DiMauro},\ and\ \citenamefont
  {Kulander}}]{Schafer1993}%
  \BibitemOpen
  \bibfield  {author} {\bibinfo {author} {\bibfnamefont {K.~J.}\ \bibnamefont
  {Schafer}}, \bibinfo {author} {\bibfnamefont {B.}~\bibnamefont {Yang}},
  \bibinfo {author} {\bibfnamefont {L.~F.}\ \bibnamefont {DiMauro}},\ and\
  \bibinfo {author} {\bibfnamefont {K.~C.}\ \bibnamefont {Kulander}},\
  }\bibfield  {title} {\bibinfo {title} {Above threshold ionization beyond the
  high harmonic cutoff},\ }\href {https://doi.org/10.1103/PhysRevLett.70.1599}
  {\bibfield  {journal} {\bibinfo  {journal} {Phys. Rev. Lett.}\ }\textbf
  {\bibinfo {volume} {70}},\ \bibinfo {pages} {1599} (\bibinfo {year}
  {1993})}\BibitemShut {NoStop}%
\bibitem [{\citenamefont {Corkum}(1993)}]{Corkum1993}%
  \BibitemOpen
  \bibfield  {author} {\bibinfo {author} {\bibfnamefont {P.~B.}\ \bibnamefont
  {Corkum}},\ }\bibfield  {title} {\bibinfo {title} {Plasma perspective on
  strong field multiphoton ionization},\ }\href
  {https://doi.org/10.1103/PhysRevLett.71.1994} {\bibfield  {journal} {\bibinfo
   {journal} {Phys. Rev. Lett.}\ }\textbf {\bibinfo {volume} {71}},\ \bibinfo
  {pages} {1994} (\bibinfo {year} {1993})}\BibitemShut {NoStop}%
\bibitem [{\citenamefont {Lewenstein}\ \emph {et~al.}(1994)\citenamefont
  {Lewenstein}, \citenamefont {Balcou}, \citenamefont {Ivanov}, \citenamefont
  {L'Huillier},\ and\ \citenamefont {~}}]{Lewenstein1994}%
  \BibitemOpen
  \bibfield  {author} {\bibinfo {author} {\bibfnamefont {M.}~\bibnamefont
  {Lewenstein}}, \bibinfo {author} {\bibfnamefont {P.}~\bibnamefont {Balcou}},
  \bibinfo {author} {\bibfnamefont {M.~Y.}\ \bibnamefont {Ivanov}}, \bibinfo
  {author} {\bibfnamefont {A.}~\bibnamefont {L'Huillier}},\ and\ \bibinfo
  {author} {\bibfnamefont {P.~B.}\ \bibnamefont {~}},\ }\bibfield  {title}
  {\bibinfo {title} {Theory of high-harmonic generation by low-frequency laser
  fields},\ }\href {https://doi.org/10.1103/PhysRevA.49.2117} {\bibfield
  {journal} {\bibinfo  {journal} {Phys. Rev. A}\ }\textbf {\bibinfo {volume}
  {49}},\ \bibinfo {pages} {2117} (\bibinfo {year} {1994})}\BibitemShut
  {NoStop}%
\bibitem [{\citenamefont {Ghimire}\ \emph {et~al.}(2011)\citenamefont
  {Ghimire}, \citenamefont {DiChiara}, \citenamefont {Sistrunk}, \citenamefont
  {Agostini}, \citenamefont {DiMauro},\ and\ \citenamefont
  {Reis}}]{Ghimire2011}%
  \BibitemOpen
  \bibfield  {author} {\bibinfo {author} {\bibfnamefont {S.}~\bibnamefont
  {Ghimire}}, \bibinfo {author} {\bibfnamefont {A.~D.}\ \bibnamefont
  {DiChiara}}, \bibinfo {author} {\bibfnamefont {E.}~\bibnamefont {Sistrunk}},
  \bibinfo {author} {\bibfnamefont {P.}~\bibnamefont {Agostini}}, \bibinfo
  {author} {\bibfnamefont {L.~F.}\ \bibnamefont {DiMauro}},\ and\ \bibinfo
  {author} {\bibfnamefont {D.~A.}\ \bibnamefont {Reis}},\ }\bibfield  {title}
  {\bibinfo {title} {Observation of high-order harmonic generation in a bulk
  crystal},\ }\href {https://doi.org/10.1038/nphys1847} {\bibfield  {journal}
  {\bibinfo  {journal} {Nature Physics}\ }\textbf {\bibinfo {volume} {7}},\
  \bibinfo {pages} {138} (\bibinfo {year} {2011})}\BibitemShut {NoStop}%
\bibitem [{\citenamefont {Ghimire}\ and\ \citenamefont
  {Reis}(2019)}]{Ghimire2019}%
  \BibitemOpen
  \bibfield  {author} {\bibinfo {author} {\bibfnamefont {S.}~\bibnamefont
  {Ghimire}}\ and\ \bibinfo {author} {\bibfnamefont {D.~A.}\ \bibnamefont
  {Reis}},\ }\bibfield  {title} {\bibinfo {title} {High-harmonic generation
  from solids},\ }\href {https://doi.org/10.1038/s41567-018-0315-5} {\bibfield
  {journal} {\bibinfo  {journal} {Nature Physics}\ }\textbf {\bibinfo {volume}
  {15}},\ \bibinfo {pages} {10} (\bibinfo {year} {2019})}\BibitemShut {NoStop}%
\bibitem [{\citenamefont {Goulielmakis}\ and\ \citenamefont
  {Brabec}(2022)}]{Goulielmakis2022}%
  \BibitemOpen
  \bibfield  {author} {\bibinfo {author} {\bibfnamefont {E.}~\bibnamefont
  {Goulielmakis}}\ and\ \bibinfo {author} {\bibfnamefont {T.}~\bibnamefont
  {Brabec}},\ }\bibfield  {title} {\bibinfo {title} {High harmonic generation
  in condensed matter},\ }\href {https://doi.org/10.1038/s41566-022-00988-y}
  {\bibfield  {journal} {\bibinfo  {journal} {Nature Photonics}\ }\textbf
  {\bibinfo {volume} {16}},\ \bibinfo {pages} {411} (\bibinfo {year}
  {2022})}\BibitemShut {NoStop}%
\bibitem [{\citenamefont {Golde}\ \emph {et~al.}(2008)\citenamefont {Golde},
  \citenamefont {Meier},\ and\ \citenamefont {Koch}}]{Golde2008}%
  \BibitemOpen
  \bibfield  {author} {\bibinfo {author} {\bibfnamefont {D.}~\bibnamefont
  {Golde}}, \bibinfo {author} {\bibfnamefont {T.}~\bibnamefont {Meier}},\ and\
  \bibinfo {author} {\bibfnamefont {S.~W.}\ \bibnamefont {Koch}},\ }\bibfield
  {title} {\bibinfo {title} {High harmonics generated in semiconductor
  nanostructures by the coupled dynamics of optical inter- and intraband
  excitations},\ }\href {https://doi.org/10.1103/PhysRevB.77.075330} {\bibfield
   {journal} {\bibinfo  {journal} {Phys. Rev. B}\ }\textbf {\bibinfo {volume}
  {77}},\ \bibinfo {pages} {075330} (\bibinfo {year} {2008})}\BibitemShut
  {NoStop}%
\bibitem [{\citenamefont {Vampa}\ \emph {et~al.}(2014)\citenamefont {Vampa},
  \citenamefont {McDonald}, \citenamefont {Orlando}, \citenamefont {Klug},
  \citenamefont {Corkum},\ and\ \citenamefont {Brabec}}]{Vampa2014}%
  \BibitemOpen
  \bibfield  {author} {\bibinfo {author} {\bibfnamefont {G.}~\bibnamefont
  {Vampa}}, \bibinfo {author} {\bibfnamefont {C.~R.}\ \bibnamefont {McDonald}},
  \bibinfo {author} {\bibfnamefont {G.}~\bibnamefont {Orlando}}, \bibinfo
  {author} {\bibfnamefont {D.~D.}\ \bibnamefont {Klug}}, \bibinfo {author}
  {\bibfnamefont {P.~B.}\ \bibnamefont {Corkum}},\ and\ \bibinfo {author}
  {\bibfnamefont {T.}~\bibnamefont {Brabec}},\ }\bibfield  {title} {\bibinfo
  {title} {Theoretical analysis of high-harmonic generation in solids},\ }\href
  {https://doi.org/10.1103/PhysRevLett.113.073901} {\bibfield  {journal}
  {\bibinfo  {journal} {Phys. Rev. Lett.}\ }\textbf {\bibinfo {volume} {113}},\
  \bibinfo {pages} {073901} (\bibinfo {year} {2014})}\BibitemShut {NoStop}%
\bibitem [{\citenamefont {Vampa}\ \emph {et~al.}(2015)\citenamefont {Vampa},
  \citenamefont {McDonald}, \citenamefont {Orlando}, \citenamefont {Corkum},\
  and\ \citenamefont {Brabec}}]{Vampa2015}%
  \BibitemOpen
  \bibfield  {author} {\bibinfo {author} {\bibfnamefont {G.}~\bibnamefont
  {Vampa}}, \bibinfo {author} {\bibfnamefont {C.~R.}\ \bibnamefont {McDonald}},
  \bibinfo {author} {\bibfnamefont {G.}~\bibnamefont {Orlando}}, \bibinfo
  {author} {\bibfnamefont {P.~B.}\ \bibnamefont {Corkum}},\ and\ \bibinfo
  {author} {\bibfnamefont {T.}~\bibnamefont {Brabec}},\ }\bibfield  {title}
  {\bibinfo {title} {Semiclassical analysis of high harmonic generation in bulk
  crystals},\ }\href {https://doi.org/10.1103/PhysRevB.91.064302} {\bibfield
  {journal} {\bibinfo  {journal} {Phys. Rev. B}\ }\textbf {\bibinfo {volume}
  {91}},\ \bibinfo {pages} {064302} (\bibinfo {year} {2015})}\BibitemShut
  {NoStop}%
\bibitem [{\citenamefont {Vampa}\ and\ \citenamefont
  {Brabec}(2017)}]{Vampa2017}%
  \BibitemOpen
  \bibfield  {author} {\bibinfo {author} {\bibfnamefont {G.}~\bibnamefont
  {Vampa}}\ and\ \bibinfo {author} {\bibfnamefont {T.}~\bibnamefont {Brabec}},\
  }\bibfield  {title} {\bibinfo {title} {Merge of high harmonic generation from
  gases and solids and its implications for attosecond science},\ }\href
  {https://doi.org/10.1088/1361-6455/aa528d} {\bibfield  {journal} {\bibinfo
  {journal} {J. Phys. B}\ }\textbf {\bibinfo {volume} {50}},\ \bibinfo {pages}
  {083001} (\bibinfo {year} {2017})}\BibitemShut {NoStop}%
\bibitem [{\citenamefont {Bogatskaya}\ \emph {et~al.}(2017)\citenamefont
  {Bogatskaya}, \citenamefont {Volkova},\ and\ \citenamefont
  {Popov}}]{Bogastkaya2017}%
  \BibitemOpen
  \bibfield  {author} {\bibinfo {author} {\bibfnamefont {A.~V.}\ \bibnamefont
  {Bogatskaya}}, \bibinfo {author} {\bibfnamefont {E.~A.}\ \bibnamefont
  {Volkova}},\ and\ \bibinfo {author} {\bibfnamefont {A.~M.}\ \bibnamefont
  {Popov}},\ }\bibfield  {title} {\bibinfo {title} {Spontaneous emission of
  atoms in a strong laser field},\ }\href
  {https://doi.org/10.1134/S1063776117090114} {\bibfield  {journal} {\bibinfo
  {journal} {Journal of Experimental and Theoretical Physics}\ }\textbf
  {\bibinfo {volume} {125}},\ \bibinfo {pages} {587} (\bibinfo {year}
  {2017})}\BibitemShut {NoStop}%
\bibitem [{\citenamefont {Gombk\"ot\ifmmode~\mbox{\H{o}}\else \H{o}\fi{}}\
  \emph {et~al.}(2021)\citenamefont {Gombk\"ot\ifmmode~\mbox{\H{o}}\else
  \H{o}\fi{}}, \citenamefont {F\"oldi},\ and\ \citenamefont
  {Varr\'o}}]{Gombkoto2021}%
  \BibitemOpen
  \bibfield  {author} {\bibinfo {author} {\bibfnamefont {A.}~\bibnamefont
  {Gombk\"ot\ifmmode~\mbox{\H{o}}\else \H{o}\fi{}}}, \bibinfo {author}
  {\bibfnamefont {P.}~\bibnamefont {F\"oldi}},\ and\ \bibinfo {author}
  {\bibfnamefont {S.}~\bibnamefont {Varr\'o}},\ }\bibfield  {title} {\bibinfo
  {title} {Quantum-optical description of photon statistics and cross
  correlations in high-order harmonic generation},\ }\href
  {https://doi.org/10.1103/PhysRevA.104.033703} {\bibfield  {journal} {\bibinfo
   {journal} {Phys. Rev. A}\ }\textbf {\bibinfo {volume} {104}},\ \bibinfo
  {pages} {033703} (\bibinfo {year} {2021})}\BibitemShut {NoStop}%
\bibitem [{\citenamefont {Gorlach}\ \emph {et~al.}(2020)\citenamefont
  {Gorlach}, \citenamefont {Neufeld}, \citenamefont {Rivera}, \citenamefont
  {Cohen},\ and\ \citenamefont {Kaminer}}]{KaminerNatCom2020}%
  \BibitemOpen
  \bibfield  {author} {\bibinfo {author} {\bibfnamefont {A.}~\bibnamefont
  {Gorlach}}, \bibinfo {author} {\bibfnamefont {O.}~\bibnamefont {Neufeld}},
  \bibinfo {author} {\bibfnamefont {N.}~\bibnamefont {Rivera}}, \bibinfo
  {author} {\bibfnamefont {O.}~\bibnamefont {Cohen}},\ and\ \bibinfo {author}
  {\bibfnamefont {I.}~\bibnamefont {Kaminer}},\ }\bibfield  {title} {\bibinfo
  {title} {The quantum-optical nature of high harmonic generation},\
  }\href@noop {} {\bibfield  {journal} {\bibinfo  {journal} {Nature
  Communications}\ }\textbf {\bibinfo {volume} {11}},\ \bibinfo {pages} {4598}
  (\bibinfo {year} {2020})}\BibitemShut {NoStop}%
\bibitem [{\citenamefont {Pizzi}\ \emph {et~al.}(2023)\citenamefont {Pizzi},
  \citenamefont {Gorlach}, \citenamefont {Rivera}, \citenamefont {Nunnenkamp},\
  and\ \citenamefont {Kaminer}}]{KaminerManyBody2023}%
  \BibitemOpen
  \bibfield  {author} {\bibinfo {author} {\bibfnamefont {A.}~\bibnamefont
  {Pizzi}}, \bibinfo {author} {\bibfnamefont {A.}~\bibnamefont {Gorlach}},
  \bibinfo {author} {\bibfnamefont {N.}~\bibnamefont {Rivera}}, \bibinfo
  {author} {\bibfnamefont {A.}~\bibnamefont {Nunnenkamp}},\ and\ \bibinfo
  {author} {\bibfnamefont {I.}~\bibnamefont {Kaminer}},\ }\bibfield  {title}
  {\bibinfo {title} {Light emission from strongly driven many-body systems},\
  }\href@noop {} {\bibfield  {journal} {\bibinfo  {journal} {Nature Physics}\
  }\textbf {\bibinfo {volume} {19}},\ \bibinfo {pages} {551} (\bibinfo {year}
  {2023})}\BibitemShut {NoStop}%
\bibitem [{\citenamefont {Stammer}\ \emph
  {et~al.}(2023{\natexlab{a}})\citenamefont {Stammer}, \citenamefont
  {Rivera-Dean}, \citenamefont {Maxwell}, \citenamefont {Lamprou},
  \citenamefont {Ord\'o\~nez}, \citenamefont {Ciappina}, \citenamefont
  {Tzallas},\ and\ \citenamefont
  {Lewenstein}}]{StammerLewenstein_tutorial_Quantum_state_engineering}%
  \BibitemOpen
  \bibfield  {author} {\bibinfo {author} {\bibfnamefont {P.}~\bibnamefont
  {Stammer}}, \bibinfo {author} {\bibfnamefont {J.}~\bibnamefont
  {Rivera-Dean}}, \bibinfo {author} {\bibfnamefont {A.}~\bibnamefont
  {Maxwell}}, \bibinfo {author} {\bibfnamefont {T.}~\bibnamefont {Lamprou}},
  \bibinfo {author} {\bibfnamefont {A.}~\bibnamefont {Ord\'o\~nez}}, \bibinfo
  {author} {\bibfnamefont {M.~F.}\ \bibnamefont {Ciappina}}, \bibinfo {author}
  {\bibfnamefont {P.}~\bibnamefont {Tzallas}},\ and\ \bibinfo {author}
  {\bibfnamefont {M.}~\bibnamefont {Lewenstein}},\ }\bibfield  {title}
  {\bibinfo {title} {Quantum electrodynamics of intense laser-matter
  interactions: A tool for quantum state engineering},\ }\href
  {https://doi.org/10.1103/PRXQuantum.4.010201} {\bibfield  {journal} {\bibinfo
   {journal} {PRX Quantum}\ }\textbf {\bibinfo {volume} {4}},\ \bibinfo {pages}
  {010201} (\bibinfo {year} {2023}{\natexlab{a}})}\BibitemShut {NoStop}%
\bibitem [{\citenamefont {Rivera-Dean}\ \emph {et~al.}(2023)\citenamefont
  {Rivera-Dean}, \citenamefont {Stammer}, \citenamefont {Maxwell},
  \citenamefont {Lamprou}, \citenamefont {Ordóñez}, \citenamefont {Pisanty},
  \citenamefont {Tzallas}, \citenamefont {Lewenstein},\ and\ \citenamefont
  {Ciappina}}]{Riveradean2023entanglementHHG_SolidState}%
  \BibitemOpen
  \bibfield  {author} {\bibinfo {author} {\bibfnamefont {J.}~\bibnamefont
  {Rivera-Dean}}, \bibinfo {author} {\bibfnamefont {P.}~\bibnamefont
  {Stammer}}, \bibinfo {author} {\bibfnamefont {A.~S.}\ \bibnamefont
  {Maxwell}}, \bibinfo {author} {\bibfnamefont {T.}~\bibnamefont {Lamprou}},
  \bibinfo {author} {\bibfnamefont {A.~F.}\ \bibnamefont {Ordóñez}}, \bibinfo
  {author} {\bibfnamefont {E.}~\bibnamefont {Pisanty}}, \bibinfo {author}
  {\bibfnamefont {P.}~\bibnamefont {Tzallas}}, \bibinfo {author} {\bibfnamefont
  {M.}~\bibnamefont {Lewenstein}},\ and\ \bibinfo {author} {\bibfnamefont
  {M.~F.}\ \bibnamefont {Ciappina}},\ }\href@noop {} {\bibinfo {title}
  {Entanglement and non-classical states of light in a strong-laser driven
  solid-state system}} (\bibinfo {year} {2023}),\ \Eprint
  {https://arxiv.org/abs/2211.00033} {arXiv:2211.00033 [quant-ph]} \BibitemShut
  {NoStop}%
\bibitem [{\citenamefont {Gonoskov}\ \emph {et~al.}(2022)\citenamefont
  {Gonoskov}, \citenamefont {Sondenheimer}, \citenamefont {Hünecke},
  \citenamefont {Kartashov}, \citenamefont {Peschel},\ and\ \citenamefont
  {Gräfe}}]{GonoskovGrafe2022nonclassical}%
  \BibitemOpen
  \bibfield  {author} {\bibinfo {author} {\bibfnamefont {I.}~\bibnamefont
  {Gonoskov}}, \bibinfo {author} {\bibfnamefont {R.}~\bibnamefont
  {Sondenheimer}}, \bibinfo {author} {\bibfnamefont {C.}~\bibnamefont
  {Hünecke}}, \bibinfo {author} {\bibfnamefont {D.}~\bibnamefont {Kartashov}},
  \bibinfo {author} {\bibfnamefont {U.}~\bibnamefont {Peschel}},\ and\ \bibinfo
  {author} {\bibfnamefont {S.}~\bibnamefont {Gräfe}},\ }\href@noop {}
  {\bibinfo {title} {Nonclassical light generation and control from
  laser-driven semiconductor intraband excitations}} (\bibinfo {year} {2022}),\
  \Eprint {https://arxiv.org/abs/2211.06177} {arXiv:2211.06177 [quant-ph]}
  \BibitemShut {NoStop}%
\bibitem [{\citenamefont {Even~Tzur}\ \emph {et~al.}(2023)\citenamefont
  {Even~Tzur}, \citenamefont {Birk}, \citenamefont {Gorlach}, \citenamefont
  {Kr{\"u}ger}, \citenamefont {Kaminer},\ and\ \citenamefont
  {Cohen}}]{Tzur2023_photon_statistics_force}%
  \BibitemOpen
  \bibfield  {author} {\bibinfo {author} {\bibfnamefont {M.}~\bibnamefont
  {Even~Tzur}}, \bibinfo {author} {\bibfnamefont {M.}~\bibnamefont {Birk}},
  \bibinfo {author} {\bibfnamefont {A.}~\bibnamefont {Gorlach}}, \bibinfo
  {author} {\bibfnamefont {M.}~\bibnamefont {Kr{\"u}ger}}, \bibinfo {author}
  {\bibfnamefont {I.}~\bibnamefont {Kaminer}},\ and\ \bibinfo {author}
  {\bibfnamefont {O.}~\bibnamefont {Cohen}},\ }\bibfield  {title} {\bibinfo
  {title} {Photon-statistics force in ultrafast electron dynamics},\ }\href
  {https://doi.org/10.1038/s41566-023-01209-w} {\bibfield  {journal} {\bibinfo
  {journal} {Nature Photonics}\ }\textbf {\bibinfo {volume} {17}},\ \bibinfo
  {pages} {501} (\bibinfo {year} {2023})}\BibitemShut {NoStop}%
\bibitem [{\citenamefont {Gorlach}\ \emph {et~al.}(2022)\citenamefont
  {Gorlach}, \citenamefont {Tzur}, \citenamefont {Birk}, \citenamefont
  {Krüger}, \citenamefont {Rivera}, \citenamefont {Cohen},\ and\ \citenamefont
  {Kaminer}}]{GolarchKaminerHHGWithQuantumLight2022}%
  \BibitemOpen
  \bibfield  {author} {\bibinfo {author} {\bibfnamefont {A.}~\bibnamefont
  {Gorlach}}, \bibinfo {author} {\bibfnamefont {M.~E.}\ \bibnamefont {Tzur}},
  \bibinfo {author} {\bibfnamefont {M.}~\bibnamefont {Birk}}, \bibinfo {author}
  {\bibfnamefont {M.}~\bibnamefont {Krüger}}, \bibinfo {author} {\bibfnamefont
  {N.}~\bibnamefont {Rivera}}, \bibinfo {author} {\bibfnamefont
  {O.}~\bibnamefont {Cohen}},\ and\ \bibinfo {author} {\bibfnamefont
  {I.}~\bibnamefont {Kaminer}},\ }\href@noop {} {\bibinfo {title} {High
  harmonic generation driven by quantum light}} (\bibinfo {year} {2022}),\
  \Eprint {https://arxiv.org/abs/2211.03188} {arXiv:2211.03188
  [physics.optics]} \BibitemShut {NoStop}%
\bibitem [{\citenamefont {Stammer}\ \emph
  {et~al.}(2023{\natexlab{b}})\citenamefont {Stammer}, \citenamefont
  {Rivera-Dean}, \citenamefont {Maxwell}, \citenamefont {Lamprou},
  \citenamefont {Argüello-Luengo}, \citenamefont {Tzallas}, \citenamefont
  {Ciappina},\ and\ \citenamefont {Lewenstein}}]{Stammer2023entanglement}%
  \BibitemOpen
  \bibfield  {author} {\bibinfo {author} {\bibfnamefont {P.}~\bibnamefont
  {Stammer}}, \bibinfo {author} {\bibfnamefont {J.}~\bibnamefont
  {Rivera-Dean}}, \bibinfo {author} {\bibfnamefont {A.~S.}\ \bibnamefont
  {Maxwell}}, \bibinfo {author} {\bibfnamefont {T.}~\bibnamefont {Lamprou}},
  \bibinfo {author} {\bibfnamefont {J.}~\bibnamefont {Argüello-Luengo}},
  \bibinfo {author} {\bibfnamefont {P.}~\bibnamefont {Tzallas}}, \bibinfo
  {author} {\bibfnamefont {M.~F.}\ \bibnamefont {Ciappina}},\ and\ \bibinfo
  {author} {\bibfnamefont {M.}~\bibnamefont {Lewenstein}},\ }\href@noop {}
  {\bibinfo {title} {Entanglement and squeezing of the optical field modes in
  high harmonic generation}} (\bibinfo {year} {2023}{\natexlab{b}}),\ \Eprint
  {https://arxiv.org/abs/2310.15030} {arXiv:2310.15030 [quant-ph]} \BibitemShut
  {NoStop}%
\bibitem [{\citenamefont {Yangaliev}\ \emph {et~al.}(2020)\citenamefont
  {Yangaliev}, \citenamefont {Krainov},\ and\ \citenamefont
  {Tolstikhin}}]{Yangaliev2020_OlegToltikhin}%
  \BibitemOpen
  \bibfield  {author} {\bibinfo {author} {\bibfnamefont {D.~N.}\ \bibnamefont
  {Yangaliev}}, \bibinfo {author} {\bibfnamefont {V.~P.}\ \bibnamefont
  {Krainov}},\ and\ \bibinfo {author} {\bibfnamefont {O.~I.}\ \bibnamefont
  {Tolstikhin}},\ }\bibfield  {title} {\bibinfo {title} {Quantum theory of
  radiation by nonstationary systems with application to high-order harmonic
  generation},\ }\href {https://doi.org/10.1103/PhysRevA.101.013410} {\bibfield
   {journal} {\bibinfo  {journal} {Phys. Rev. A}\ }\textbf {\bibinfo {volume}
  {101}},\ \bibinfo {pages} {013410} (\bibinfo {year} {2020})}\BibitemShut
  {NoStop}%
\bibitem [{\citenamefont {Stammer}(2022)}]{Stammer2022_mmt_paper}%
  \BibitemOpen
  \bibfield  {author} {\bibinfo {author} {\bibfnamefont {P.}~\bibnamefont
  {Stammer}},\ }\bibfield  {title} {\bibinfo {title} {Theory of entanglement
  and measurement in high-order harmonic generation},\ }\href
  {https://doi.org/10.1103/PhysRevA.106.L050402} {\bibfield  {journal}
  {\bibinfo  {journal} {Phys. Rev. A}\ }\textbf {\bibinfo {volume} {106}},\
  \bibinfo {pages} {L050402} (\bibinfo {year} {2022})}\BibitemShut {NoStop}%
\bibitem [{\citenamefont {Tzur}\ \emph {et~al.}(2023)\citenamefont {Tzur},
  \citenamefont {Birk}, \citenamefont {Gorlach}, \citenamefont {Kaminer},
  \citenamefont {Krueger},\ and\ \citenamefont
  {Cohen}}]{Tzur2023generation_of_squeezed_HHG}%
  \BibitemOpen
  \bibfield  {author} {\bibinfo {author} {\bibfnamefont {M.~E.}\ \bibnamefont
  {Tzur}}, \bibinfo {author} {\bibfnamefont {M.}~\bibnamefont {Birk}}, \bibinfo
  {author} {\bibfnamefont {A.}~\bibnamefont {Gorlach}}, \bibinfo {author}
  {\bibfnamefont {I.}~\bibnamefont {Kaminer}}, \bibinfo {author} {\bibfnamefont
  {M.}~\bibnamefont {Krueger}},\ and\ \bibinfo {author} {\bibfnamefont
  {O.}~\bibnamefont {Cohen}},\ }\href@noop {} {\bibinfo {title} {Generation of
  squeezed high-order harmonics}} (\bibinfo {year} {2023}),\ \Eprint
  {https://arxiv.org/abs/2311.11257} {arXiv:2311.11257 [quant-ph]} \BibitemShut
  {NoStop}%
\bibitem [{\citenamefont {Lewenstein}\ \emph {et~al.}(2021)\citenamefont
  {Lewenstein}, \citenamefont {Ciappina}, \citenamefont {Pisanty},
  \citenamefont {Rivera-Dean}, \citenamefont {Stammer}, \citenamefont
  {Lamprou},\ and\ \citenamefont {Tzallas}}]{LewensteinCatState2021}%
  \BibitemOpen
  \bibfield  {author} {\bibinfo {author} {\bibfnamefont {M.}~\bibnamefont
  {Lewenstein}}, \bibinfo {author} {\bibfnamefont {M.~F.}\ \bibnamefont
  {Ciappina}}, \bibinfo {author} {\bibfnamefont {E.}~\bibnamefont {Pisanty}},
  \bibinfo {author} {\bibfnamefont {J.}~\bibnamefont {Rivera-Dean}}, \bibinfo
  {author} {\bibfnamefont {P.}~\bibnamefont {Stammer}}, \bibinfo {author}
  {\bibfnamefont {T.}~\bibnamefont {Lamprou}},\ and\ \bibinfo {author}
  {\bibfnamefont {P.}~\bibnamefont {Tzallas}},\ }\bibfield  {title} {\bibinfo
  {title} {Generation of optical schr{\"o}dinger cat states in intense
  laser--matter interactions},\ }\href@noop {} {\bibfield  {journal} {\bibinfo
  {journal} {Nature Physics}\ }\textbf {\bibinfo {volume} {17}},\ \bibinfo
  {pages} {1104} (\bibinfo {year} {2021})}\BibitemShut {NoStop}%
\bibitem [{\citenamefont {Tsatrafyllis}\ \emph {et~al.}(2017)\citenamefont
  {Tsatrafyllis}, \citenamefont {Kominis}, \citenamefont {Gonoskov},\ and\
  \citenamefont {Tzallas}}]{Tsatrafyllis2017}%
  \BibitemOpen
  \bibfield  {author} {\bibinfo {author} {\bibfnamefont {N.}~\bibnamefont
  {Tsatrafyllis}}, \bibinfo {author} {\bibfnamefont {I.~K.}\ \bibnamefont
  {Kominis}}, \bibinfo {author} {\bibfnamefont {I.~A.}\ \bibnamefont
  {Gonoskov}},\ and\ \bibinfo {author} {\bibfnamefont {P.}~\bibnamefont
  {Tzallas}},\ }\bibfield  {title} {\bibinfo {title} {High-order harmonics
  measured by the photon statistics of the infrared driving-field exiting the
  atomic medium},\ }\href {https://doi.org/10.1038/ncomms15170} {\bibfield
  {journal} {\bibinfo  {journal} {Nature Communications}\ }\textbf {\bibinfo
  {volume} {8}},\ \bibinfo {pages} {15170} (\bibinfo {year}
  {2017})}\BibitemShut {NoStop}%
\bibitem [{\citenamefont {Tsatrafyllis}\ \emph {et~al.}(2019)\citenamefont
  {Tsatrafyllis}, \citenamefont {K\"uhn}, \citenamefont {Dumergue},
  \citenamefont {Foldi}, \citenamefont {Kahaly}, \citenamefont {Cormier},
  \citenamefont {Gonoskov}, \citenamefont {Kiss}, \citenamefont {Varju},
  \citenamefont {Varro},\ and\ \citenamefont {Tzallas}}]{Tsatrafyllis2019}%
  \BibitemOpen
  \bibfield  {author} {\bibinfo {author} {\bibfnamefont {N.}~\bibnamefont
  {Tsatrafyllis}}, \bibinfo {author} {\bibfnamefont {S.}~\bibnamefont
  {K\"uhn}}, \bibinfo {author} {\bibfnamefont {M.}~\bibnamefont {Dumergue}},
  \bibinfo {author} {\bibfnamefont {P.}~\bibnamefont {Foldi}}, \bibinfo
  {author} {\bibfnamefont {S.}~\bibnamefont {Kahaly}}, \bibinfo {author}
  {\bibfnamefont {E.}~\bibnamefont {Cormier}}, \bibinfo {author} {\bibfnamefont
  {I.~A.}\ \bibnamefont {Gonoskov}}, \bibinfo {author} {\bibfnamefont
  {B.}~\bibnamefont {Kiss}}, \bibinfo {author} {\bibfnamefont {K.}~\bibnamefont
  {Varju}}, \bibinfo {author} {\bibfnamefont {S.}~\bibnamefont {Varro}},\ and\
  \bibinfo {author} {\bibfnamefont {P.}~\bibnamefont {Tzallas}},\ }\bibfield
  {title} {\bibinfo {title} {Quantum optical signatures in a strong laser pulse
  after interaction with semiconductors},\ }\href
  {https://doi.org/10.1103/PhysRevLett.122.193602} {\bibfield  {journal}
  {\bibinfo  {journal} {Phys. Rev. Lett.}\ }\textbf {\bibinfo {volume} {122}},\
  \bibinfo {pages} {193602} (\bibinfo {year} {2019})}\BibitemShut {NoStop}%
\bibitem [{\citenamefont {Rivera-Dean}\ \emph {et~al.}(2022)\citenamefont
  {Rivera-Dean}, \citenamefont {Lamprou}, \citenamefont {Pisanty},
  \citenamefont {Stammer}, \citenamefont {Ord\'o\~nez}, \citenamefont
  {Maxwell}, \citenamefont {Ciappina}, \citenamefont {Lewenstein},\ and\
  \citenamefont {Tzallas}}]{RiveraDean2022}%
  \BibitemOpen
  \bibfield  {author} {\bibinfo {author} {\bibfnamefont {J.}~\bibnamefont
  {Rivera-Dean}}, \bibinfo {author} {\bibfnamefont {T.}~\bibnamefont
  {Lamprou}}, \bibinfo {author} {\bibfnamefont {E.}~\bibnamefont {Pisanty}},
  \bibinfo {author} {\bibfnamefont {P.}~\bibnamefont {Stammer}}, \bibinfo
  {author} {\bibfnamefont {A.~F.}\ \bibnamefont {Ord\'o\~nez}}, \bibinfo
  {author} {\bibfnamefont {A.~S.}\ \bibnamefont {Maxwell}}, \bibinfo {author}
  {\bibfnamefont {M.~F.}\ \bibnamefont {Ciappina}}, \bibinfo {author}
  {\bibfnamefont {M.}~\bibnamefont {Lewenstein}},\ and\ \bibinfo {author}
  {\bibfnamefont {P.}~\bibnamefont {Tzallas}},\ }\bibfield  {title} {\bibinfo
  {title} {Strong laser fields and their power to generate controllable
  high-photon-number coherent-state superpositions},\ }\href
  {https://doi.org/10.1103/PhysRevA.105.033714} {\bibfield  {journal} {\bibinfo
   {journal} {Phys. Rev. A}\ }\textbf {\bibinfo {volume} {105}},\ \bibinfo
  {pages} {033714} (\bibinfo {year} {2022})}\BibitemShut {NoStop}%
\bibitem [{\citenamefont {Ourjoumtsev}\ \emph {et~al.}(2006)\citenamefont
  {Ourjoumtsev}, \citenamefont {Tualle-Brouri}, \citenamefont {Laurat},\ and\
  \citenamefont {Grangier}}]{Ourjoumtsev2006}%
  \BibitemOpen
  \bibfield  {author} {\bibinfo {author} {\bibfnamefont {A.}~\bibnamefont
  {Ourjoumtsev}}, \bibinfo {author} {\bibfnamefont {R.}~\bibnamefont
  {Tualle-Brouri}}, \bibinfo {author} {\bibfnamefont {J.}~\bibnamefont
  {Laurat}},\ and\ \bibinfo {author} {\bibfnamefont {P.}~\bibnamefont
  {Grangier}},\ }\bibfield  {title} {\bibinfo {title} {Generating optical
  schrödinger kittens for quantum information processing},\ }\href
  {https://doi.org/10.1126/science.1122858} {\bibfield  {journal} {\bibinfo
  {journal} {Science}\ }\textbf {\bibinfo {volume} {312}},\ \bibinfo {pages}
  {83} (\bibinfo {year} {2006})},\ \Eprint
  {https://arxiv.org/abs/https://www.science.org/doi/pdf/10.1126/science.1122858}
  {https://www.science.org/doi/pdf/10.1126/science.1122858} \BibitemShut
  {NoStop}%
\bibitem [{\citenamefont {Ourjoumtsev}\ \emph {et~al.}(2007)\citenamefont
  {Ourjoumtsev}, \citenamefont {Jeong}, \citenamefont {Tualle-Brouri},\ and\
  \citenamefont {Grangier}}]{Ourjoumtsev2007}%
  \BibitemOpen
  \bibfield  {author} {\bibinfo {author} {\bibfnamefont {A.}~\bibnamefont
  {Ourjoumtsev}}, \bibinfo {author} {\bibfnamefont {H.}~\bibnamefont {Jeong}},
  \bibinfo {author} {\bibfnamefont {R.}~\bibnamefont {Tualle-Brouri}},\ and\
  \bibinfo {author} {\bibfnamefont {P.}~\bibnamefont {Grangier}},\ }\bibfield
  {title} {\bibinfo {title} {Generation of optical `schr{\"o}dinger cats'from
  photon number states},\ }\href {https://doi.org/10.1038/nature06054}
  {\bibfield  {journal} {\bibinfo  {journal} {Nature}\ }\textbf {\bibinfo
  {volume} {448}},\ \bibinfo {pages} {784} (\bibinfo {year}
  {2007})}\BibitemShut {NoStop}%
\bibitem [{\citenamefont {Zavatta}\ \emph {et~al.}(2004)\citenamefont
  {Zavatta}, \citenamefont {Viciani},\ and\ \citenamefont
  {Bellini}}]{Zavatta2004}%
  \BibitemOpen
  \bibfield  {author} {\bibinfo {author} {\bibfnamefont {A.}~\bibnamefont
  {Zavatta}}, \bibinfo {author} {\bibfnamefont {S.}~\bibnamefont {Viciani}},\
  and\ \bibinfo {author} {\bibfnamefont {M.}~\bibnamefont {Bellini}},\
  }\bibfield  {title} {\bibinfo {title} {Quantum-to-classical transition with
  single-photon-added coherent states of light},\ }\href
  {https://doi.org/10.1126/science.1103190} {\bibfield  {journal} {\bibinfo
  {journal} {Science}\ }\textbf {\bibinfo {volume} {306}},\ \bibinfo {pages}
  {660} (\bibinfo {year} {2004})},\ \Eprint
  {https://arxiv.org/abs/https://www.science.org/doi/pdf/10.1126/science.1103190}
  {https://www.science.org/doi/pdf/10.1126/science.1103190} \BibitemShut
  {NoStop}%
\bibitem [{\citenamefont {Acín}\ \emph {et~al.}(2018)\citenamefont {Acín},
  \citenamefont {Bloch}, \citenamefont {Buhrman}, \citenamefont {Calarco},
  \citenamefont {Eichler}, \citenamefont {Eisert}, \citenamefont {Esteve},
  \citenamefont {Gisin}, \citenamefont {Glaser}, \citenamefont {Jelezko},
  \citenamefont {Kuhr}, \citenamefont {Lewenstein}, \citenamefont {Riedel},
  \citenamefont {Schmidt}, \citenamefont {Thew}, \citenamefont {Wallraff},
  \citenamefont {Walmsley},\ and\ \citenamefont
  {Wilhelm}}]{Acin2018_QTRoadmap}%
  \BibitemOpen
  \bibfield  {author} {\bibinfo {author} {\bibfnamefont {A.}~\bibnamefont
  {Acín}}, \bibinfo {author} {\bibfnamefont {I.}~\bibnamefont {Bloch}},
  \bibinfo {author} {\bibfnamefont {H.}~\bibnamefont {Buhrman}}, \bibinfo
  {author} {\bibfnamefont {T.}~\bibnamefont {Calarco}}, \bibinfo {author}
  {\bibfnamefont {C.}~\bibnamefont {Eichler}}, \bibinfo {author} {\bibfnamefont
  {J.}~\bibnamefont {Eisert}}, \bibinfo {author} {\bibfnamefont
  {D.}~\bibnamefont {Esteve}}, \bibinfo {author} {\bibfnamefont
  {N.}~\bibnamefont {Gisin}}, \bibinfo {author} {\bibfnamefont {S.~J.}\
  \bibnamefont {Glaser}}, \bibinfo {author} {\bibfnamefont {F.}~\bibnamefont
  {Jelezko}}, \bibinfo {author} {\bibfnamefont {S.}~\bibnamefont {Kuhr}},
  \bibinfo {author} {\bibfnamefont {M.}~\bibnamefont {Lewenstein}}, \bibinfo
  {author} {\bibfnamefont {M.~F.}\ \bibnamefont {Riedel}}, \bibinfo {author}
  {\bibfnamefont {P.~O.}\ \bibnamefont {Schmidt}}, \bibinfo {author}
  {\bibfnamefont {R.}~\bibnamefont {Thew}}, \bibinfo {author} {\bibfnamefont
  {A.}~\bibnamefont {Wallraff}}, \bibinfo {author} {\bibfnamefont
  {I.}~\bibnamefont {Walmsley}},\ and\ \bibinfo {author} {\bibfnamefont
  {F.~K.}\ \bibnamefont {Wilhelm}},\ }\bibfield  {title} {\bibinfo {title} {The
  quantum technologies roadmap: a european community view},\ }\href
  {https://doi.org/10.1088/1367-2630/aad1ea} {\bibfield  {journal} {\bibinfo
  {journal} {New Journal of Physics}\ }\textbf {\bibinfo {volume} {20}},\
  \bibinfo {pages} {080201} (\bibinfo {year} {2018})}\BibitemShut {NoStop}%
\bibitem [{\citenamefont {Deutsch}(2020)}]{DeutschSecondQuantumRevolution}%
  \BibitemOpen
  \bibfield  {author} {\bibinfo {author} {\bibfnamefont {I.~H.}\ \bibnamefont
  {Deutsch}},\ }\bibfield  {title} {\bibinfo {title} {Harnessing the power of
  the second quantum revolution},\ }\href
  {https://doi.org/10.1103/PRXQuantum.1.020101} {\bibfield  {journal} {\bibinfo
   {journal} {PRX Quantum}\ }\textbf {\bibinfo {volume} {1}},\ \bibinfo {pages}
  {020101} (\bibinfo {year} {2020})}\BibitemShut {NoStop}%
\bibitem [{\citenamefont {Polino}\ \emph {et~al.}(2020)\citenamefont {Polino},
  \citenamefont {Valeri}, \citenamefont {Spagnolo},\ and\ \citenamefont
  {Sciarrino}}]{Polino2020}%
  \BibitemOpen
  \bibfield  {author} {\bibinfo {author} {\bibfnamefont {E.}~\bibnamefont
  {Polino}}, \bibinfo {author} {\bibfnamefont {M.}~\bibnamefont {Valeri}},
  \bibinfo {author} {\bibfnamefont {N.}~\bibnamefont {Spagnolo}},\ and\
  \bibinfo {author} {\bibfnamefont {F.}~\bibnamefont {Sciarrino}},\ }\bibfield
  {title} {\bibinfo {title} {{Photonic quantum metrology}},\ }\href
  {https://doi.org/10.1116/5.0007577} {\bibfield  {journal} {\bibinfo
  {journal} {AVS Quantum Science}\ }\textbf {\bibinfo {volume} {2}},\ \bibinfo
  {pages} {024703} (\bibinfo {year} {2020})},\ \Eprint
  {https://arxiv.org/abs/https://pubs.aip.org/avs/aqs/article-pdf/doi/10.1116/5.0007577/13961558/024703\_1\_online.pdf}
  {https://pubs.aip.org/avs/aqs/article-pdf/doi/10.1116/5.0007577/13961558/024703\_1\_online.pdf}
  \BibitemShut {NoStop}%
\bibitem [{\citenamefont {Barbieri}(2022)}]{Barbieri2022}%
  \BibitemOpen
  \bibfield  {author} {\bibinfo {author} {\bibfnamefont {M.}~\bibnamefont
  {Barbieri}},\ }\bibfield  {title} {\bibinfo {title} {Optical quantum
  metrology},\ }\href {https://doi.org/10.1103/PRXQuantum.3.010202} {\bibfield
  {journal} {\bibinfo  {journal} {PRX Quantum}\ }\textbf {\bibinfo {volume}
  {3}},\ \bibinfo {pages} {010202} (\bibinfo {year} {2022})}\BibitemShut
  {NoStop}%
\bibitem [{\citenamefont {Silva}\ \emph {et~al.}(2018)\citenamefont {Silva},
  \citenamefont {Blinov}, \citenamefont {Rubtsov}, \citenamefont {Smirnova},\
  and\ \citenamefont {Ivanov}}]{Silva2018}%
  \BibitemOpen
  \bibfield  {author} {\bibinfo {author} {\bibfnamefont {R.~E.~F.}\
  \bibnamefont {Silva}}, \bibinfo {author} {\bibfnamefont {I.~V.}\ \bibnamefont
  {Blinov}}, \bibinfo {author} {\bibfnamefont {A.~N.}\ \bibnamefont {Rubtsov}},
  \bibinfo {author} {\bibfnamefont {O.}~\bibnamefont {Smirnova}},\ and\
  \bibinfo {author} {\bibfnamefont {M.}~\bibnamefont {Ivanov}},\ }\bibfield
  {title} {\bibinfo {title} {High-harmonic spectroscopy of ultrafast many-body
  dynamics in strongly correlated systems},\ }\href
  {https://doi.org/10.1038/s41566-018-0129-0} {\bibfield  {journal} {\bibinfo
  {journal} {Nature Photonics}\ }\textbf {\bibinfo {volume} {12}},\ \bibinfo
  {pages} {266} (\bibinfo {year} {2018})}\BibitemShut {NoStop}%
\bibitem [{\citenamefont {Murakami}\ \emph {et~al.}(2018)\citenamefont
  {Murakami}, \citenamefont {Eckstein},\ and\ \citenamefont
  {Werner}}]{Murakami2018}%
  \BibitemOpen
  \bibfield  {author} {\bibinfo {author} {\bibfnamefont {Y.}~\bibnamefont
  {Murakami}}, \bibinfo {author} {\bibfnamefont {M.}~\bibnamefont {Eckstein}},\
  and\ \bibinfo {author} {\bibfnamefont {P.}~\bibnamefont {Werner}},\
  }\bibfield  {title} {\bibinfo {title} {High-harmonic generation in {{M}}ott
  insulators},\ }\href {https://doi.org/10.1103/PhysRevLett.121.057405}
  {\bibfield  {journal} {\bibinfo  {journal} {Phys. Rev. Lett.}\ }\textbf
  {\bibinfo {volume} {121}},\ \bibinfo {pages} {057405} (\bibinfo {year}
  {2018})}\BibitemShut {NoStop}%
\bibitem [{\citenamefont {Murakami}\ and\ \citenamefont
  {Werner}(2018)}]{Murakami2018_2}%
  \BibitemOpen
  \bibfield  {author} {\bibinfo {author} {\bibfnamefont {Y.}~\bibnamefont
  {Murakami}}\ and\ \bibinfo {author} {\bibfnamefont {P.}~\bibnamefont
  {Werner}},\ }\bibfield  {title} {\bibinfo {title} {Nonequilibrium steady
  states of electric field driven $\text{M}$ott insulators},\ }\href
  {https://doi.org/10.1103/PhysRevB.98.075102} {\bibfield  {journal} {\bibinfo
  {journal} {Phys. Rev. B}\ }\textbf {\bibinfo {volume} {98}},\ \bibinfo
  {pages} {075102} (\bibinfo {year} {2018})}\BibitemShut {NoStop}%
\bibitem [{\citenamefont {Hansen}\ \emph {et~al.}(2022)\citenamefont {Hansen},
  \citenamefont {Jensen},\ and\ \citenamefont {Madsen}}]{Hansen22}%
  \BibitemOpen
  \bibfield  {author} {\bibinfo {author} {\bibfnamefont {T.}~\bibnamefont
  {Hansen}}, \bibinfo {author} {\bibfnamefont {S.~V.~B.}\ \bibnamefont
  {Jensen}},\ and\ \bibinfo {author} {\bibfnamefont {L.~B.}\ \bibnamefont
  {Madsen}},\ }\bibfield  {title} {\bibinfo {title} {Correlation effects in
  high-order harmonic generation from finite systems},\ }\href
  {https://doi.org/10.1103/PhysRevA.105.053118} {\bibfield  {journal} {\bibinfo
   {journal} {Phys. Rev. A}\ }\textbf {\bibinfo {volume} {105}},\ \bibinfo
  {pages} {053118} (\bibinfo {year} {2022})}\BibitemShut {NoStop}%
\bibitem [{\citenamefont {Hansen}\ and\ \citenamefont
  {Madsen}(2022)}]{Hansen22_2}%
  \BibitemOpen
  \bibfield  {author} {\bibinfo {author} {\bibfnamefont {T.}~\bibnamefont
  {Hansen}}\ and\ \bibinfo {author} {\bibfnamefont {L.~B.}\ \bibnamefont
  {Madsen}},\ }\bibfield  {title} {\bibinfo {title} {Doping effects in
  high-harmonic generation from correlated systems},\ }\href
  {https://doi.org/10.1103/PhysRevB.106.235142} {\bibfield  {journal} {\bibinfo
   {journal} {Phys. Rev. B}\ }\textbf {\bibinfo {volume} {106}},\ \bibinfo
  {pages} {235142} (\bibinfo {year} {2022})}\BibitemShut {NoStop}%
\bibitem [{\citenamefont {Murakami}\ \emph {et~al.}(2021)\citenamefont
  {Murakami}, \citenamefont {Takayoshi}, \citenamefont {Koga},\ and\
  \citenamefont {Werner}}]{Murakami2021}%
  \BibitemOpen
  \bibfield  {author} {\bibinfo {author} {\bibfnamefont {Y.}~\bibnamefont
  {Murakami}}, \bibinfo {author} {\bibfnamefont {S.}~\bibnamefont {Takayoshi}},
  \bibinfo {author} {\bibfnamefont {A.}~\bibnamefont {Koga}},\ and\ \bibinfo
  {author} {\bibfnamefont {P.}~\bibnamefont {Werner}},\ }\bibfield  {title}
  {\bibinfo {title} {High-harmonic generation in one-dimensional {{M}}ott
  insulators},\ }\href {https://doi.org/10.1103/PhysRevB.103.035110} {\bibfield
   {journal} {\bibinfo  {journal} {Phys. Rev. B}\ }\textbf {\bibinfo {volume}
  {103}},\ \bibinfo {pages} {035110} (\bibinfo {year} {2021})}\BibitemShut
  {NoStop}%
\bibitem [{\citenamefont {Lysne}\ \emph {et~al.}(2020)\citenamefont {Lysne},
  \citenamefont {Murakami},\ and\ \citenamefont {Werner}}]{Lysne2020}%
  \BibitemOpen
  \bibfield  {author} {\bibinfo {author} {\bibfnamefont {M.}~\bibnamefont
  {Lysne}}, \bibinfo {author} {\bibfnamefont {Y.}~\bibnamefont {Murakami}},\
  and\ \bibinfo {author} {\bibfnamefont {P.}~\bibnamefont {Werner}},\
  }\bibfield  {title} {\bibinfo {title} {Signatures of bosonic excitations in
  high-harmonic spectra of $\text{M}$ott insulators},\ }\href
  {https://doi.org/10.1103/PhysRevB.101.195139} {\bibfield  {journal} {\bibinfo
   {journal} {Phys. Rev. B}\ }\textbf {\bibinfo {volume} {101}},\ \bibinfo
  {pages} {195139} (\bibinfo {year} {2020})}\BibitemShut {NoStop}%
\bibitem [{\citenamefont {Tancogne-Dejean}\ \emph {et~al.}(2018)\citenamefont
  {Tancogne-Dejean}, \citenamefont {Sentef},\ and\ \citenamefont
  {Rubio}}]{Tancogne-Dejean2018}%
  \BibitemOpen
  \bibfield  {author} {\bibinfo {author} {\bibfnamefont {N.}~\bibnamefont
  {Tancogne-Dejean}}, \bibinfo {author} {\bibfnamefont {M.~A.}\ \bibnamefont
  {Sentef}},\ and\ \bibinfo {author} {\bibfnamefont {A.}~\bibnamefont
  {Rubio}},\ }\bibfield  {title} {\bibinfo {title} {Ultrafast modification of
  hubbard ${{U}}$ in a strongly correlated material: Ab initio high-harmonic
  generation in {{N}}i{{O}}},\ }\href
  {https://doi.org/10.1103/PhysRevLett.121.097402} {\bibfield  {journal}
  {\bibinfo  {journal} {Phys. Rev. Lett.}\ }\textbf {\bibinfo {volume} {121}},\
  \bibinfo {pages} {097402} (\bibinfo {year} {2018})}\BibitemShut {NoStop}%
\bibitem [{\citenamefont {Imai}\ \emph {et~al.}(2020)\citenamefont {Imai},
  \citenamefont {Ono},\ and\ \citenamefont {Ishihara}}]{Imai2020}%
  \BibitemOpen
  \bibfield  {author} {\bibinfo {author} {\bibfnamefont {S.}~\bibnamefont
  {Imai}}, \bibinfo {author} {\bibfnamefont {A.}~\bibnamefont {Ono}},\ and\
  \bibinfo {author} {\bibfnamefont {S.}~\bibnamefont {Ishihara}},\ }\bibfield
  {title} {\bibinfo {title} {High harmonic generation in a correlated electron
  system},\ }\href {https://doi.org/10.1103/PhysRevLett.124.157404} {\bibfield
  {journal} {\bibinfo  {journal} {Phys. Rev. Lett.}\ }\textbf {\bibinfo
  {volume} {124}},\ \bibinfo {pages} {157404} (\bibinfo {year}
  {2020})}\BibitemShut {NoStop}%
\bibitem [{\citenamefont {Chinzei}\ and\ \citenamefont
  {Ikeda}(2020)}]{Chinzei2020}%
  \BibitemOpen
  \bibfield  {author} {\bibinfo {author} {\bibfnamefont {K.}~\bibnamefont
  {Chinzei}}\ and\ \bibinfo {author} {\bibfnamefont {T.~N.}\ \bibnamefont
  {Ikeda}},\ }\bibfield  {title} {\bibinfo {title} {Disorder effects on the
  origin of high-order harmonic generation in solids},\ }\href
  {https://doi.org/10.1103/PhysRevResearch.2.013033} {\bibfield  {journal}
  {\bibinfo  {journal} {Phys. Rev. Research}\ }\textbf {\bibinfo {volume}
  {2}},\ \bibinfo {pages} {013033} (\bibinfo {year} {2020})}\BibitemShut
  {NoStop}%
\bibitem [{\citenamefont {Orthodoxou}\ \emph {et~al.}(2021)\citenamefont
  {Orthodoxou}, \citenamefont {Za{\"i}r},\ and\ \citenamefont
  {Booth}}]{Orthodoxou2021}%
  \BibitemOpen
  \bibfield  {author} {\bibinfo {author} {\bibfnamefont {C.}~\bibnamefont
  {Orthodoxou}}, \bibinfo {author} {\bibfnamefont {A.}~\bibnamefont
  {Za{\"i}r}},\ and\ \bibinfo {author} {\bibfnamefont {G.~H.}\ \bibnamefont
  {Booth}},\ }\bibfield  {title} {\bibinfo {title} {High harmonic generation in
  two-dimensional $\text{M}$ott insulators},\ }\href
  {https://doi.org/10.1038/s41535-021-00377-8} {\bibfield  {journal} {\bibinfo
  {journal} {npj Quantum Materials}\ }\textbf {\bibinfo {volume} {6}},\
  \bibinfo {pages} {76} (\bibinfo {year} {2021})}\BibitemShut {NoStop}%
\bibitem [{\citenamefont {Shao}\ \emph {et~al.}(2022)\citenamefont {Shao},
  \citenamefont {Lu}, \citenamefont {Zhang}, \citenamefont {Yu}, \citenamefont
  {Tohyama},\ and\ \citenamefont {Lu}}]{Shao2022}%
  \BibitemOpen
  \bibfield  {author} {\bibinfo {author} {\bibfnamefont {C.}~\bibnamefont
  {Shao}}, \bibinfo {author} {\bibfnamefont {H.}~\bibnamefont {Lu}}, \bibinfo
  {author} {\bibfnamefont {X.}~\bibnamefont {Zhang}}, \bibinfo {author}
  {\bibfnamefont {C.}~\bibnamefont {Yu}}, \bibinfo {author} {\bibfnamefont
  {T.}~\bibnamefont {Tohyama}},\ and\ \bibinfo {author} {\bibfnamefont
  {R.}~\bibnamefont {Lu}},\ }\bibfield  {title} {\bibinfo {title}
  {High-harmonic generation approaching the quantum critical point of strongly
  correlated systems},\ }\href {https://doi.org/10.1103/PhysRevLett.128.047401}
  {\bibfield  {journal} {\bibinfo  {journal} {Phys. Rev. Lett.}\ }\textbf
  {\bibinfo {volume} {128}},\ \bibinfo {pages} {047401} (\bibinfo {year}
  {2022})}\BibitemShut {NoStop}%
\bibitem [{\citenamefont {Masur}\ \emph {et~al.}(2022)\citenamefont {Masur},
  \citenamefont {Bondar},\ and\ \citenamefont {McCaul}}]{Masur2022}%
  \BibitemOpen
  \bibfield  {author} {\bibinfo {author} {\bibfnamefont {J.}~\bibnamefont
  {Masur}}, \bibinfo {author} {\bibfnamefont {D.~I.}\ \bibnamefont {Bondar}},\
  and\ \bibinfo {author} {\bibfnamefont {G.}~\bibnamefont {McCaul}},\
  }\bibfield  {title} {\bibinfo {title} {Optical distinguishability of
  $\text{M}$ott insulators in the time versus frequency domain},\ }\href
  {https://doi.org/10.1103/PhysRevA.106.013110} {\bibfield  {journal} {\bibinfo
   {journal} {Phys. Rev. A}\ }\textbf {\bibinfo {volume} {106}},\ \bibinfo
  {pages} {013110} (\bibinfo {year} {2022})}\BibitemShut {NoStop}%
\bibitem [{\citenamefont {Udono}\ \emph {et~al.}(2022)\citenamefont {Udono},
  \citenamefont {Sugimoto}, \citenamefont {Kaneko},\ and\ \citenamefont
  {Ohta}}]{Udono22}%
  \BibitemOpen
  \bibfield  {author} {\bibinfo {author} {\bibfnamefont {M.}~\bibnamefont
  {Udono}}, \bibinfo {author} {\bibfnamefont {K.}~\bibnamefont {Sugimoto}},
  \bibinfo {author} {\bibfnamefont {T.}~\bibnamefont {Kaneko}},\ and\ \bibinfo
  {author} {\bibfnamefont {Y.}~\bibnamefont {Ohta}},\ }\bibfield  {title}
  {\bibinfo {title} {Excitonic effects on high-harmonic generation in
  $\text{M}$ott insulators},\ }\href
  {https://doi.org/10.1103/PhysRevB.105.L241108} {\bibfield  {journal}
  {\bibinfo  {journal} {Phys. Rev. B}\ }\textbf {\bibinfo {volume} {105}},\
  \bibinfo {pages} {L241108} (\bibinfo {year} {2022})}\BibitemShut {NoStop}%
\bibitem [{\citenamefont {Murakami}\ \emph {et~al.}(2022)\citenamefont
  {Murakami}, \citenamefont {Uchida}, \citenamefont {Koga}, \citenamefont
  {Tanaka},\ and\ \citenamefont {Werner}}]{Murakami2022}%
  \BibitemOpen
  \bibfield  {author} {\bibinfo {author} {\bibfnamefont {Y.}~\bibnamefont
  {Murakami}}, \bibinfo {author} {\bibfnamefont {K.}~\bibnamefont {Uchida}},
  \bibinfo {author} {\bibfnamefont {A.}~\bibnamefont {Koga}}, \bibinfo {author}
  {\bibfnamefont {K.}~\bibnamefont {Tanaka}},\ and\ \bibinfo {author}
  {\bibfnamefont {P.}~\bibnamefont {Werner}},\ }\bibfield  {title} {\bibinfo
  {title} {Anomalous temperature dependence of high-harmonic generation in
  $\text{M}$ott insulators},\ }\href
  {https://doi.org/10.1103/PhysRevLett.129.157401} {\bibfield  {journal}
  {\bibinfo  {journal} {Phys. Rev. Lett.}\ }\textbf {\bibinfo {volume} {129}},\
  \bibinfo {pages} {157401} (\bibinfo {year} {2022})}\BibitemShut {NoStop}%
\bibitem [{\citenamefont {Uchida}\ \emph {et~al.}(2022)\citenamefont {Uchida},
  \citenamefont {Mattoni}, \citenamefont {Yonezawa}, \citenamefont {Nakamura},
  \citenamefont {Maeno},\ and\ \citenamefont {Tanaka}}]{Uchida2022}%
  \BibitemOpen
  \bibfield  {author} {\bibinfo {author} {\bibfnamefont {K.}~\bibnamefont
  {Uchida}}, \bibinfo {author} {\bibfnamefont {G.}~\bibnamefont {Mattoni}},
  \bibinfo {author} {\bibfnamefont {S.}~\bibnamefont {Yonezawa}}, \bibinfo
  {author} {\bibfnamefont {F.}~\bibnamefont {Nakamura}}, \bibinfo {author}
  {\bibfnamefont {Y.}~\bibnamefont {Maeno}},\ and\ \bibinfo {author}
  {\bibfnamefont {K.}~\bibnamefont {Tanaka}},\ }\bibfield  {title} {\bibinfo
  {title} {High-order harmonic generation and its unconventional scaling law in
  the $\text{M}$ott-insulating $\text{Ca}_{2}\text{RuO}_{4}$},\ }\href
  {https://doi.org/10.1103/PhysRevLett.128.127401} {\bibfield  {journal}
  {\bibinfo  {journal} {Phys. Rev. Lett.}\ }\textbf {\bibinfo {volume} {128}},\
  \bibinfo {pages} {127401} (\bibinfo {year} {2022})}\BibitemShut {NoStop}%
\bibitem [{\citenamefont {Grånäs}\ \emph {et~al.}(2020)\citenamefont
  {Grånäs}, \citenamefont {Vaskivsky}, \citenamefont {Wang}, \citenamefont
  {Thunström}, \citenamefont {Ghimire}, \citenamefont {Knut}, \citenamefont
  {Söderström}, \citenamefont {Kjellsson}, \citenamefont {Turenne},
  \citenamefont {Engel}, \citenamefont {Beye}, \citenamefont {Lu},
  \citenamefont {Reid}, \citenamefont {Schlotter}, \citenamefont {Coslovich},
  \citenamefont {Hoffmann}, \citenamefont {Kolesov}, \citenamefont
  {Schüßler-Langeheine}, \citenamefont {Styervoyedov}, \citenamefont
  {Tancogne-Dejean}, \citenamefont {Sentef}, \citenamefont {Reis},
  \citenamefont {Rubio}, \citenamefont {Parkin}, \citenamefont {Karis},
  \citenamefont {Nordgren}, \citenamefont {Rubensson}, \citenamefont
  {Eriksson},\ and\ \citenamefont {Dürr}}]{Granas2022}%
  \BibitemOpen
  \bibfield  {author} {\bibinfo {author} {\bibfnamefont {O.}~\bibnamefont
  {Grånäs}}, \bibinfo {author} {\bibfnamefont {I.}~\bibnamefont {Vaskivsky}},
  \bibinfo {author} {\bibfnamefont {X.}~\bibnamefont {Wang}}, \bibinfo {author}
  {\bibfnamefont {P.}~\bibnamefont {Thunström}}, \bibinfo {author}
  {\bibfnamefont {S.}~\bibnamefont {Ghimire}}, \bibinfo {author} {\bibfnamefont
  {R.}~\bibnamefont {Knut}}, \bibinfo {author} {\bibfnamefont {J.}~\bibnamefont
  {Söderström}}, \bibinfo {author} {\bibfnamefont {L.}~\bibnamefont
  {Kjellsson}}, \bibinfo {author} {\bibfnamefont {D.}~\bibnamefont {Turenne}},
  \bibinfo {author} {\bibfnamefont {R.~Y.}\ \bibnamefont {Engel}}, \bibinfo
  {author} {\bibfnamefont {M.}~\bibnamefont {Beye}}, \bibinfo {author}
  {\bibfnamefont {J.}~\bibnamefont {Lu}}, \bibinfo {author} {\bibfnamefont
  {A.~H.}\ \bibnamefont {Reid}}, \bibinfo {author} {\bibfnamefont
  {W.}~\bibnamefont {Schlotter}}, \bibinfo {author} {\bibfnamefont
  {G.}~\bibnamefont {Coslovich}}, \bibinfo {author} {\bibfnamefont
  {M.}~\bibnamefont {Hoffmann}}, \bibinfo {author} {\bibfnamefont
  {G.}~\bibnamefont {Kolesov}}, \bibinfo {author} {\bibfnamefont
  {C.}~\bibnamefont {Schüßler-Langeheine}}, \bibinfo {author} {\bibfnamefont
  {A.}~\bibnamefont {Styervoyedov}}, \bibinfo {author} {\bibfnamefont
  {N.}~\bibnamefont {Tancogne-Dejean}}, \bibinfo {author} {\bibfnamefont
  {M.~A.}\ \bibnamefont {Sentef}}, \bibinfo {author} {\bibfnamefont {D.~A.}\
  \bibnamefont {Reis}}, \bibinfo {author} {\bibfnamefont {A.}~\bibnamefont
  {Rubio}}, \bibinfo {author} {\bibfnamefont {S.~S.~P.}\ \bibnamefont
  {Parkin}}, \bibinfo {author} {\bibfnamefont {O.}~\bibnamefont {Karis}},
  \bibinfo {author} {\bibfnamefont {J.}~\bibnamefont {Nordgren}}, \bibinfo
  {author} {\bibfnamefont {J.~E.}\ \bibnamefont {Rubensson}}, \bibinfo {author}
  {\bibfnamefont {O.}~\bibnamefont {Eriksson}},\ and\ \bibinfo {author}
  {\bibfnamefont {H.~A.}\ \bibnamefont {Dürr}},\ }\href
  {https://doi.org/10.48550/ARXIV.2008.11115} {\bibinfo {title} {Ultrafast
  modification of the electronic structure of a correlated insulator}}
  (\bibinfo {year} {2020})\BibitemShut {NoStop}%
\bibitem [{\citenamefont {Bionta}\ \emph {et~al.}(2021)\citenamefont {Bionta},
  \citenamefont {Haddad}, \citenamefont {Leblanc}, \citenamefont {Gruson},
  \citenamefont {Lassonde}, \citenamefont {Ibrahim}, \citenamefont {Chaillou},
  \citenamefont {\'Emond}, \citenamefont {Otto}, \citenamefont
  {Jim\'enez-Gal\'an}, \citenamefont {Silva}, \citenamefont {Ivanov},
  \citenamefont {Siwick}, \citenamefont {Chaker},\ and\ \citenamefont
  {L\'egar\'e}}]{Bionta2021}%
  \BibitemOpen
  \bibfield  {author} {\bibinfo {author} {\bibfnamefont {M.~R.}\ \bibnamefont
  {Bionta}}, \bibinfo {author} {\bibfnamefont {E.}~\bibnamefont {Haddad}},
  \bibinfo {author} {\bibfnamefont {A.}~\bibnamefont {Leblanc}}, \bibinfo
  {author} {\bibfnamefont {V.}~\bibnamefont {Gruson}}, \bibinfo {author}
  {\bibfnamefont {P.}~\bibnamefont {Lassonde}}, \bibinfo {author}
  {\bibfnamefont {H.}~\bibnamefont {Ibrahim}}, \bibinfo {author} {\bibfnamefont
  {J.}~\bibnamefont {Chaillou}}, \bibinfo {author} {\bibfnamefont
  {N.}~\bibnamefont {\'Emond}}, \bibinfo {author} {\bibfnamefont {M.~R.}\
  \bibnamefont {Otto}}, \bibinfo {author} {\bibfnamefont {A.}~\bibnamefont
  {Jim\'enez-Gal\'an}}, \bibinfo {author} {\bibfnamefont {R.~E.~F.}\
  \bibnamefont {Silva}}, \bibinfo {author} {\bibfnamefont {M.}~\bibnamefont
  {Ivanov}}, \bibinfo {author} {\bibfnamefont {B.~J.}\ \bibnamefont {Siwick}},
  \bibinfo {author} {\bibfnamefont {M.}~\bibnamefont {Chaker}},\ and\ \bibinfo
  {author} {\bibfnamefont {F.}~\bibnamefont {L\'egar\'e}},\ }\bibfield  {title}
  {\bibinfo {title} {Tracking ultrafast solid-state dynamics using high
  harmonic spectroscopy},\ }\href
  {https://doi.org/10.1103/PhysRevResearch.3.023250} {\bibfield  {journal}
  {\bibinfo  {journal} {Phys. Rev. Research}\ }\textbf {\bibinfo {volume}
  {3}},\ \bibinfo {pages} {023250} (\bibinfo {year} {2021})}\BibitemShut
  {NoStop}%
\bibitem [{\citenamefont {Sloan}\ \emph {et~al.}(2023)\citenamefont {Sloan},
  \citenamefont {Gorlach}, \citenamefont {Tzur}, \citenamefont {Rivera},
  \citenamefont {Cohen}, \citenamefont {Kaminer},\ and\ \citenamefont
  {Soljačić}}]{Sloan2023entangling}%
  \BibitemOpen
  \bibfield  {author} {\bibinfo {author} {\bibfnamefont {J.}~\bibnamefont
  {Sloan}}, \bibinfo {author} {\bibfnamefont {A.}~\bibnamefont {Gorlach}},
  \bibinfo {author} {\bibfnamefont {M.~E.}\ \bibnamefont {Tzur}}, \bibinfo
  {author} {\bibfnamefont {N.}~\bibnamefont {Rivera}}, \bibinfo {author}
  {\bibfnamefont {O.}~\bibnamefont {Cohen}}, \bibinfo {author} {\bibfnamefont
  {I.}~\bibnamefont {Kaminer}},\ and\ \bibinfo {author} {\bibfnamefont
  {M.}~\bibnamefont {Soljačić}},\ }\href@noop {} {\bibinfo {title}
  {Entangling extreme ultraviolet photons through strong field pair
  generation}} (\bibinfo {year} {2023}),\ \Eprint
  {https://arxiv.org/abs/2309.16466} {arXiv:2309.16466 [quant-ph]} \BibitemShut
  {NoStop}%
\bibitem [{\citenamefont {Essler}\ \emph {et~al.}(2005)\citenamefont {Essler},
  \citenamefont {Frahm}, \citenamefont {Göhmann}, \citenamefont {Klümper},\
  and\ \citenamefont {Korepin}}]{Hubbard_Essler}%
  \BibitemOpen
  \bibfield  {author} {\bibinfo {author} {\bibfnamefont {F.~H.~L.}\
  \bibnamefont {Essler}}, \bibinfo {author} {\bibfnamefont {H.}~\bibnamefont
  {Frahm}}, \bibinfo {author} {\bibfnamefont {F.}~\bibnamefont {Göhmann}},
  \bibinfo {author} {\bibfnamefont {A.}~\bibnamefont {Klümper}},\ and\
  \bibinfo {author} {\bibfnamefont {V.~E.}\ \bibnamefont {Korepin}},\ }\href
  {https://doi.org/10.1017/CBO9780511534843} {\emph {\bibinfo {title} {The
  One-Dimensional {{H}}ubbard Model}}}\ (\bibinfo  {publisher} {Cambridge
  University Press},\ \bibinfo {address} {Cambridge},\ \bibinfo {year}
  {2005})\BibitemShut {NoStop}%
\bibitem [{\citenamefont {Lee}\ \emph {et~al.}(2006)\citenamefont {Lee},
  \citenamefont {Nagaosa},\ and\ \citenamefont {Wen}}]{Lee06}%
  \BibitemOpen
  \bibfield  {author} {\bibinfo {author} {\bibfnamefont {P.~A.}\ \bibnamefont
  {Lee}}, \bibinfo {author} {\bibfnamefont {N.}~\bibnamefont {Nagaosa}},\ and\
  \bibinfo {author} {\bibfnamefont {X.-G.}\ \bibnamefont {Wen}},\ }\bibfield
  {title} {\bibinfo {title} {Doping a $\text{M}$ott insulator: Physics of
  high-temperature superconductivity},\ }\href
  {https://doi.org/10.1103/RevModPhys.78.17} {\bibfield  {journal} {\bibinfo
  {journal} {Rev. Mod. Phys.}\ }\textbf {\bibinfo {volume} {78}},\ \bibinfo
  {pages} {17} (\bibinfo {year} {2006})}\BibitemShut {NoStop}%
\bibitem [{\citenamefont {Imada}\ \emph {et~al.}(1998)\citenamefont {Imada},
  \citenamefont {Fujimori},\ and\ \citenamefont {Tokura}}]{Imada98}%
  \BibitemOpen
  \bibfield  {author} {\bibinfo {author} {\bibfnamefont {M.}~\bibnamefont
  {Imada}}, \bibinfo {author} {\bibfnamefont {A.}~\bibnamefont {Fujimori}},\
  and\ \bibinfo {author} {\bibfnamefont {Y.}~\bibnamefont {Tokura}},\
  }\bibfield  {title} {\bibinfo {title} {Metal-insulator transitions},\ }\href
  {https://doi.org/10.1103/RevModPhys.70.1039} {\bibfield  {journal} {\bibinfo
  {journal} {Rev. Mod. Phys.}\ }\textbf {\bibinfo {volume} {70}},\ \bibinfo
  {pages} {1039} (\bibinfo {year} {1998})}\BibitemShut {NoStop}%
\bibitem [{\citenamefont {Mollow}(1975)}]{Mollow1975}%
  \BibitemOpen
  \bibfield  {author} {\bibinfo {author} {\bibfnamefont {B.~R.}\ \bibnamefont
  {Mollow}},\ }\bibfield  {title} {\bibinfo {title} {Pure-state analysis of
  resonant light scattering: Radiative damping, saturation, and multiphoton
  effects},\ }\href {https://doi.org/10.1103/PhysRevA.12.1919} {\bibfield
  {journal} {\bibinfo  {journal} {Phys. Rev. A}\ }\textbf {\bibinfo {volume}
  {12}},\ \bibinfo {pages} {1919} (\bibinfo {year} {1975})}\BibitemShut
  {NoStop}%
\bibitem [{\citenamefont {{Cohen-Tannoudji}}\ \emph {et~al.}(1998)\citenamefont
  {{Cohen-Tannoudji}}, \citenamefont {{Grynberg}},\ and\ \citenamefont
  {{Dupont-Roc}}}]{Cohen_Tannoudji1998_atomphoton}%
  \BibitemOpen
  \bibfield  {author} {\bibinfo {author} {\bibfnamefont {C.}~\bibnamefont
  {{Cohen-Tannoudji}}}, \bibinfo {author} {\bibfnamefont {G.}~\bibnamefont
  {{Grynberg}}},\ and\ \bibinfo {author} {\bibfnamefont {J.}~\bibnamefont
  {{Dupont-Roc}}},\ }\href@noop {} {\emph {\bibinfo {title} {Atom-Photon
  Interactions: Basic Processes and Applications}}}\ (\bibinfo  {publisher}
  {Wiley},\ \bibinfo {address} {New York},\ \bibinfo {year} {1998})\BibitemShut
  {NoStop}%
\bibitem [{\citenamefont {Tomita}\ and\ \citenamefont
  {Nasu}(2001)}]{Tomita2001}%
  \BibitemOpen
  \bibfield  {author} {\bibinfo {author} {\bibfnamefont {N.}~\bibnamefont
  {Tomita}}\ and\ \bibinfo {author} {\bibfnamefont {K.}~\bibnamefont {Nasu}},\
  }\bibfield  {title} {\bibinfo {title} {Quantum fluctuation effects on light
  absorption spectra of the one-dimensional extended {{H}}ubbard model},\
  }\href {https://doi.org/10.1103/PhysRevB.63.085107} {\bibfield  {journal}
  {\bibinfo  {journal} {Phys. Rev. B}\ }\textbf {\bibinfo {volume} {63}},\
  \bibinfo {pages} {085107} (\bibinfo {year} {2001})}\BibitemShut {NoStop}%
\bibitem [{\citenamefont {Park}\ and\ \citenamefont {Light}(1986)}]{Park1986}%
  \BibitemOpen
  \bibfield  {author} {\bibinfo {author} {\bibfnamefont {T.~J.}\ \bibnamefont
  {Park}}\ and\ \bibinfo {author} {\bibfnamefont {J.~C.}\ \bibnamefont
  {Light}},\ }\bibfield  {title} {\bibinfo {title} {Unitary quantum time
  evolution by iterative $\text{L}$anczos reduction},\ }\href
  {https://doi.org/10.1063/1.451548} {\bibfield  {journal} {\bibinfo  {journal}
  {J. Chem. Phys.}\ }\textbf {\bibinfo {volume} {85}},\ \bibinfo {pages} {5870}
  (\bibinfo {year} {1986})}\BibitemShut {NoStop}%
\bibitem [{\citenamefont {Smyth}\ \emph {et~al.}(1998)\citenamefont {Smyth},
  \citenamefont {Parker},\ and\ \citenamefont {Taylor}}]{Smyth1998}%
  \BibitemOpen
  \bibfield  {author} {\bibinfo {author} {\bibfnamefont {E.~S.}\ \bibnamefont
  {Smyth}}, \bibinfo {author} {\bibfnamefont {J.~S.}\ \bibnamefont {Parker}},\
  and\ \bibinfo {author} {\bibfnamefont {K.}~\bibnamefont {Taylor}},\
  }\bibfield  {title} {\bibinfo {title} {Numerical integration of the
  time-dependent $\text{S}$chrödinger equation for laser-driven helium},\
  }\href {https://doi.org/https://doi.org/10.1016/S0010-4655(98)00083-6}
  {\bibfield  {journal} {\bibinfo  {journal} {Comput. Phys. Commun.}\ }\textbf
  {\bibinfo {volume} {114}},\ \bibinfo {pages} {1} (\bibinfo {year}
  {1998})}\BibitemShut {NoStop}%
\bibitem [{\citenamefont {Guan}\ \emph {et~al.}(2007)\citenamefont {Guan},
  \citenamefont {Zatsarinny}, \citenamefont {Bartschat}, \citenamefont
  {Schneider}, \citenamefont {Feist},\ and\ \citenamefont {Noble}}]{Guan2007}%
  \BibitemOpen
  \bibfield  {author} {\bibinfo {author} {\bibfnamefont {X.}~\bibnamefont
  {Guan}}, \bibinfo {author} {\bibfnamefont {O.}~\bibnamefont {Zatsarinny}},
  \bibinfo {author} {\bibfnamefont {K.}~\bibnamefont {Bartschat}}, \bibinfo
  {author} {\bibfnamefont {B.~I.}\ \bibnamefont {Schneider}}, \bibinfo {author}
  {\bibfnamefont {J.}~\bibnamefont {Feist}},\ and\ \bibinfo {author}
  {\bibfnamefont {C.~J.}\ \bibnamefont {Noble}},\ }\bibfield  {title} {\bibinfo
  {title} {General approach to few-cycle intense laser interactions with
  complex atoms},\ }\href {https://doi.org/10.1103/PhysRevA.76.053411}
  {\bibfield  {journal} {\bibinfo  {journal} {Phys. Rev. A}\ }\textbf {\bibinfo
  {volume} {76}},\ \bibinfo {pages} {053411} (\bibinfo {year}
  {2007})}\BibitemShut {NoStop}%
\bibitem [{\citenamefont {Frapiccini}\ \emph {et~al.}(2014)\citenamefont
  {Frapiccini}, \citenamefont {Hamido}, \citenamefont {Schr\"oter},
  \citenamefont {Pyke}, \citenamefont {Mota-Furtado}, \citenamefont {O'Mahony},
  \citenamefont {Madro\~nero}, \citenamefont {Eiglsperger},\ and\ \citenamefont
  {Piraux}}]{Frapiccini2014}%
  \BibitemOpen
  \bibfield  {author} {\bibinfo {author} {\bibfnamefont {A.~L.}\ \bibnamefont
  {Frapiccini}}, \bibinfo {author} {\bibfnamefont {A.}~\bibnamefont {Hamido}},
  \bibinfo {author} {\bibfnamefont {S.}~\bibnamefont {Schr\"oter}}, \bibinfo
  {author} {\bibfnamefont {D.}~\bibnamefont {Pyke}}, \bibinfo {author}
  {\bibfnamefont {F.}~\bibnamefont {Mota-Furtado}}, \bibinfo {author}
  {\bibfnamefont {P.~F.}\ \bibnamefont {O'Mahony}}, \bibinfo {author}
  {\bibfnamefont {J.}~\bibnamefont {Madro\~nero}}, \bibinfo {author}
  {\bibfnamefont {J.}~\bibnamefont {Eiglsperger}},\ and\ \bibinfo {author}
  {\bibfnamefont {B.}~\bibnamefont {Piraux}},\ }\bibfield  {title} {\bibinfo
  {title} {Explicit schemes for time propagating many-body wave functions},\
  }\href {https://doi.org/10.1103/PhysRevA.89.023418} {\bibfield  {journal}
  {\bibinfo  {journal} {Phys. Rev. A}\ }\textbf {\bibinfo {volume} {89}},\
  \bibinfo {pages} {023418} (\bibinfo {year} {2014})}\BibitemShut {NoStop}%
\bibitem [{\citenamefont {Baggesen}\ and\ \citenamefont
  {Madsen}(2011)}]{Baggesen2011}%
  \BibitemOpen
  \bibfield  {author} {\bibinfo {author} {\bibfnamefont {J.~C.}\ \bibnamefont
  {Baggesen}}\ and\ \bibinfo {author} {\bibfnamefont {L.~B.}\ \bibnamefont
  {Madsen}},\ }\bibfield  {title} {\bibinfo {title} {On the dipole, velocity
  and acceleration forms in high-order harmonic generation from a single atom
  or molecule},\ }\href {https://doi.org/10.1088/0953-4075/44/11/115601}
  {\bibfield  {journal} {\bibinfo  {journal} {J. Phys. B}\ }\textbf {\bibinfo
  {volume} {44}},\ \bibinfo {pages} {115601} (\bibinfo {year}
  {2011})}\BibitemShut {NoStop}%
\bibitem [{\citenamefont {Gerry}\ and\ \citenamefont
  {Knight}(2004)}]{Gerry_knight_2004}%
  \BibitemOpen
  \bibfield  {author} {\bibinfo {author} {\bibfnamefont {C.}~\bibnamefont
  {Gerry}}\ and\ \bibinfo {author} {\bibfnamefont {P.}~\bibnamefont {Knight}},\
  }\href {https://doi.org/10.1017/CBO9780511791239} {\emph {\bibinfo {title}
  {Introductory Quantum Optics}}}\ (\bibinfo  {publisher} {Cambridge University
  Press},\ \bibinfo {year} {2004})\BibitemShut {NoStop}%
\bibitem [{\citenamefont {Scully}\ and\ \citenamefont
  {Zubairy}(1997)}]{Scully_zubairy_1997}%
  \BibitemOpen
  \bibfield  {author} {\bibinfo {author} {\bibfnamefont {M.~O.}\ \bibnamefont
  {Scully}}\ and\ \bibinfo {author} {\bibfnamefont {M.~S.}\ \bibnamefont
  {Zubairy}},\ }\href {https://doi.org/10.1017/CBO9780511813993} {\emph
  {\bibinfo {title} {Quantum Optics}}}\ (\bibinfo  {publisher} {Cambridge
  University Press},\ \bibinfo {year} {1997})\BibitemShut {NoStop}%
\bibitem [{\citenamefont {Braunstein}\ and\ \citenamefont {van
  Loock}(2005)}]{Braunstein2005}%
  \BibitemOpen
  \bibfield  {author} {\bibinfo {author} {\bibfnamefont {S.~L.}\ \bibnamefont
  {Braunstein}}\ and\ \bibinfo {author} {\bibfnamefont {P.}~\bibnamefont {van
  Loock}},\ }\bibfield  {title} {\bibinfo {title} {Quantum information with
  continuous variables},\ }\href {https://doi.org/10.1103/RevModPhys.77.513}
  {\bibfield  {journal} {\bibinfo  {journal} {Rev. Mod. Phys.}\ }\textbf
  {\bibinfo {volume} {77}},\ \bibinfo {pages} {513} (\bibinfo {year}
  {2005})}\BibitemShut {NoStop}%
\bibitem [{\citenamefont {Slusher}\ \emph {et~al.}(1985)\citenamefont
  {Slusher}, \citenamefont {Hollberg}, \citenamefont {Yurke}, \citenamefont
  {Mertz},\ and\ \citenamefont {Valley}}]{Slusher1985_squeezed_light_homodyne}%
  \BibitemOpen
  \bibfield  {author} {\bibinfo {author} {\bibfnamefont {R.~E.}\ \bibnamefont
  {Slusher}}, \bibinfo {author} {\bibfnamefont {L.~W.}\ \bibnamefont
  {Hollberg}}, \bibinfo {author} {\bibfnamefont {B.}~\bibnamefont {Yurke}},
  \bibinfo {author} {\bibfnamefont {J.~C.}\ \bibnamefont {Mertz}},\ and\
  \bibinfo {author} {\bibfnamefont {J.~F.}\ \bibnamefont {Valley}},\ }\bibfield
   {title} {\bibinfo {title} {Observation of squeezed states generated by
  four-wave mixing in an optical cavity},\ }\href
  {https://doi.org/10.1103/PhysRevLett.55.2409} {\bibfield  {journal} {\bibinfo
   {journal} {Phys. Rev. Lett.}\ }\textbf {\bibinfo {volume} {55}},\ \bibinfo
  {pages} {2409} (\bibinfo {year} {1985})}\BibitemShut {NoStop}%
\bibitem [{\citenamefont {Breitenbach}\ \emph {et~al.}(1997)\citenamefont
  {Breitenbach}, \citenamefont {Schiller},\ and\ \citenamefont
  {Mlynek}}]{Breitenbach1997_squeezed_light_homodyne}%
  \BibitemOpen
  \bibfield  {author} {\bibinfo {author} {\bibfnamefont {G.}~\bibnamefont
  {Breitenbach}}, \bibinfo {author} {\bibfnamefont {S.}~\bibnamefont
  {Schiller}},\ and\ \bibinfo {author} {\bibfnamefont {J.}~\bibnamefont
  {Mlynek}},\ }\bibfield  {title} {\bibinfo {title} {Measurement of the quantum
  states of squeezed light},\ }\href {https://doi.org/10.1038/387471a0}
  {\bibfield  {journal} {\bibinfo  {journal} {Nature}\ }\textbf {\bibinfo
  {volume} {387}},\ \bibinfo {pages} {471} (\bibinfo {year}
  {1997})}\BibitemShut {NoStop}%
\bibitem [{\citenamefont {Short}\ and\ \citenamefont
  {Mandel}(1983)}]{Short1983_photon_counting_experiment}%
  \BibitemOpen
  \bibfield  {author} {\bibinfo {author} {\bibfnamefont {R.}~\bibnamefont
  {Short}}\ and\ \bibinfo {author} {\bibfnamefont {L.}~\bibnamefont {Mandel}},\
  }\bibfield  {title} {\bibinfo {title} {Observation of sub-poissonian photon
  statistics},\ }\href {https://doi.org/10.1103/PhysRevLett.51.384} {\bibfield
  {journal} {\bibinfo  {journal} {Phys. Rev. Lett.}\ }\textbf {\bibinfo
  {volume} {51}},\ \bibinfo {pages} {384} (\bibinfo {year} {1983})}\BibitemShut
  {NoStop}%
\bibitem [{\citenamefont {Bartley}\ \emph {et~al.}(2013)\citenamefont
  {Bartley}, \citenamefont {Donati}, \citenamefont {Jin}, \citenamefont
  {Datta}, \citenamefont {Barbieri},\ and\ \citenamefont
  {Walmsley}}]{Bartley2013_sub_binom_light}%
  \BibitemOpen
  \bibfield  {author} {\bibinfo {author} {\bibfnamefont {T.~J.}\ \bibnamefont
  {Bartley}}, \bibinfo {author} {\bibfnamefont {G.}~\bibnamefont {Donati}},
  \bibinfo {author} {\bibfnamefont {X.-M.}\ \bibnamefont {Jin}}, \bibinfo
  {author} {\bibfnamefont {A.}~\bibnamefont {Datta}}, \bibinfo {author}
  {\bibfnamefont {M.}~\bibnamefont {Barbieri}},\ and\ \bibinfo {author}
  {\bibfnamefont {I.~A.}\ \bibnamefont {Walmsley}},\ }\bibfield  {title}
  {\bibinfo {title} {Direct observation of sub-binomial light},\ }\href
  {https://doi.org/10.1103/PhysRevLett.110.173602} {\bibfield  {journal}
  {\bibinfo  {journal} {Phys. Rev. Lett.}\ }\textbf {\bibinfo {volume} {110}},\
  \bibinfo {pages} {173602} (\bibinfo {year} {2013})}\BibitemShut {NoStop}%
\bibitem [{\citenamefont {Gebhard}(1997)}]{MMIT97}%
  \BibitemOpen
  \bibfield  {author} {\bibinfo {author} {\bibfnamefont {F.}~\bibnamefont
  {Gebhard}},\ }\href {https://link.springer.com/book/10.1007/3-540-14858-2}
  {\emph {\bibinfo {title} {The $\text{M}$ott Metal-Insulator Transition:
  Models and Methods}}},\ Vol.\ \bibinfo {volume} {137}\ (\bibinfo  {publisher}
  {Springer Berlin, Heidelberg},\ \bibinfo {year} {1997})\BibitemShut {NoStop}%
\bibitem [{\citenamefont {Maekawa}(2004)}]{TMO04}%
  \BibitemOpen
  \bibfield  {author} {\bibinfo {author} {\bibfnamefont {S.}~\bibnamefont
  {Maekawa}},\ }\href@noop {} {\emph {\bibinfo {title} {Physics of Transition
  Metal Oxides}}},\ \bibinfo {edition} {1st}\ ed.,\ Springer Series in
  Solid-State Sciences, 144\ (\bibinfo  {publisher} {Springer Berlin,
  Heidelberg},\ \bibinfo {year} {2004})\BibitemShut {NoStop}%
\bibitem [{\citenamefont {Oka}(2012)}]{Oka2012}%
  \BibitemOpen
  \bibfield  {author} {\bibinfo {author} {\bibfnamefont {T.}~\bibnamefont
  {Oka}},\ }\bibfield  {title} {\bibinfo {title} {Nonlinear doublon production
  in a mott insulator: Landau-dykhne method applied to an integrable model},\
  }\href {https://doi.org/10.1103/PhysRevB.86.075148} {\bibfield  {journal}
  {\bibinfo  {journal} {Phys. Rev. B}\ }\textbf {\bibinfo {volume} {86}},\
  \bibinfo {pages} {075148} (\bibinfo {year} {2012})}\BibitemShut {NoStop}%
\bibitem [{\citenamefont {Hansen}\ and\ \citenamefont
  {Madsen}(2023)}]{hansen2023effects_of_lattice_imperfections}%
  \BibitemOpen
  \bibfield  {author} {\bibinfo {author} {\bibfnamefont {T.}~\bibnamefont
  {Hansen}}\ and\ \bibinfo {author} {\bibfnamefont {L.~B.}\ \bibnamefont
  {Madsen}},\ }\href@noop {} {\bibinfo {title} {Effects of lattice
  imperfections on high-harmonic generation from correlated systems}} (\bibinfo
  {year} {2023}),\ \Eprint {https://arxiv.org/abs/2306.08379} {arXiv:2306.08379
  [cond-mat.str-el]} \BibitemShut {NoStop}%
\bibitem [{\citenamefont {Walmsley}(2015)}]{Walmsley2015}%
  \BibitemOpen
  \bibfield  {author} {\bibinfo {author} {\bibfnamefont {I.~A.}\ \bibnamefont
  {Walmsley}},\ }\bibfield  {title} {\bibinfo {title} {Quantum optics: Science
  and technology in a new light},\ }\href
  {https://doi.org/10.1126/science.aab0097} {\bibfield  {journal} {\bibinfo
  {journal} {Science}\ }\textbf {\bibinfo {volume} {348}},\ \bibinfo {pages}
  {525} (\bibinfo {year} {2015})},\ \Eprint
  {https://arxiv.org/abs/https://www.science.org/doi/pdf/10.1126/science.aab0097}
  {https://www.science.org/doi/pdf/10.1126/science.aab0097} \BibitemShut
  {NoStop}%
\bibitem [{\citenamefont {Mahan}(2000)}]{Mahan2000}%
  \BibitemOpen
  \bibfield  {author} {\bibinfo {author} {\bibfnamefont {G.~D.}\ \bibnamefont
  {Mahan}},\ }\href@noop {} {\emph {\bibinfo {title} {Many-Particle Physics}}}\
  (\bibinfo  {publisher} {Springer New York, New York},\ \bibinfo {year}
  {2000})\BibitemShut {NoStop}%
\end{thebibliography}%

\end{document}